\documentclass[preprintnumbers,article,amsmath,amssymb,floatfix,10pt,prd,onecolumn,
superscriptaddress,nofootinbib]{revtex4}
\usepackage{bbm}
\usepackage{amsfonts}
\usepackage{mathrsfs}
\usepackage{latexsym}
\usepackage[cp1251]{inputenc}
\usepackage{epsfig}
\usepackage{epstopdf}
\usepackage{graphicx}
\usepackage{amssymb}
\usepackage{amsmath}
\usepackage{dcolumn}
\usepackage{bm}
\usepackage{color}
\usepackage{xcolor}
\usepackage{amsmath,graphicx}
\usepackage{palatino}
\usepackage{changes}

\usepackage[colorlinks=true,
            linkcolor=blue,
            urlcolor=blue,
            citecolor=blue]{hyperref}
\begin{document}

\title{\bf Quasi Periodic Oscillations of Test Particles and Red-Blue Shifts of the Photons Emitted By the Charged Test Particles Orbiting the Charged Black Hole in the Presence of Quintessence and Clouds of Strings}

\author{G. Mustafa}
\email{gmustafa3828@gmail.com}\affiliation{Department of Mathematics, Shanghai University,
Shanghai, 200444, Shanghai, People's Republic of China}

\author{Ibrar Hussain}
\email{ibrar.hussain@seecs.nust.edu.pk}\affiliation{School of Electrical Engineering and Computer Science,
National University of Sciences and Technology, H-12, Islamabad, Pakistan}

\author{Wu-Ming Liu}
\email{wliu@iphy.ac.cn}\affiliation{Beijing National Laboratory for Condensed Matter
Physics, Institute of Physics, Chinese Academy of Sciences, Beijing 100190, China}
\affiliation{School of Physical Sciences, University of Chinese
Academy of Sciences, Beijing 100190, China}
\affiliation{Songshan Lake Materials Laboratory, Dongguan 523808,
Guangdong, China}

\begin{abstract}
Here we examine the circular motion of test particles and photons in the spacetime geometry of charged black hole surrounded by quintessence and clouds of strings for the equation of state parameter $\omega_q=-2/3$. We observe that there exist stable circular orbits in this geometry for very small values of the quintessence and string cloud parameters, i.e., $0<\gamma<<1$ and $0<\alpha<<1$. We observe that if the values of $\gamma$ and $\alpha$ increase, the test particle can more easily escape the gravitational field of the black hole. While the effect of the charge $Q$ of the black hole on the effective potential is just opposite to that of the $\gamma$ and $\alpha$. Further, we investigate the quasi-periodic oscillations of test particles near the stable circular orbits. With the increasing values of $Q$, the stable circular orbits get away from the central object; therefore, one can observe lower epicyclic frequencies away from the central gravitating source with the increase in the values of $Q$. The stable circular orbits get close to the central object as $\gamma$ and $\alpha$ increases. Thus, one can observe the epicyclic frequencies relative to the central object and would be higher as compared to the epicyclic frequencies observed in the case of pure Riessner-Nordstrom black hole, i.e., without quintessence. The redshift parameter $z$ of the photons emitted by the charged test particles moving in the stable circular orbits around the central source increases with an increase in the parameter $\alpha$ and decreases with an increase in the values of the charge $Q$. For different values of the parameter $\gamma$, it stays constant. In the Banados-Silk-West (BSW) process study, we note that the centre of mass-energy at the horizon of this Riessner-Nordstrom black hole with quintessence and string clouds increases indefinitely if the charge of one of the colliding particles attains its critical value. It is observed that the effective force on the test particles decreases as the values of the parameters $\gamma$ and $\alpha$ increase, and it increases with the increasing values of the charge $Q$. For a better understanding of the study, we show the dependence of the radii of the circular orbits, energy and angular momentum of the particles, effective potential, effective force, quasi-periodic oscillations and red-blue shifts of photons of the test particles in the circular orbits on the parameters $\alpha$, $\gamma$ and $Q$ graphically. \\

\end{abstract}

\maketitle

\date{\today}

%%%%%%%%%%%%%%%%%%%%%%%%%%%%%%%%%%%%%%%%%%%%%%%%%%%%%%%%%%%%%%%%%%%%%%
%%%%%%%%%%%%%        Introduction        %%%%%%%%%%%%%%%%%%%%%%%%%%%%%
%%%%%%%%%%%%%%%%%%%%%%%%%%%%%%%%%%%%%%%%%%%%%%%%%%%%%%%%%%%%%%%%%%%%%%
\section{Introduction}
The study of null and timelike circular geodesics is of prime interest to the researchers working in this field. This study of geodesic motion can provide a better understanding of the gravitational field and the geometric structure around massive compact objects such as neutron stars and black holes. The binary motion of black holes produces gravitational waves and recently reported by the observatories working on it \cite{1}. The recent imaging of the black hole shadow of M87*, reported by the Event Horizon Telescope (EHT) \cite{1a}, has further increased the interest of researcher in the black hole Physics. The accretion of plasma on black holes is another appealing phenomena to prob into the dynamics of particles in the vicinity of black holes. The circular null geodesics around black holes may be helpful in the comprehension of gravitational lensing \cite{2} and shadows of black holes \cite{3}, which is quite relevant to study the image of the background galaxies that can also amplify the light coming from them. In the literature timelike and null circular geodesics in black hole spacetimes are extensively studied \cite{4,5,6,7,8,9,9a,9b,9c}. In this regard charged and neutral particle motion in the spacetime field of different types of static and rotating black holes with and without charge has been analysed \cite{10,11,12,13,14,15}. The effect of some external fields such as magnetic field and quintessence field has been examined on the dynamics of timelike and null particles in the vicinity of black holes \cite{16,17,18,19,20,20a,20b,20c,20z}. Besides, the influence of perfect fluid and string clouds on particle dynamics have also been discussed in different theories of gravity and some constraints have been obtained on the parameters of the theories (see for example \cite{20x,20p,20d,20e,20f,20g,20h})\\

An interesting phenomenon known as the quasi periodic oscillations (QPOs) detected in the X-ray radiation of microquasars, which are binary systems of black holes surrounded by some accretion disc of matter that flowing from companion stars, is related with the stable circular orbits of test particles in the vicinity of black hole spacetimes \cite{21}. Friction in the disc close to the inner most circular orbit of the test particles is very strong and the matter in the accretion disk starts to emit X-rays \cite{22}. For the study of accretion discs in astrophysics, the QPOs are of great interest as they are considered as an efficient test of highly compact stellar models and useful tool for an accurate measurement of black hole parameters, such as mass , charge and spin. The techniques of spectroscopy which is the frequency distribution of photons, and timing that is photon number time dependence, can be applied to particular microquasars, to extract some useful information about the binary system \cite{MS1}. To comprehend the strong gravitational fields, the QPOs from the accreting matter around black holes, have been investigated in the literature \cite{Rezzolla03,TKSS,SKT,SKT2}. Various models including disc-seismic model, hot-spot model, resonance model and warped disk model have been proposed to understand the nature of QPOs \cite{Rezzolla13b}. Till today the mechanism responsible for the production of the QPOs is not exactly known, since none of the above mentioned models can fit the data observed from different astrophysical sources \cite{MS}.\\

The presence of gravitational field makes it difficult for the orbiting particles around black holes to remain stable that results in the collision of particle and different astrophysical phenomena can take place. One such idea of particle collision of test particles near the horizon of rotating black hole was presented by Banados-Silk-West and in the literature it is known as the BSM mechanism \cite{23}. The BSW mechanism which gives an infinite energy in the centre-of-mass frame of the colliding particles coming from infinity, at some point arbitrarily close to the horizon of black holes may be related with the energy at the Planck-scale at which gravity and quantum mechanics may meet each other \cite{24}. This very high scale of energy is also of interest for physicists to study the extra dimensions of spacetime and to achieve the theory of the grand unification of the fundamental forces in nature \cite{25}. The largest and the most powerful particle accelerator on the earth is the large hadron collider (LHC), that could accelerate particles to a collision energy of 10 TeV, which is still very small in comparison to the Planck-scale energy of $10^{16}$ TeV \cite{26}. For probing the Physics at the Planck-scale some new mechanisms should be proposed. Therefore, The BSW mechanism of the collision of particles near the horizon of black holes may provide such a possible approach. The BSW mechanism has been studied for different rotating and charged black holes in General Relativity (GR) and other modified theories of gravity \cite{27,28,29,30,31,32}.\\

In GR the gravitational redshift is the phenomenon of photons moving away from a gravitational well and hence lose energy. As a result of the lose of energy a decrease in the frequency occurs and therefore, the wavelength increases, this is known as the frequency redshift. When a photon is coming into a gravitational well the opposite effect of gain of energy takes place. Due to this the frequency of the ingoing photon increases and consequently its wavelength decreases. This effect is known as the gravitational bullshit. The gravitational red-blue shifts of photons from test particles moving in stable circular orbits around black holes is of great interest to astrophysicists. In this regard Herrera et. al have investigated the red-blue shift of photons coming from the timelike particles in the stable circular orbits to estimate the mass $M$ and the rotation parameter $a$ of the Kerr black hole \cite{33}. Further, they have also estimated the distance of earth from the black hole. Kuniyal et. al have studied the red-blue shifts of light emitted from the test particles in the vicinity of Schwarzschild black hole in the noncommutative geometry and have obtained an estimate of the mass of the black hole from the frequency shit \cite{34}. The study of red-blue shifts of photons from geodesic particles in stable circular orbits in the spacetime field of Kerr-Newman (anti) de-Sitter black hole has been carried out by Kraniotis \cite{35}. Becerril  et. al have obtained the mass parameter of Bardeen, Hayward and Ay\'{o}n-Beato-Garc\'{i}a black holes in terms of red-blue shifts of photons emitted by test particles moving in orbits around these black holes \cite{36}. In a recent work L\'{o}pez and Breton \cite{LB}, have analysed the effects of strong magnetic field on the red-blue shifts of the photons emitted by neutral as well as charged particles orbiting the Ernst black hole.\\

The Schwarzschild black hole solution of the Einstein field equations was generalized by Kiselev in the presence of quintessence field, which is considered as a dark energy candidate and is responsible for the current accelerated expansion of the observable Universe \cite{37}. The charged version, which is the dark energy counterpart of the Reissner-Nordstom black hole has also been presented \cite{37}. Particle dynamics in these dark energy black holes have been studied extensively \cite{38,39,40}. The effects of the quintessence field on the motion of null and timelike particles in these black hole spacetimes have been observed and a comparison with the motion of particles in the Schwarzschild and the Reissner-Nordstom black holes has been presented \cite{9c,39,40a}. The effects of the string clouds, which are assumed to be formed in the early stages of the structure formation of our Universe due to the symmetry breaking \cite{41}, have also been explored on the motion of null and timelike particles in the vicinity of black hole spacetimes \cite{9b}. A black hole solution both in the presence of string clouds and quintessence field has been obtained in the literature \cite{42}. In a recent work the null and timelike radial and circular geodesics have been analysed in the spacetime field of the Schwarzschild black hole in the presence of both the quintessence field and string clouds \cite{9c}. It was shown that how the presence of the quintessence field and the string clouds affect the null and time like geodesic motion in the spacetime field of the black hole \cite{9c}. The charged version of the black hole with string clouds and quintessence was then presented \cite{43}. Here in the present work we are   keen to analyse the dynamics of particle in this charged black hole with string clouds and quintessence to look at the effects of charge on the motion of particles. We observe that the existence of the stable circular orbits is granted for very small value of the string cloud parameter $\alpha$ and quintessence parameter $\gamma$. Further we noticed that the stable circular orbits get shrink, in the presence of $\alpha$ and $\gamma$, with charge as compared to the un charged counterpart \cite{9c}. Further we have studied the QPOs of test particles near the stable circular orbits, the red-blue shifts of the photons emitted by the timelike particles in the circular orbits, the effective force and the centre-of-mass energy of the colliding particles near the horizon of the charged black hole in the presence of the string clouds and quintessence. The details are given in the subsequent sections.\\

The paper is structured as follows: In the next Section we discuss the horizon structure of the the charged black hole in the presence of the string clouds and quintessence and show that how the existence of the horizon get effected with the variation of values of $\alpha$, $\gamma$ and charge $Q$ of the black hole. In the same Section we present the basic equations of particle motion along with the effective potential of the charged particles. In the Section 3 we give the details of the null and timelike circular geodesics motion. QPOs of the timelike particles close to the circular orbits are discussed in the Section 4. In the Section 5 we analyse the red-blue shifts of the light emitted by the test particles in the circular motion. We study the acceleration of radially falling particles from infinity, at the horizon of the black hole in the Section 6. The give the discussion of the effective force in the Section 7. Finally we give a summery of our work in the Section 8.\\

\section{Basic calculation for particle trajectories}

The charged spherically symmetric and static spacetime with quintessence and clouds of strings is expressed as \cite{43}
\begin{equation}\label{1}
ds^{2}=-\left(1-\alpha-\frac{2 M}{r}-\frac{\gamma}{r^{3\omega_{q}+1}}+\frac{Q^{2}}{r^2}\right)dt^{2}+\left(1-\alpha-\frac{2 M}{r}-\frac{\gamma}{r^{3\omega_{q}+1}}+\frac{Q^{2}}{r^2}\right)^{-1}dr^2+r^{2}(d\theta^{2}+sin^{2}\theta d\phi^{2}),
\end{equation}
where $\omega_{q}$, $M$, $\alpha$, $\gamma$, and $Q$ represent the equation of state parameter for quintessence field, mass of the black hole, clouds of strings parameter, quintessence parameter and charge of the spacetime respectively. Further, the important restriction on equation of state parameter is expressed as: $-1<\omega_{q}<-\frac{1}{3}$.  The metric has naked singularity at $r=0$. The spacetime given by Eq.(\ref{1}) can be reduced to the Schwarzschild solution by taking $\alpha=\gamma=Q=0$. For the current analysis, we take $\omega_{q}=\frac{-2}{3}$ that corresponds to the quintessence field. The lapse function for the Eq.(\ref{1}) is realized as
\begin{equation}\label{2}
f(r)=1-\alpha-\frac{2M}{r}+\frac{Q^2}{r^2}-\frac{\gamma}{r^{-1}},
\end{equation}
where \cite{43}
\begin{equation*}
0<\gamma<\frac{\alpha^2-2 \alpha+1}{8 M},\;\;\;\;\;\;\;\;\;0<\alpha<1,\;\;\;\;\;\;\;\;0<Q <1.
\end{equation*}

\begin{figure}
\centering \epsfig{file=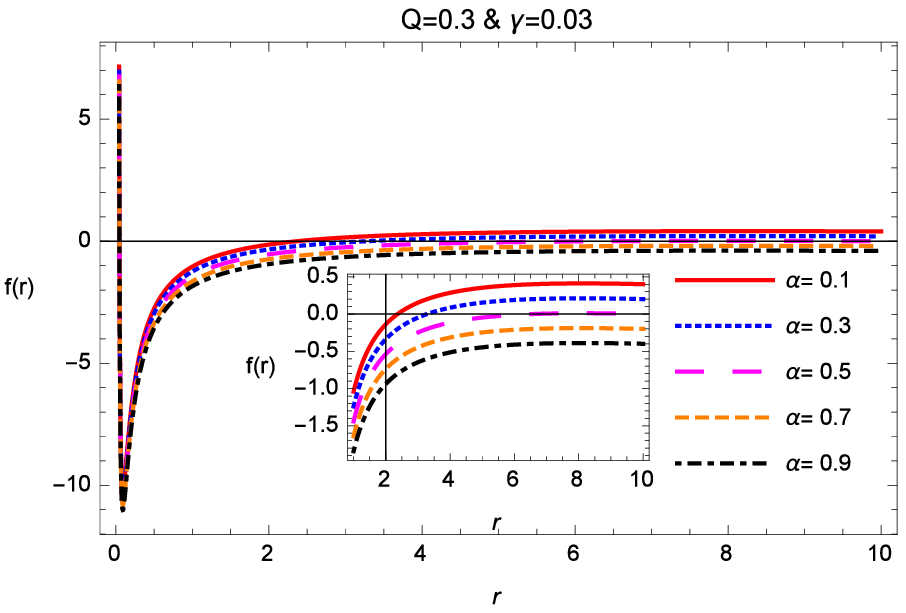, width=.32\linewidth,
height=2.02in}\epsfig{file=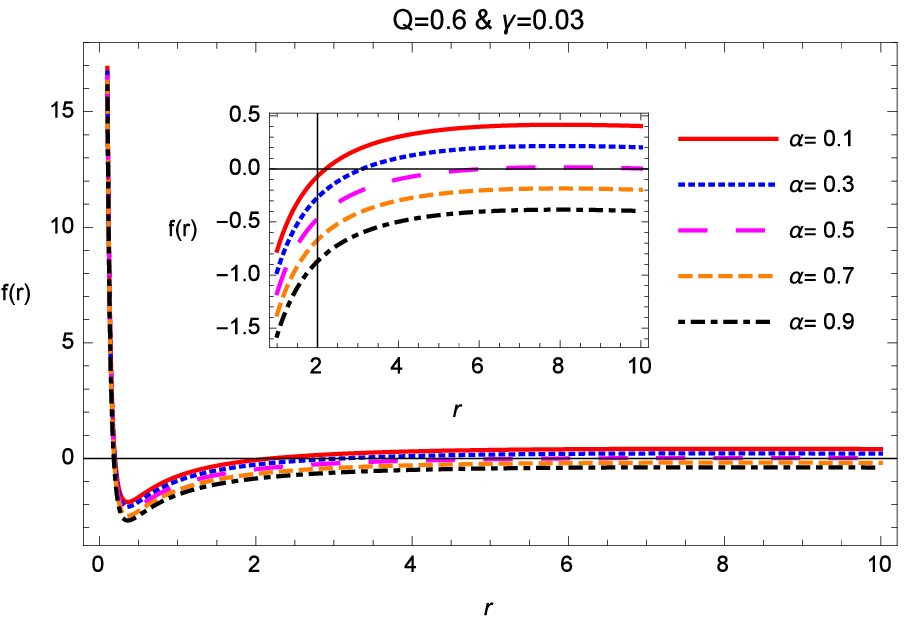, width=.32\linewidth,
height=2.02in}\epsfig{file=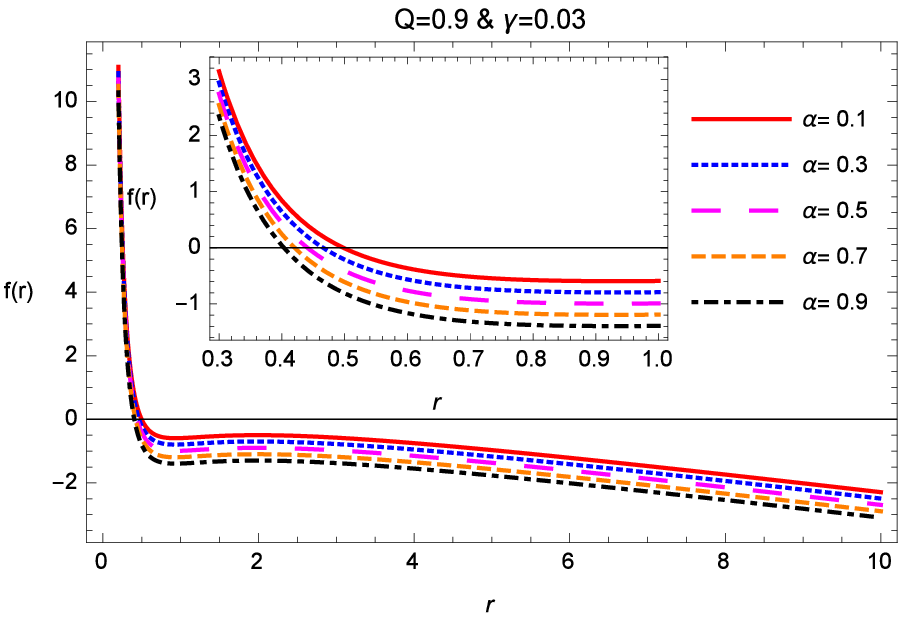, width=.32\linewidth,
height=2.02in}
\centering \epsfig{file=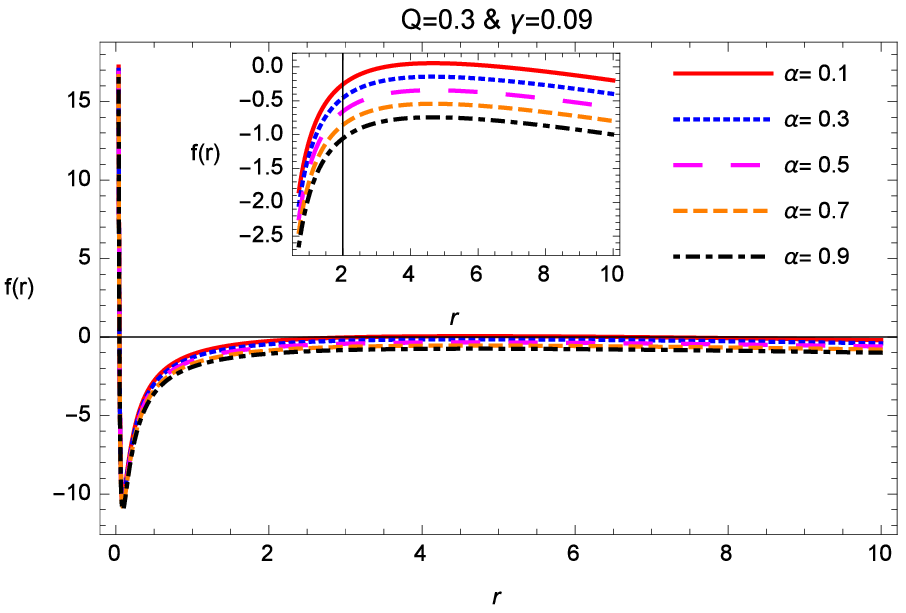, width=.32\linewidth,
height=2.02in}\epsfig{file=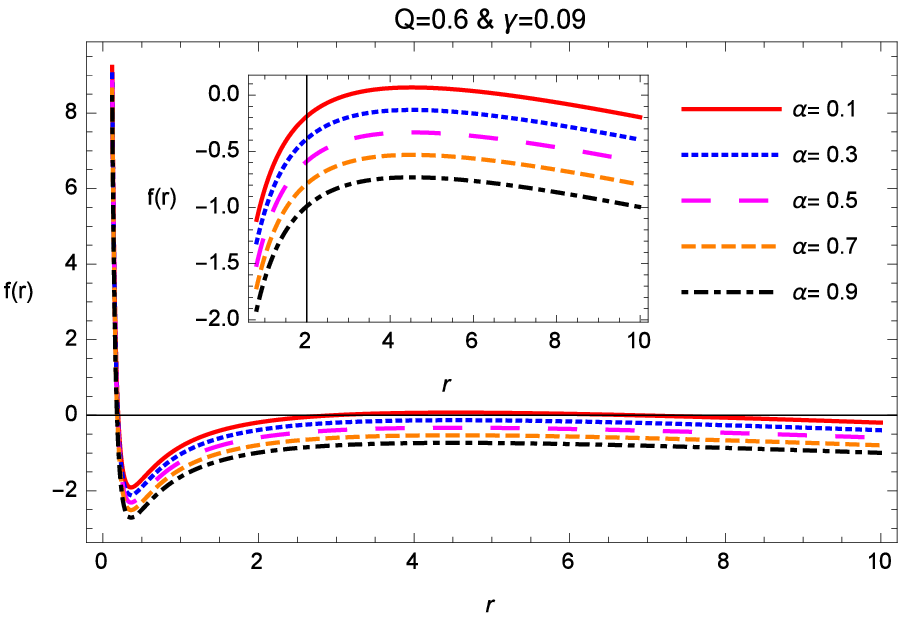, width=.32\linewidth,
height=2.02in}\epsfig{file=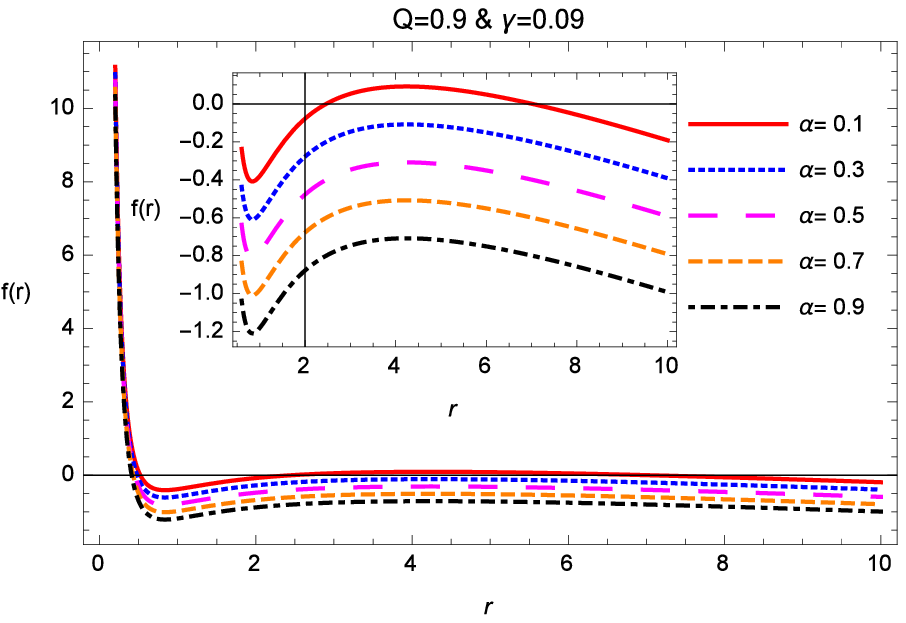, width=.32\linewidth,
height=2.02in} \caption{\label{fig1} Shows the behavior of $f(r)$.}
\end{figure}

\begin{figure}
\centering \epsfig{file=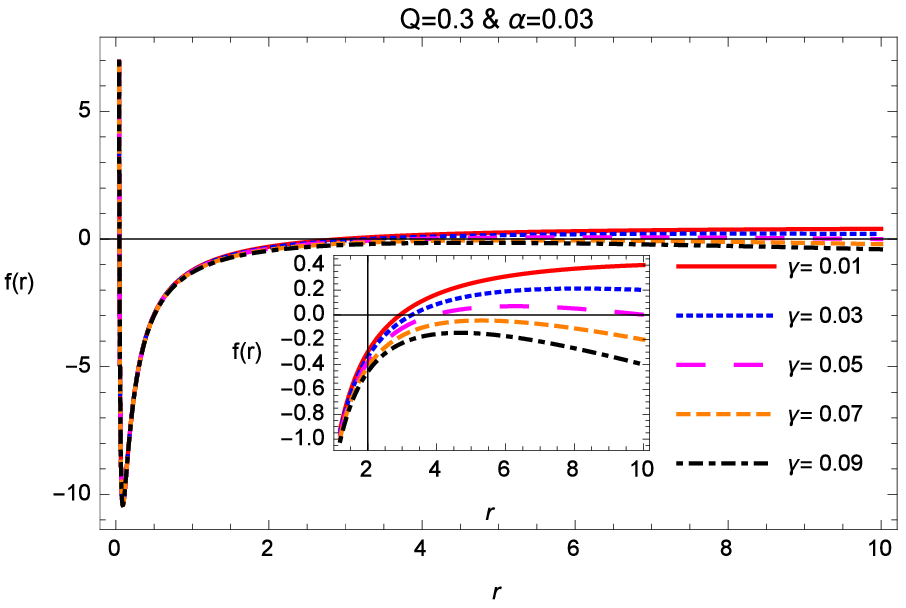, width=.32\linewidth,
height=2.02in}\epsfig{file=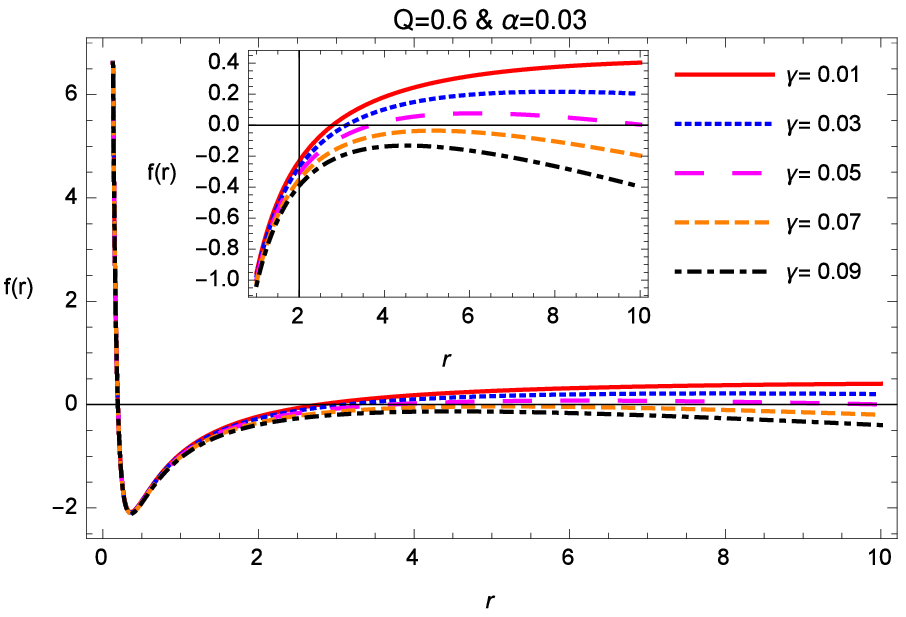, width=.32\linewidth,
height=2.02in}\epsfig{file=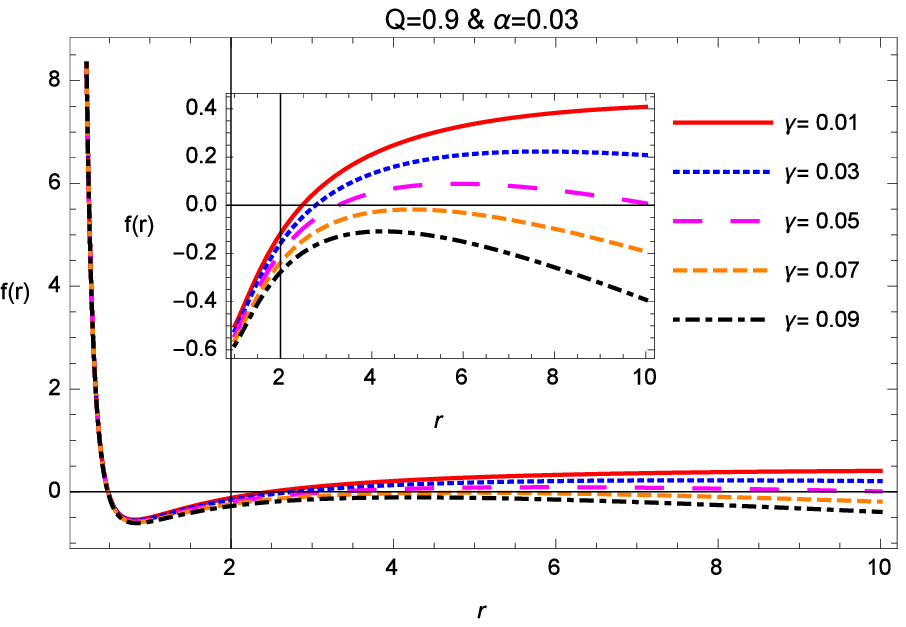, width=.32\linewidth,
height=2.02in}
\centering \epsfig{file=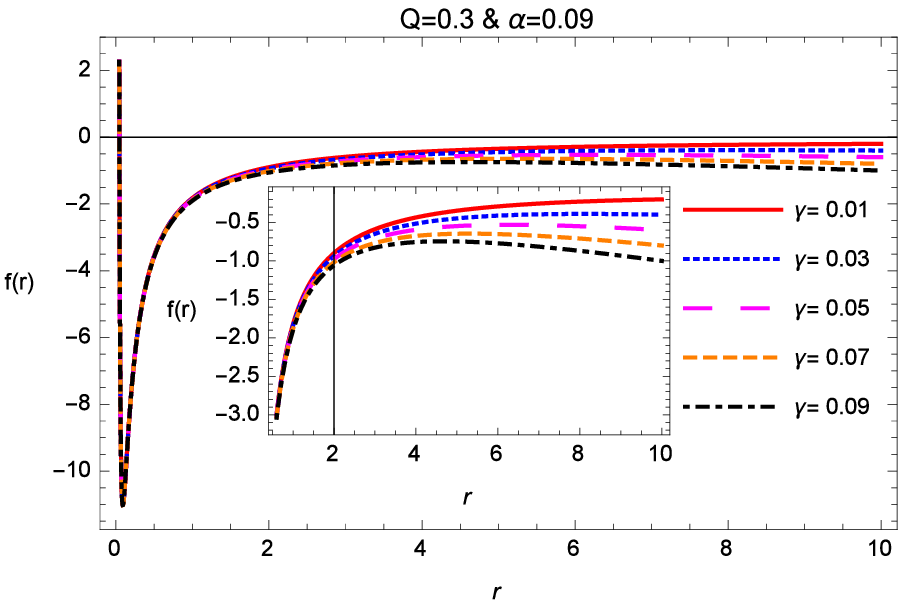, width=.32\linewidth,
height=2.02in}\epsfig{file=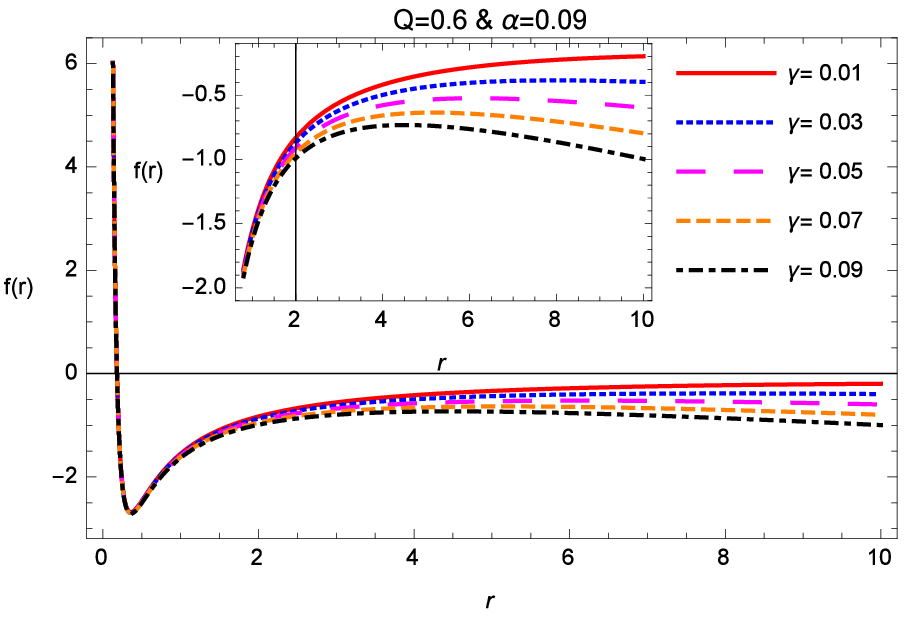, width=.32\linewidth,
height=2.02in}\epsfig{file=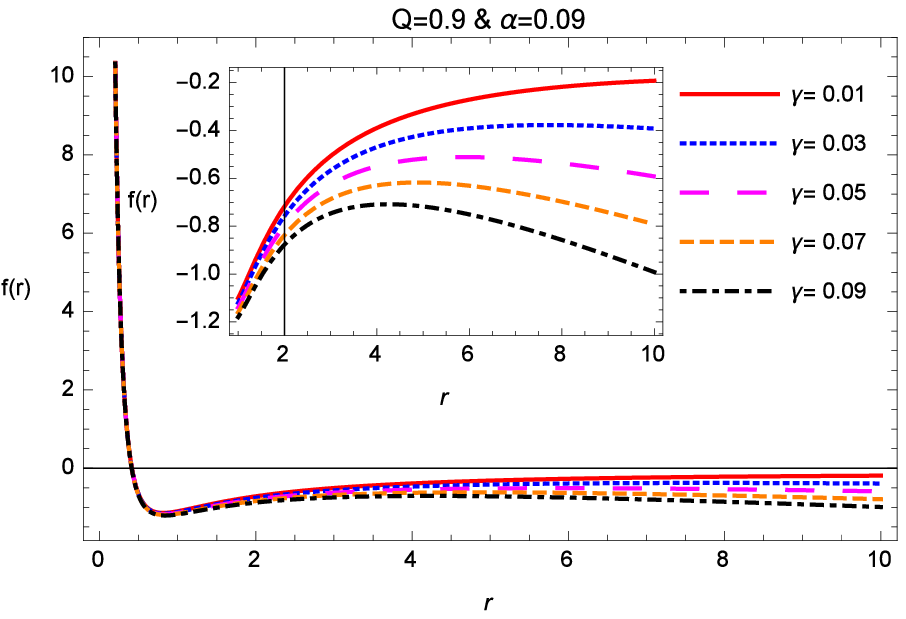, width=.32\linewidth,
height=2.02in} \caption{\label{fig2} Shows the behavior of $f(r)$.}
\end{figure}

\begin{figure}
\centering \epsfig{file=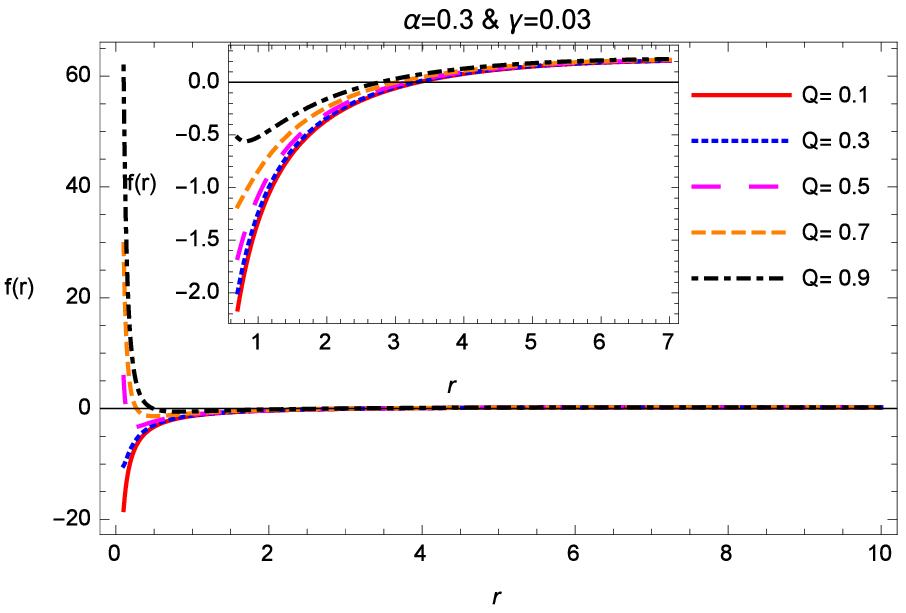, width=.32\linewidth,
height=2.02in}\epsfig{file=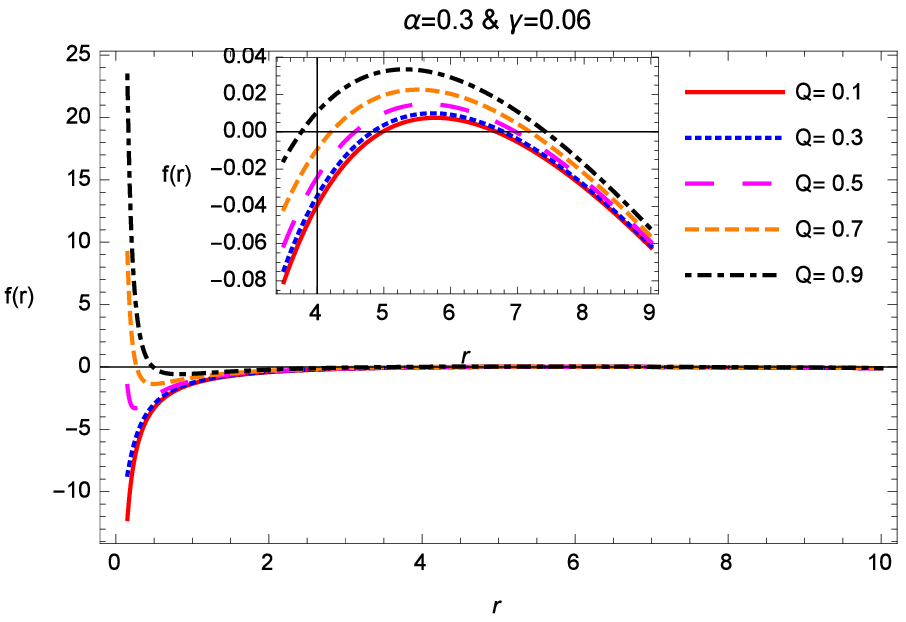, width=.32\linewidth,
height=2.02in}\epsfig{file=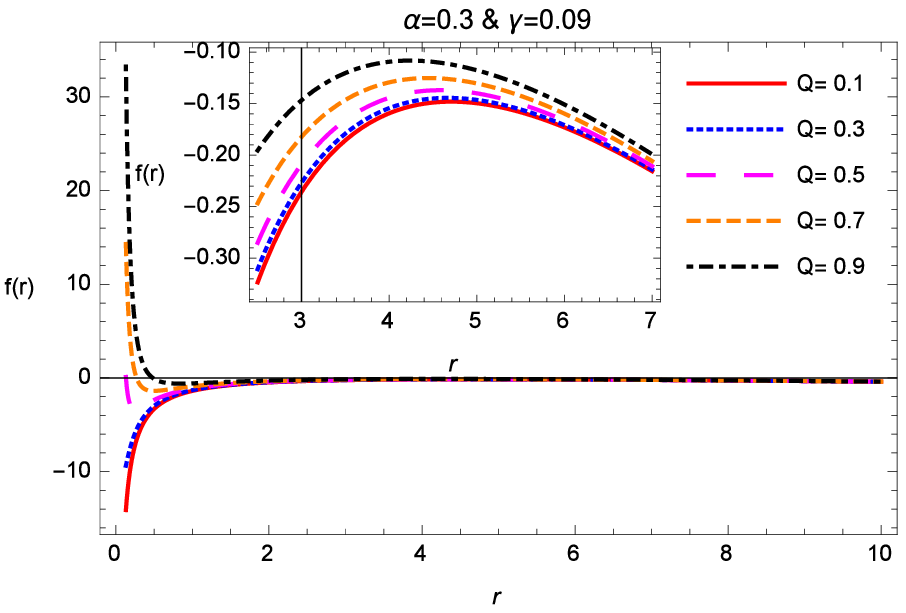, width=.32\linewidth,
height=2.02in}
\centering \epsfig{file=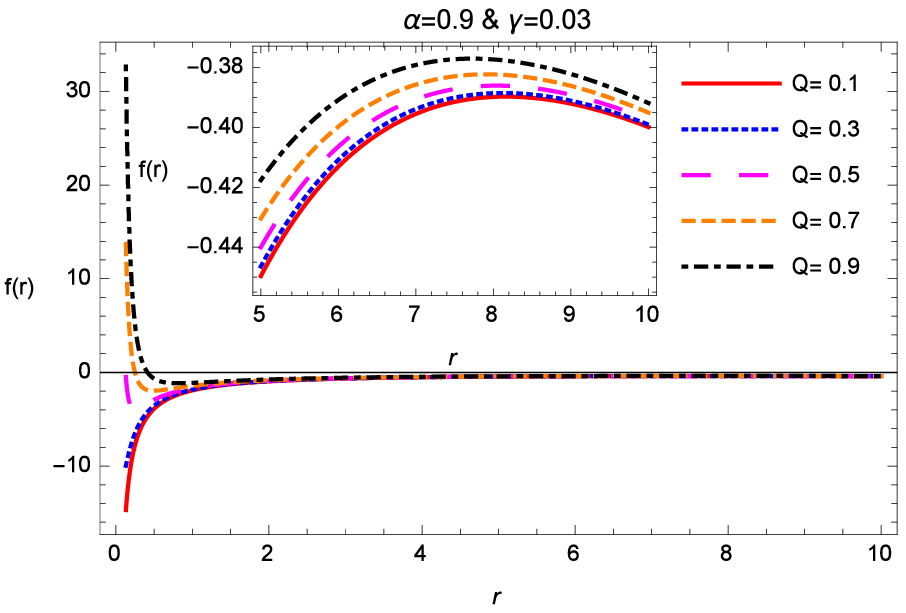, width=.32\linewidth,
height=2.02in}\epsfig{file=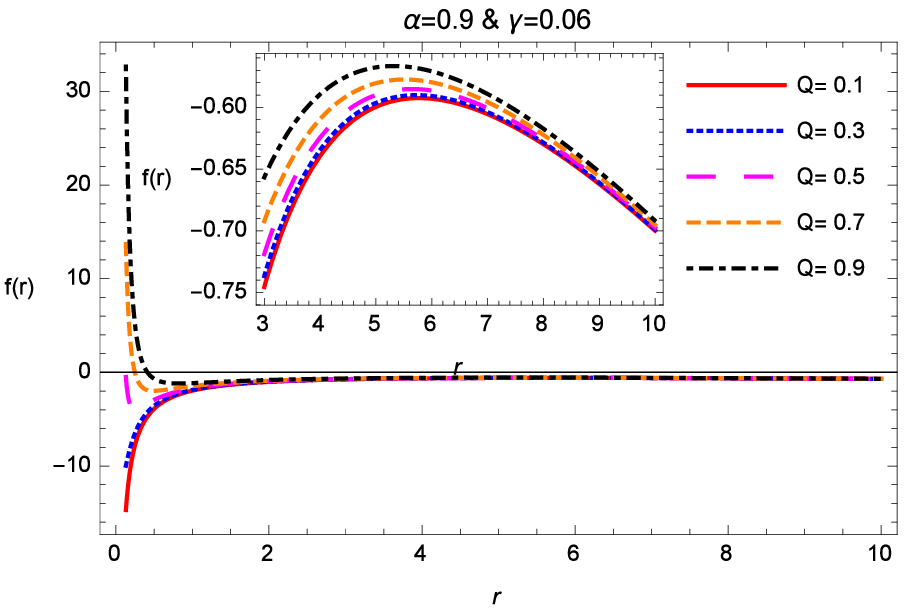, width=.32\linewidth,
height=2.02in}\epsfig{file=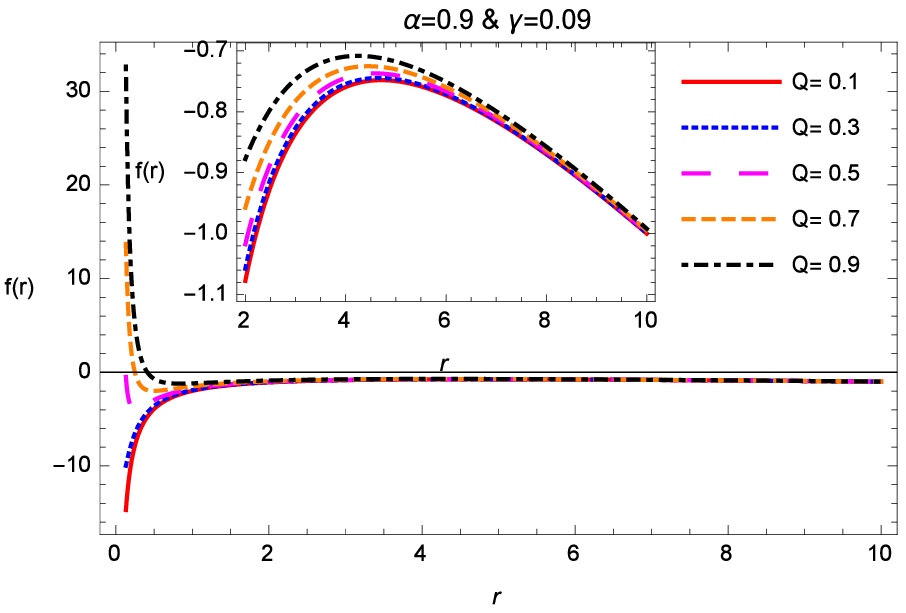, width=.32\linewidth,
height=2.02in} \caption{\label{fig3} Shows the behavior of $f(r)$.}
\end{figure}
Here we give a review of the structure for black hole horizons \cite{43}. If $r\rightarrow 0$, the charge term, i.e., $Q^2$ dominates the behavior of
the lapse function and then it become positive. If $r\rightarrow \infty$, then quintessential term $\gamma$ will dominate the lapse function and one gets the negative values. The black hole horizons given by the roots of the following equation
\begin{equation}\label{3}
h(r)=\gamma r^3-(1-\alpha)r^2+2Mr-Q^2=0.
\end{equation}
The above equation can be reduced to the following expression:
\begin{equation}\label{4}
\gamma(r-r_\gamma)(r-r_-)(r-r_+)=0,
\end{equation}
where $r_\gamma$ represents the horizon due the quintessence term (the cosmological horizon), $r_-$, and $r_+$ are the black hole horizon. From Eq. (\ref{3}), it can be seen that the number of horizon depends on the involved parameters values like, $M$, $\alpha$, $\gamma$, and $Q$. For the current analysis we take $M=1$. If we set, $Q=0$, we get the maximal value of the quintessence parameter, i.e., $\gamma$ as:
\begin{equation}\label{5}
\gamma_m(Q=0)=\frac{(1-\alpha)^2}{8}.
\end{equation}
Eq. (\ref{4}) can be reduced to
\begin{equation}\label{6}
Q^2=Q^2(r;\gamma)=\gamma r^3-(1-\alpha)r^2+2r.
\end{equation}
The local maximum for the quintessence can be obtained by using the following relation:
\begin{equation}\label{7}
\gamma_c=\frac{2[r(1-\alpha)-1]}{3r^2}.
\end{equation}
The local maxima for the quintessence and charge parameter occurs, by setting $r=2/1-\alpha$, we get
\begin{equation}\label{8}
Q^2\leq Q_c^2=\frac{4}{3(1-\alpha)},\,\,\,\,\,\,\,\,\,\,\,\,\,\,\,\gamma\leq \gamma_c=\frac{(1-\alpha)^2}{6}.
\end{equation}
The local extrema of function $h(r)$, provides
\begin{equation}\label{9}
r_{min}=\frac{1-\alpha-\sqrt{(1-\alpha)^2-6\,\gamma}}{3\,\gamma},\,\,\,\,\,\,\,\,\,r_{max}=\frac{1-\alpha+\sqrt{(1-\alpha)^2-6\,\gamma}}{3\,\gamma}.
\end{equation}
The extrema  can be classified into three different types. For the first type the black hole
horizons should be coincided as $r_- = r_+$. In this case the minimum value of $h(r)$ corresponds to the
black hole horizon as
\begin{equation}\label{10}
Q^2=Q_{c_{1}}^2=\frac{\left[-(1-\alpha)+9\,\gamma+\sqrt{(1-\alpha)^2-6\,\gamma}\right]^3}{27\,\gamma^2}.
\end{equation}
The horizon for quintessence parameter $\gamma$ is defined as
\begin{equation}\label{11}
r_q=\frac{1-\alpha+2\sqrt{(1-\alpha)-6\,\gamma}}{3\,\gamma}.
\end{equation}
In the second case the external black hole horizon coincides as $r_+ = r_q$. The final relation is expressed as
\begin{equation}\label{12}
Q^2=Q_{c_{2}}^2=\frac{\left[-(1-\alpha)+9\,\gamma-\sqrt{(1-\alpha)^2-6\,\gamma}\right]^3}{27\,\gamma^2}.
\end{equation}
The radius of the interior horizon is realized as
\begin{equation}\label{13}
r_-=\frac{1-\alpha-2\sqrt{(1-\alpha)-6\,\gamma}}{3\,\gamma}.
\end{equation}
For the third type, the relation is
\begin{equation}\label{14}
\gamma^2\leq \gamma_c^2=\frac{4}{3(1-\alpha)},\,\,\,\,\,\,\,\,\,\,\,\,\,\,\,\gamma\leq \gamma_c=\frac{(1-\alpha)^2}{6}.
\end{equation}
The dependence of the horizon on the parameters $\alpha$, $\gamma$ and $Q$ can be inferred from Figs. (\ref{1}-\ref{3}).

The Lagrangian for the charged spacetime given by Eq. (\ref{1}) is
\begin{equation}\label{15}
\mathbf{L}=\frac{1}{2}g_{\mu\nu}\dot{y}^{\mu}\dot{y}^{\nu}+\frac{q A_{\mu}}{\mathfrak{m}}\dot{y}^{\mu},
\end{equation}
where $A_{\mu}$ is the electromagnetic four-potential and $\dot{y}^{\mu}=\frac{dy^{\mu}}{d\tau}$. The Lagrangian for the metric given by Eq. (\ref{1}) is expressed as
\begin{equation}\label{16}
\mathbf{L}=\mathfrak{m}\bigg(-\left(1-\alpha-\frac{2M}{r}+\frac{Q^2}{r^2}-\frac{\gamma}{r^{-1}}\right)
\frac{\dot{t}^{2}}{2}+\frac{\dot{r}^{2}}{2\left(1-\alpha-\frac{2M}{r}+\frac{Q^2}{r^2}-\frac{\gamma}{r^{-1}}\right)}+\frac{r^{2}}{2}(\dot{\theta}^{2}+sin^{2}\theta \dot{\phi}^{2})\bigg)+\frac{q Q}{r}\dot{t}.
\end{equation}
For this Lagrangian we have the following two conserved quantities, the specific energy $E$ and the specific angular momentum $L$ of the particle
\begin{eqnarray}
\dot{t}&&=\left(E-\frac{q Q}{r}\right)\frac{1}{m\left(1-\alpha-\frac{2M}{r}+\frac{Q^2}{r^2}-\frac{\gamma}{r^{-1}}\right)},\label{16a}\\
\dot{\phi}&&=\frac{L}{r^{2}sin^{2}\theta},\label{17}
\end{eqnarray}
In the equatorial plane, i.e we take $\theta=\frac{\pi}{2}$, Eq.(\ref{17}) becomes
\begin{equation}\label{18}
-\left(1-\alpha-\frac{2M}{r}+\frac{Q^2}{r^2}-\frac{\gamma}{r^{-1}}\right)\dot{t}^{2}+\frac{\dot{r}^{2}}
{\left(1-\alpha-\frac{2M}{r}+\frac{Q^2}{r^2}-\frac{\gamma}{r^{-1}}\right)}+r^{2}\dot{\phi}^{2}=-\zeta,
\end{equation}
where $\zeta=0,1$ is used to define the null and timelike geodesics respectively. Using the normalization condition $u^a u_a =-1$, the equation of motion is
\begin{equation}\label{19}
\bigg(\frac{dr}{dQ}\bigg)^{2}+ V_{eff}(r)=E^{2},
\end{equation}
where
\begin{equation}\label{20}
V_{eff}(r)= \left(-\alpha -\frac{2 M}{r}+\frac{Q ^2}{r^2}-\frac{\gamma }{r^{-1}}+1\right)\left(\zeta+\frac{L^2}{r^2}\right)+\frac{q Q }{\mathfrak{m}r},
\end{equation}
 represents the effective potential.

\section{Circular Motion}

In this section we examine the circular motion of massive particles and photons. The expressions for energy $E^2$ and angular momentum $L^2$ are given as:
\begin{eqnarray}
E^{2}&&=\frac{2 r^2}{r^3 \left(-\gamma -\frac{2 r}{r (2 M+r (\alpha +\gamma  r-1))-Q ^2}\right)+2 M r-2 Q ^2},\label{21}\\
L^{2}&&=\frac{r^5 \left(\frac{2 \left(M r-Q ^2\right)}{r^3}-\gamma \right)}{r^3 \left(\gamma +\frac{2 r}{r (2 M+r (\alpha +\gamma  r-1))-Q ^2}\right)-2 M r+2 Q ^2}. \label{22}
\end{eqnarray}
The graphical behavior of $E^2$ and $L^2$ is provided in Figs. (\ref{4}) and (\ref{5}) for different values of the parameters $\alpha$, $\gamma$, and $Q$.
\begin{figure}
\centering \epsfig{file=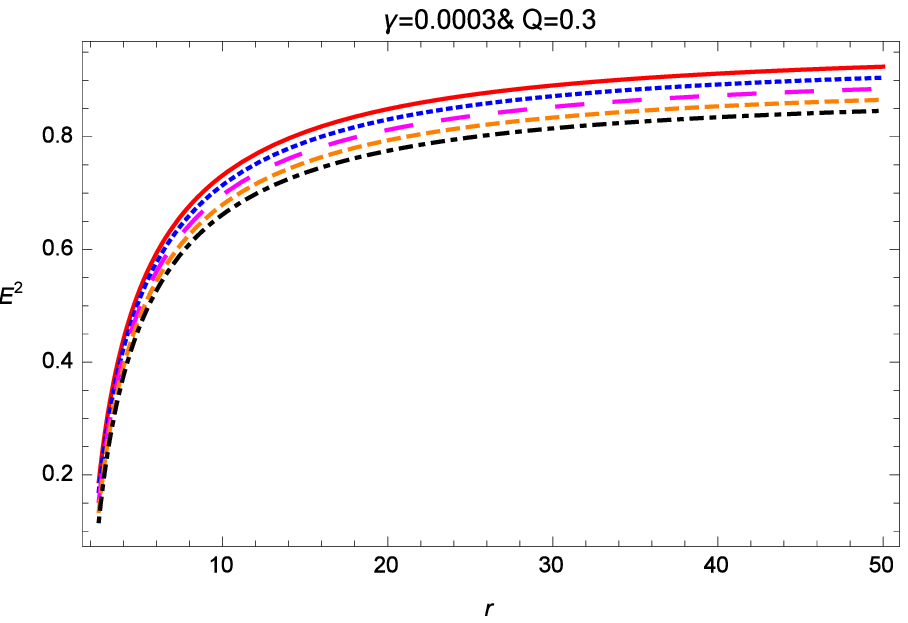, width=.32\linewidth,
height=2.02in}\epsfig{file=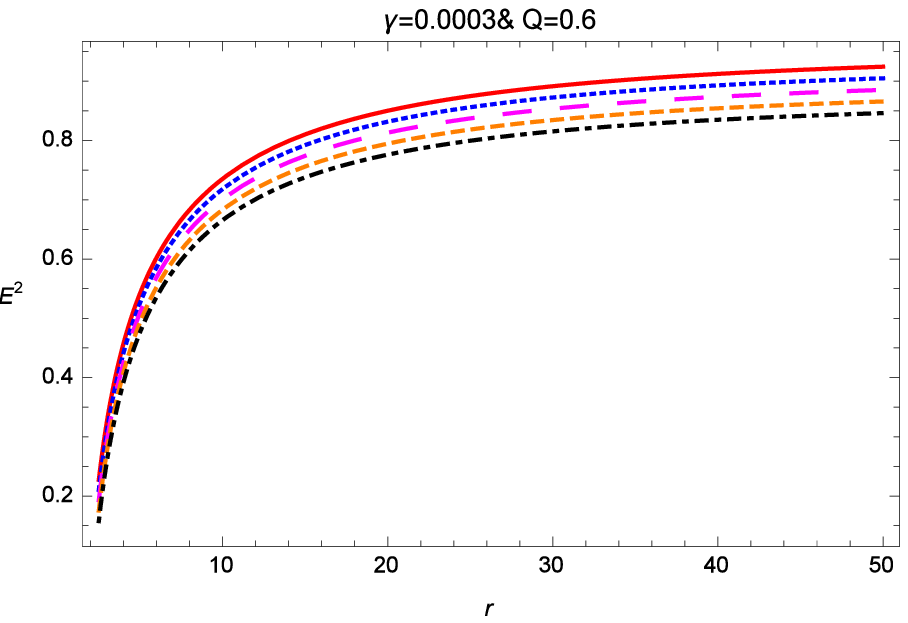, width=.32\linewidth,
height=2.02in}\epsfig{file=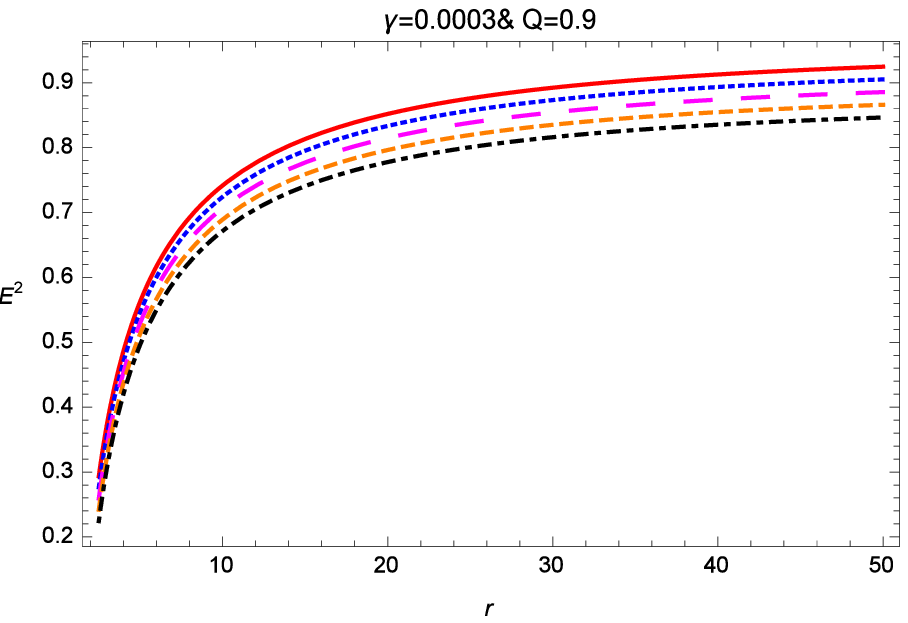, width=.32\linewidth,
height=2.02in}
\centering \epsfig{file=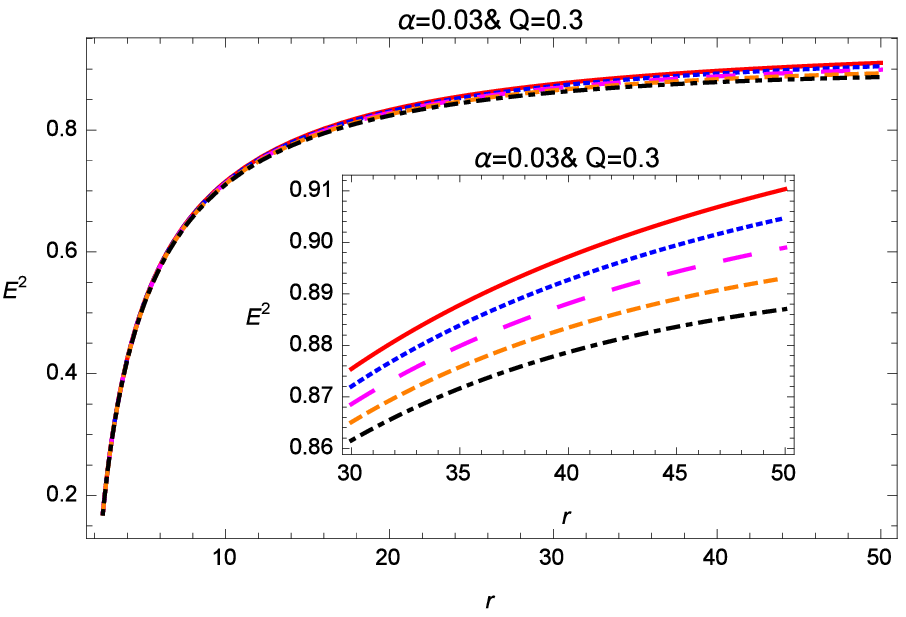, width=.32\linewidth,
height=2.02in}\epsfig{file=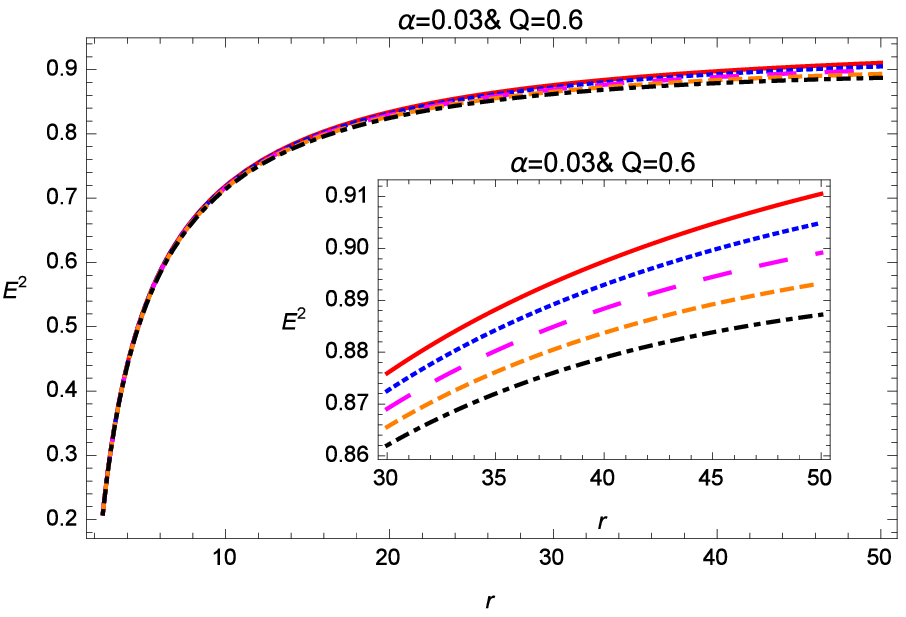, width=.32\linewidth,
height=2.02in}\epsfig{file=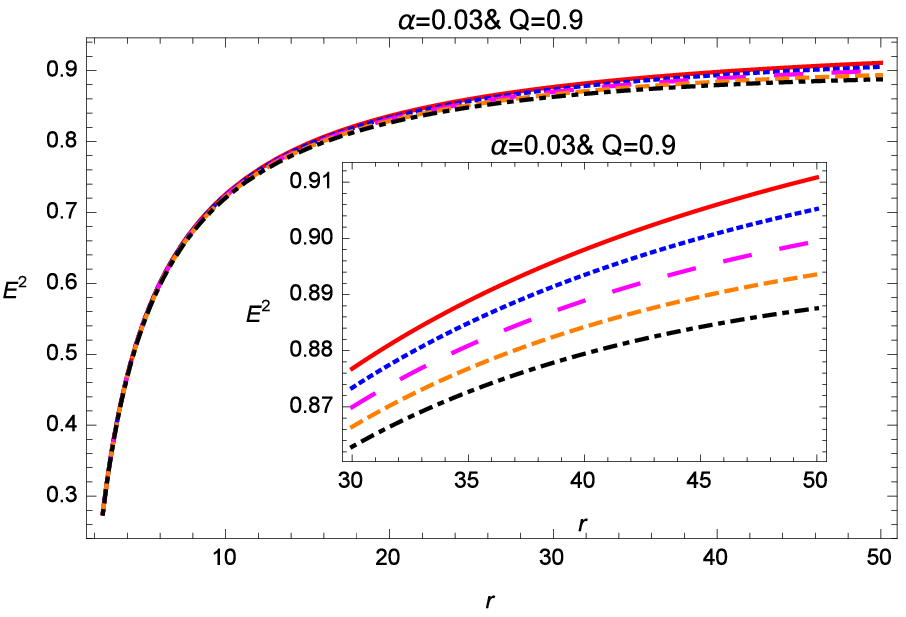, width=.32\linewidth,
height=2.02in}
\centering \epsfig{file=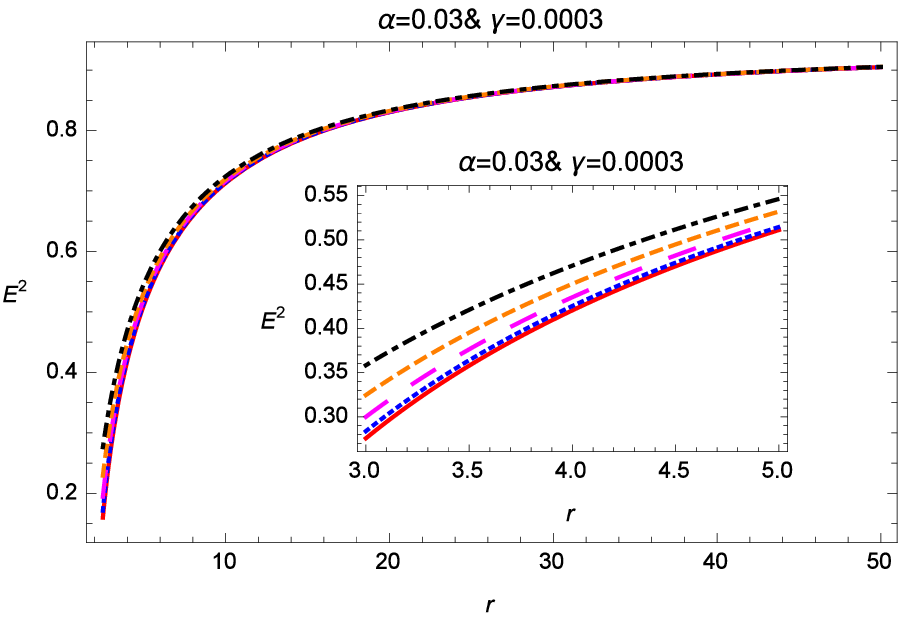, width=.32\linewidth,
height=2.02in}\epsfig{file=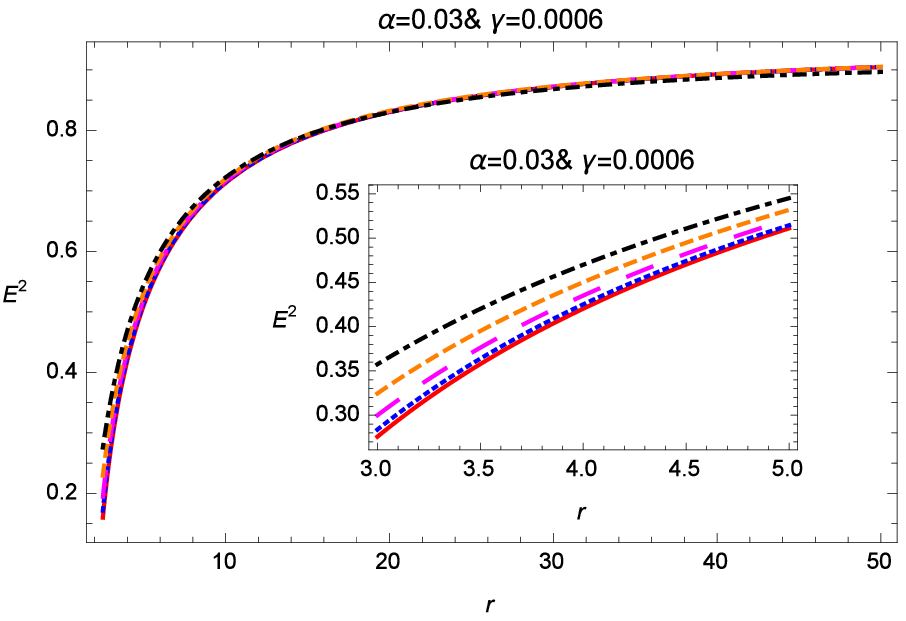, width=.32\linewidth,
height=2.02in}\epsfig{file=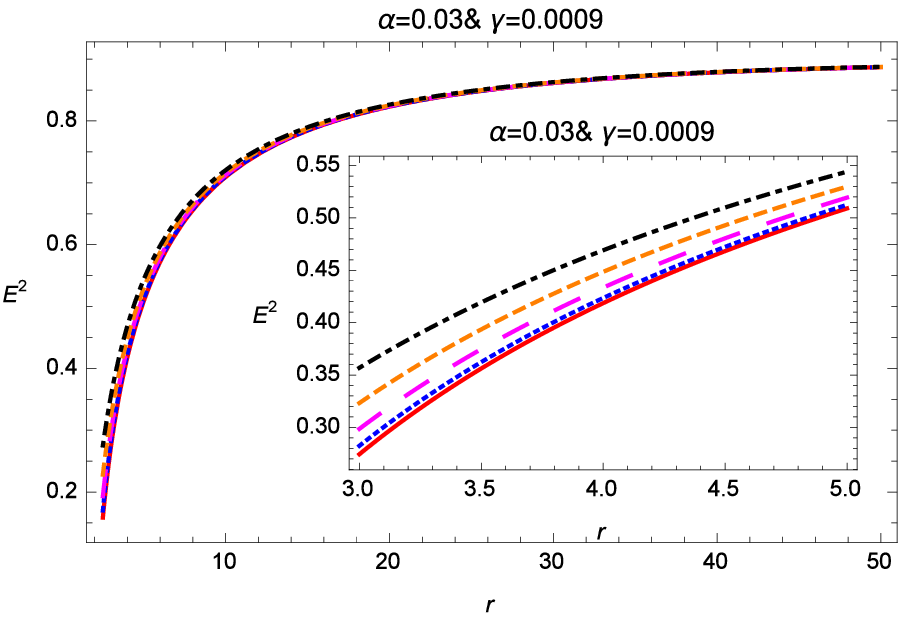, width=.32\linewidth,
height=2.02in} \caption{\label{fig4} Shows the behavior of $E^2$. In the first row, we take different five values of cloud parameter, i.e., $\alpha=0.01(\textcolor{red}{\bigstar})$, $\alpha=0.03(\textcolor{blue}{\bigstar})$, $\alpha=0.05(\textcolor{magenta}{\bigstar})$, $\alpha=0.07(\textcolor{orange}{\bigstar})$, and $\alpha=0.09(\textcolor{black}{\bigstar})$. In the second row, we take different five values of quintessential parameter, i.e., $\gamma=0.0001(\textcolor{red}{\bigstar})$, $\gamma=0.0003(\textcolor{blue}{\bigstar})$, $\gamma=0.0005(\textcolor{magenta}{\bigstar})$, $\gamma=0.0007(\textcolor{orange}{\bigstar})$, and $\gamma=0.0009(\textcolor{black}{\bigstar})$. In the third row, we take different five values of charge parameter, i.e., $Q=0.1(\textcolor{red}{\bigstar})$, $Q=0.3(\textcolor{blue}{\bigstar})$, $Q=0.5(\textcolor{magenta}{\bigstar})$, $Q=0.7(\textcolor{orange}{\bigstar})$, and $Q=0.9(\textcolor{black}{\bigstar})$.}
\end{figure}
\begin{figure}
\centering \epsfig{file=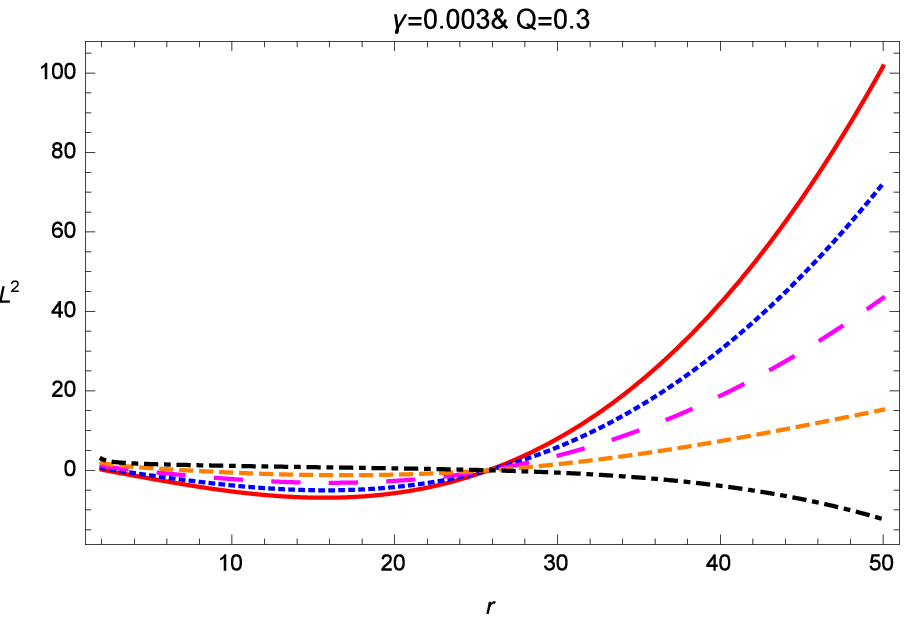, width=.32\linewidth,
height=2.02in}\epsfig{file=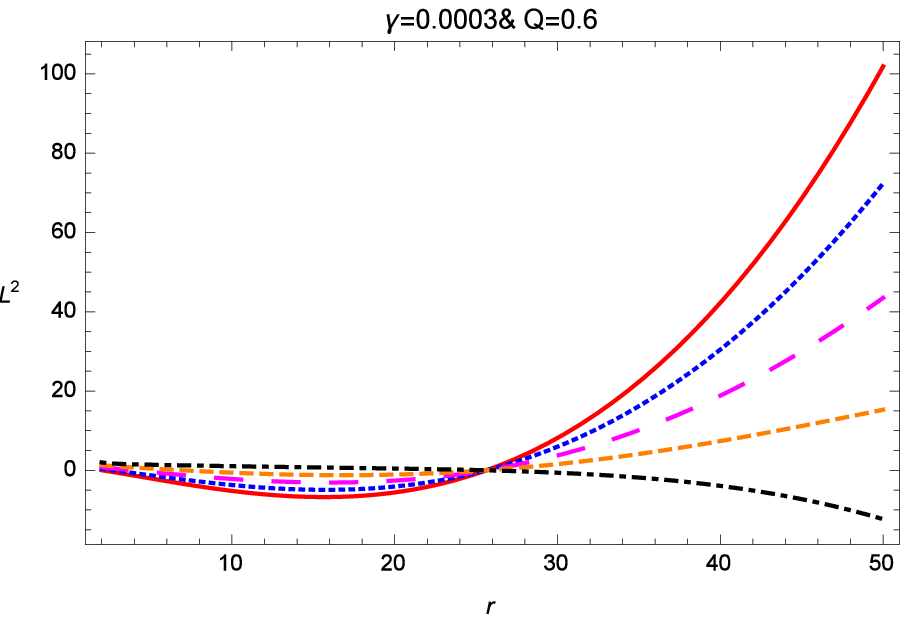, width=.32\linewidth,
height=2.02in}\epsfig{file=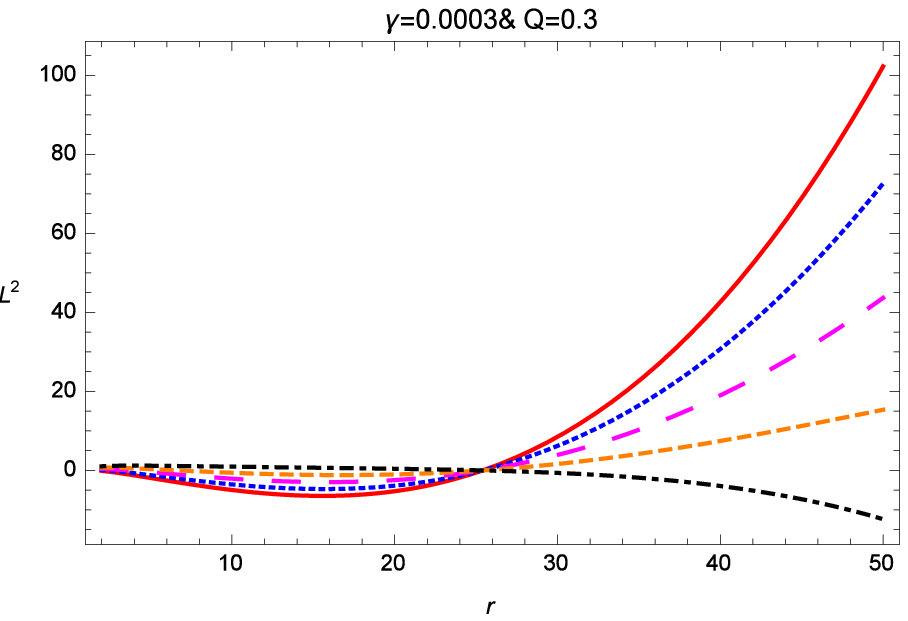, width=.32\linewidth,
height=2.02in}
\centering \epsfig{file=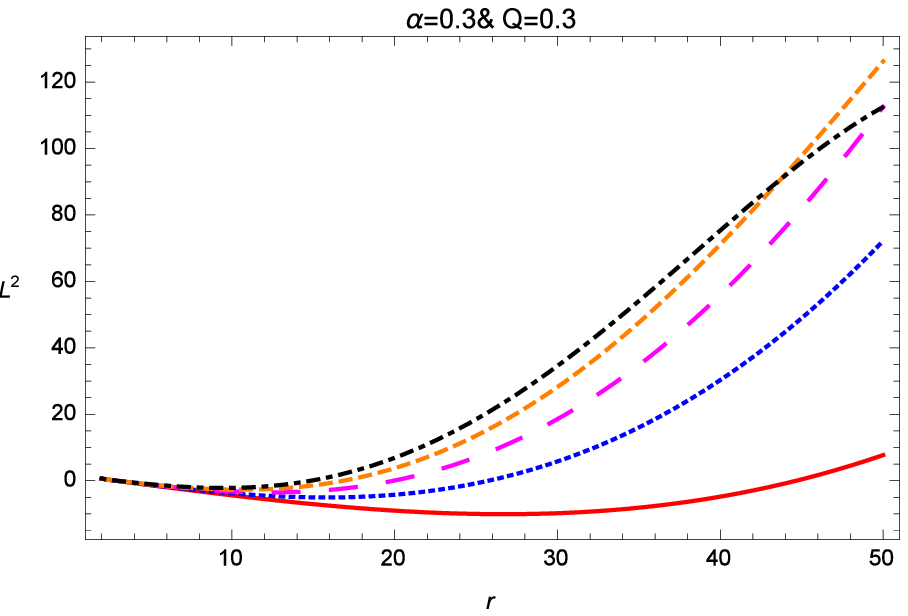, width=.32\linewidth,
height=2.02in}\epsfig{file=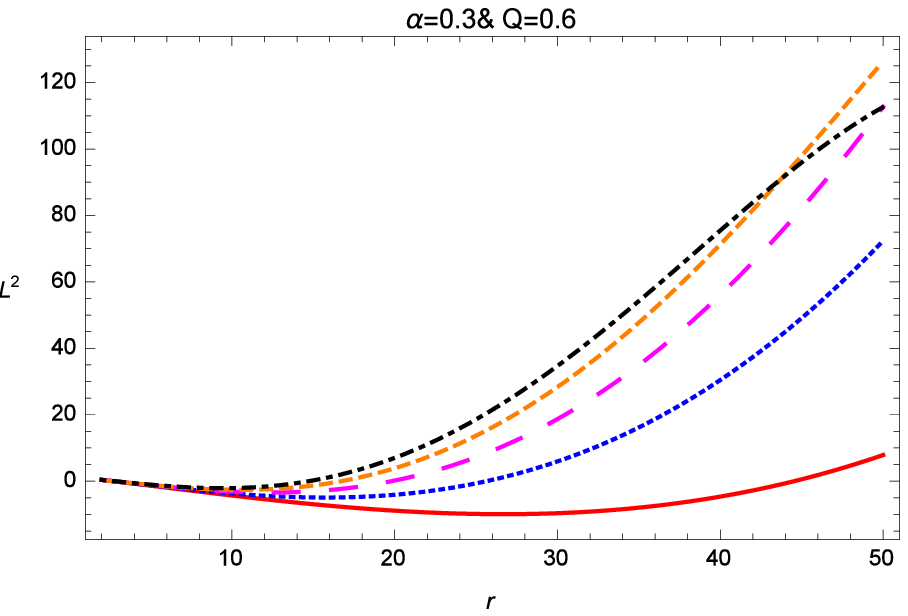, width=.32\linewidth,
height=2.02in}\epsfig{file=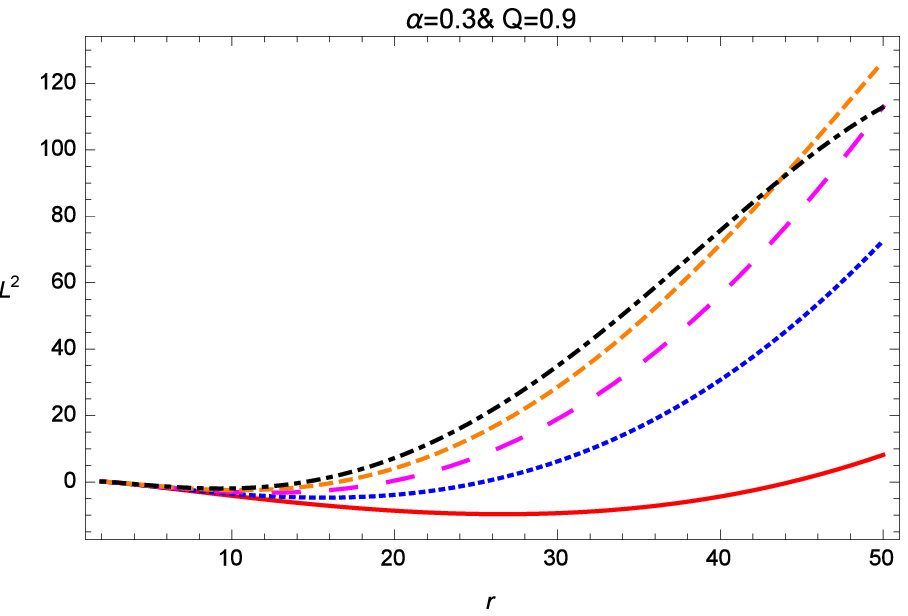, width=.32\linewidth,
height=2.02in}
\centering \epsfig{file=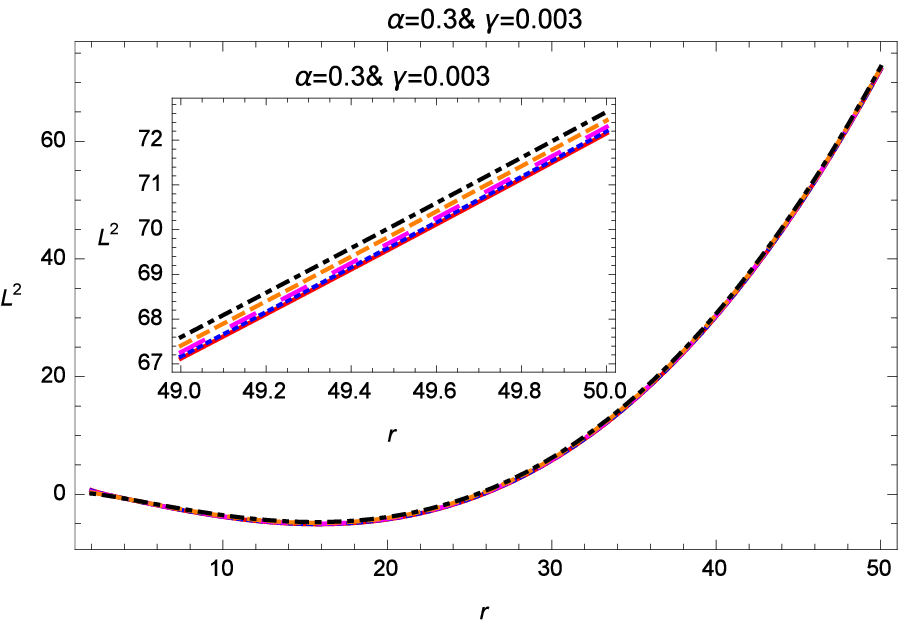, width=.32\linewidth,
height=2.02in}\epsfig{file=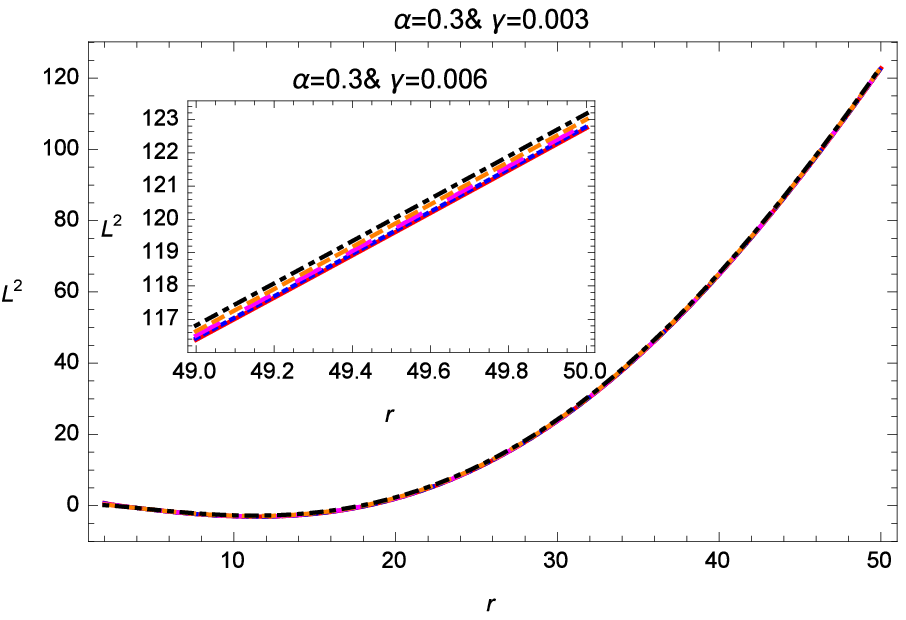, width=.32\linewidth,
height=2.02in}\epsfig{file=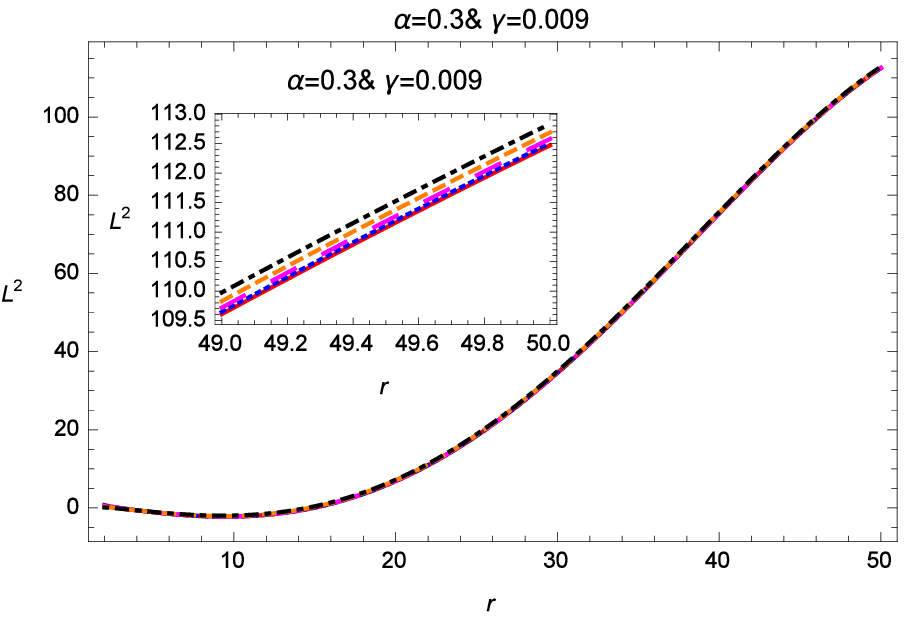, width=.32\linewidth,
height=2.02in} \caption{\label{fig5} Shows the behavior of $L^2$. In the first row, we take different five values of cloud parameter, i.e., $\alpha=0.01(\textcolor{red}{\bigstar})$, $\alpha=0.03(\textcolor{blue}{\bigstar})$, $\alpha=0.05(\textcolor{magenta}{\bigstar})$, $\alpha=0.07(\textcolor{orange}{\bigstar})$, and $\alpha=0.09(\textcolor{black}{\bigstar})$. In the second row, we take different five values of quintessential parameter, i.e., $\gamma=0.0001(\textcolor{red}{\bigstar})$, $\gamma=0.0003(\textcolor{blue}{\bigstar})$, $\gamma=0.0005(\textcolor{magenta}{\bigstar})$, $\gamma=0.0007(\textcolor{orange}{\bigstar})$, and $\gamma=0.0009(\textcolor{black}{\bigstar})$. In the third row, we take different five values of charge parameter, i.e., $Q=0.1(\textcolor{red}{\bigstar})$, $Q=0.3(\textcolor{blue}{\bigstar})$, $Q=0.5(\textcolor{magenta}{\bigstar})$, $Q=0.7(\textcolor{orange}{\bigstar})$, and $Q=0.9(\textcolor{black}{\bigstar})$.}
\end{figure}
\subsection{Null Case}

The effective potential for the null particle case is defined as:
\begin{equation}\label{23}
V_{eff}(r)= \left(-\alpha -\frac{2 M}{r}+\frac{Q ^2}{r^2}-\frac{\gamma }{r^{-1}}+1\right)\left(\frac{L^2}{r^2}\right)+\frac{q Q }{\mathfrak{m}r}.
\end{equation}
\begin{figure}
\centering \epsfig{file=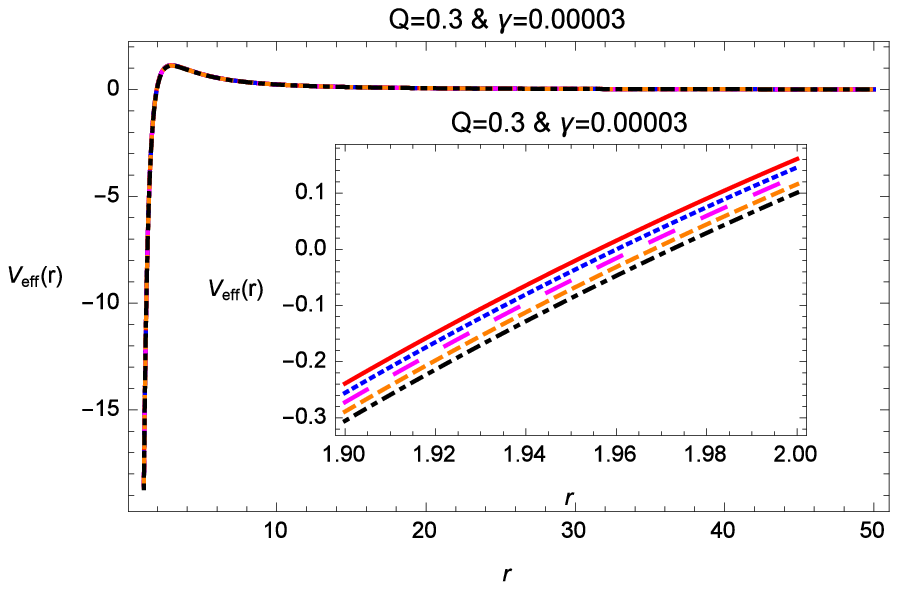, width=.32\linewidth,
height=2.02in}\epsfig{file=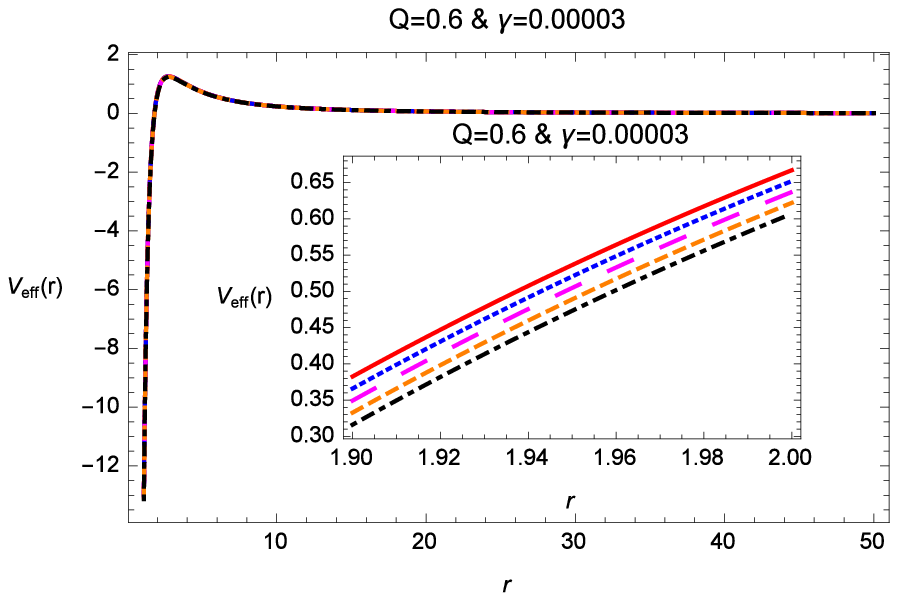, width=.32\linewidth,
height=2.02in}\epsfig{file=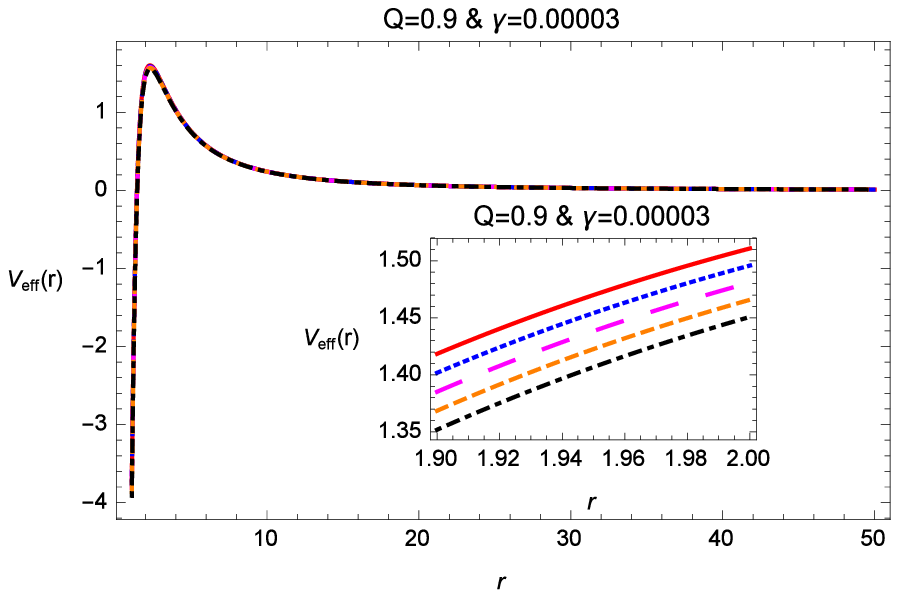, width=.32\linewidth,
height=2.02in}
\centering \epsfig{file=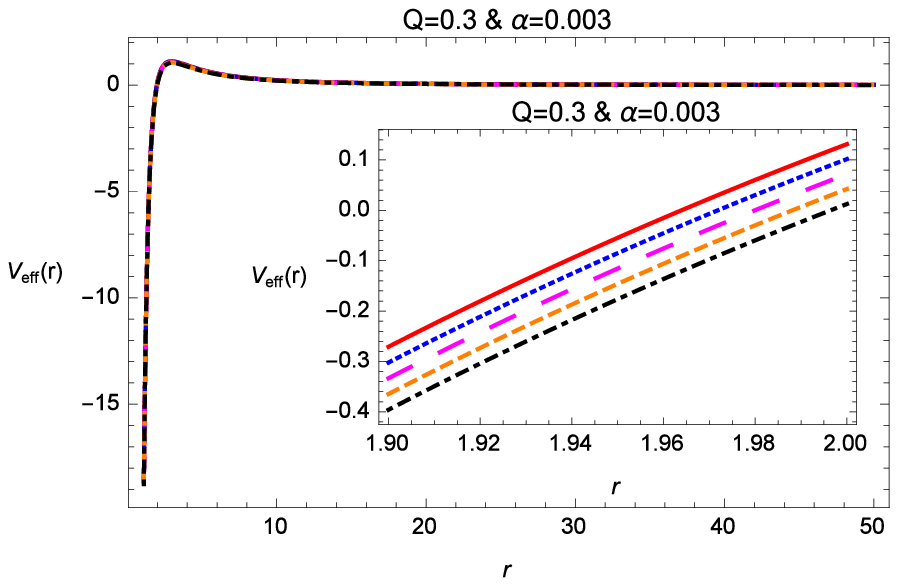, width=.32\linewidth,
height=2.02in}\epsfig{file=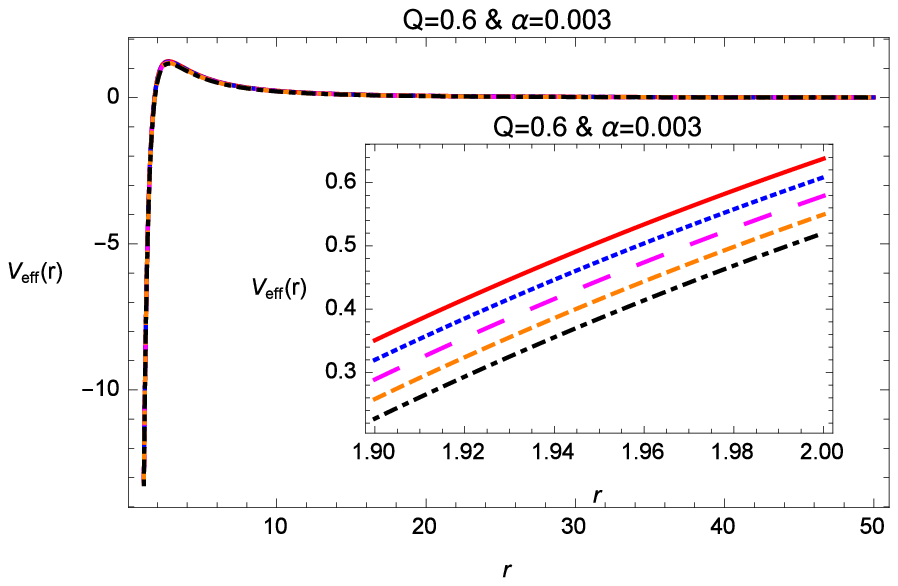, width=.32\linewidth,
height=2.02in}\epsfig{file=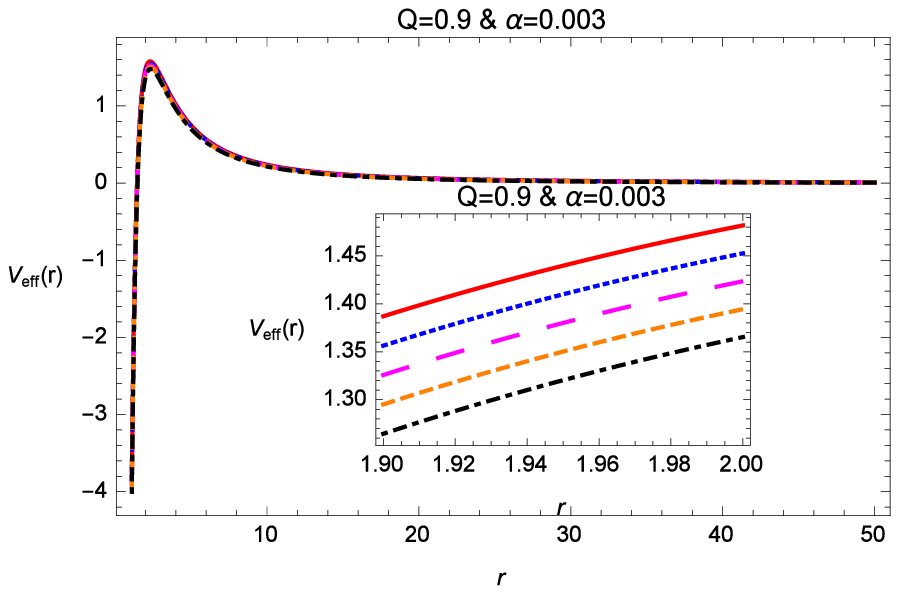, width=.32\linewidth,
height=2.02in}
\centering \epsfig{file=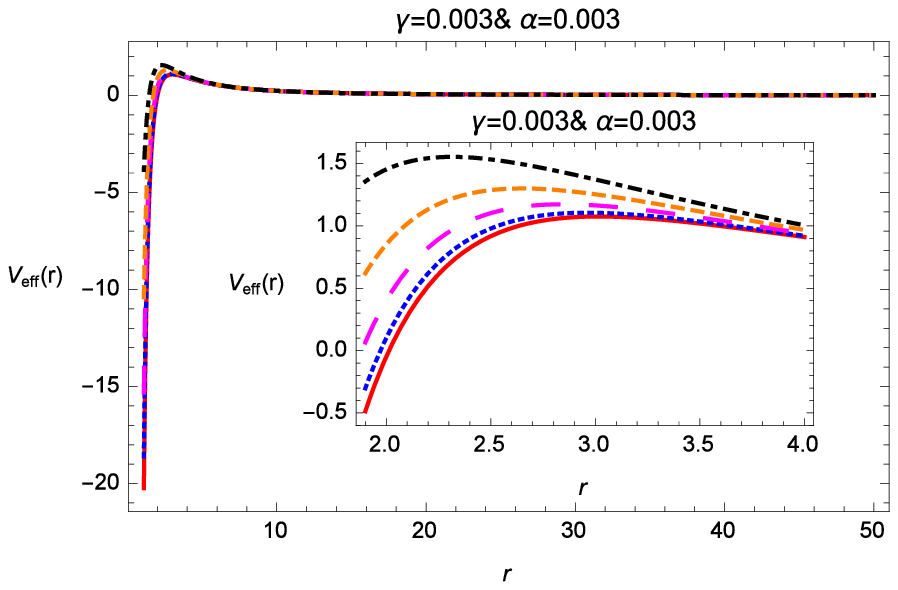, width=.32\linewidth,
height=2.02in}\epsfig{file=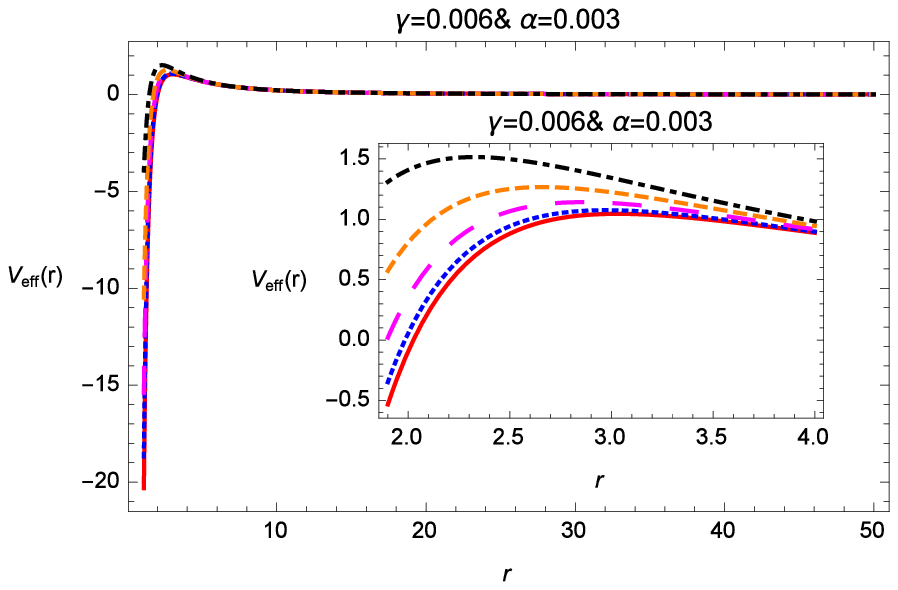, width=.32\linewidth,
height=2.02in}\epsfig{file=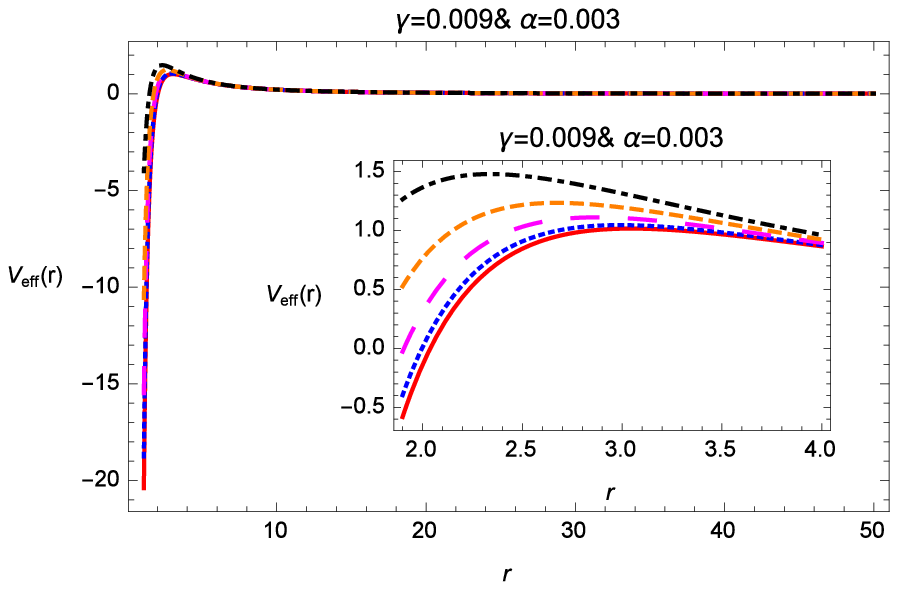, width=.32\linewidth,
height=2.02in} \caption{\label{fig6} Shows the behavior of $V_{eff}(r)$. In the first row, we take different five values of cloud parameter, i.e., $\alpha=0.01(\textcolor{red}{\bigstar})$, $\alpha=0.03(\textcolor{blue}{\bigstar})$, $\alpha=0.05(\textcolor{magenta}{\bigstar})$, $\alpha=0.07(\textcolor{orange}{\bigstar})$, and $\alpha=0.09(\textcolor{black}{\bigstar})$. In the second row, we take different five values of quintessential parameter, i.e., $\gamma=0.0001(\textcolor{red}{\bigstar})$, $\gamma=0.0003(\textcolor{blue}{\bigstar})$, $\gamma=0.0005(\textcolor{magenta}{\bigstar})$, $\gamma=0.0007(\textcolor{orange}{\bigstar})$, and $\gamma=0.0009(\textcolor{black}{\bigstar})$. In the third row, we take different five values of charge parameter, i.e., $Q=0.1(\textcolor{red}{\bigstar})$, $Q=0.3(\textcolor{blue}{\bigstar})$, $Q=0.5(\textcolor{magenta}{\bigstar})$, $Q=0.7(\textcolor{orange}{\bigstar})$, and $Q=0.9(\textcolor{black}{\bigstar})$.}
\end{figure}
The graphical behavior of the effective potential for null case is provided in Fig. (\ref{6}). The radii of the unstable photon orbit can be derived from Eq. (\ref{23}), and are calculated as:
\begin{eqnarray}\label{24}
r_{bN}&&=\bigg(-L^4 \left(\alpha ^2-2 \alpha -4 \gamma  M+1\right)-\left(L^6 \left(\alpha ^3-3 \alpha ^2-5 \gamma ^2 Q ^2+\alpha  (3-6 \gamma  M)+6 \gamma  M-1\right)+2 \gamma  L^4 Q  \left(5 \gamma  Q ^2\right.\right.\nonumber\\&&+\left.\left.3 (\alpha -1) M\right)-5 \gamma ^2 L^2 Q ^4+(\gamma ^2 \left(L^4-L^2 Q \right)^2 \left(25 \gamma ^2 Q ^6+L^4 \left(5 Q ^2 \left(-2 \alpha ^3+6 \alpha ^2-6 \alpha +5 \gamma ^2 Q ^2+2\right)\right.\right.\right.\nonumber\\&&+\left.\left.\left.64 \gamma  M^3-12 (\alpha -1)^2 M^2+60 (\alpha -1) \gamma  M Q ^2\right)-2 \gamma  L^2 Q  \left(25 \gamma  Q ^4+32 M^3+30 (\alpha -1) M Q ^2\right)\right))^{1/2}\right)^{2/3}\nonumber\\&&-L^2 \left((\alpha -1) (L^6 \left(\alpha ^3-3 \alpha ^2-5 \gamma ^2 Q ^2+\alpha  (3-6 \gamma  M)+6 \gamma  M-1\right)+2 \gamma  L^4 Q  \left(5 \gamma  Q ^2+3 (\alpha -1) M\right)-5 \right.\nonumber\\&&\times\left.\gamma ^2 L^2 Q ^4+(\gamma ^2 \left(L^4-L^2 Q \right)^2 \left(25 \gamma ^2 Q ^6+L^4 \left(5 Q ^2 \left(-2 \alpha ^3+6 \alpha ^2-6 \alpha +5 \gamma ^2 Q ^2+2\right)+64 \gamma  M^3-12 \right.\right.\right.\nonumber\\&&\times\left.\left.\left.(\alpha -1)^2 M^2+60 (\alpha -1) \gamma  M Q ^2\right)-2 \gamma  L^2 Q  \left(25 \gamma  Q ^4+32 M^3+30 (\alpha -1) M Q ^2\right)\right))^{1/2})^{1/3}+4 \gamma  M Q \right)\bigg)\nonumber\\&&\times \bigg(\gamma  \left(L^2-Q \right) (L^6 \left(\alpha ^3-3 \alpha ^2-5 \gamma ^2 Q ^2+\alpha  (3-6 \gamma  M)+6 \gamma  M-1\right)+2 \gamma  L^4 Q  \left(5 \gamma  Q ^2+3 (\alpha -1) M\right)\nonumber\\&&-5 \gamma ^2 L^2 Q ^4+(\gamma ^2 \left(L^4-L^2 Q \right)^2 \left(25 \gamma ^2 Q ^6+L^4 \left(5 Q ^2 \left(-2 \alpha ^3+6 \alpha ^2-6 \alpha +5 \gamma ^2 Q ^2+2\right)+64 \gamma  M^3-12 \right.\right.\nonumber\\&&\times\left.\left.(\alpha -1)^2 M^2+60 (\alpha -1) \gamma  M Q ^2\right)-2 \gamma  L^2 Q  \left(25 \gamma  Q ^4+32 M^3+30 (\alpha -1) M Q ^2\right)\right))^{1/2})^{1/3}\bigg)^{-1}.
\end{eqnarray}
\begin{figure}
\centering \epsfig{file=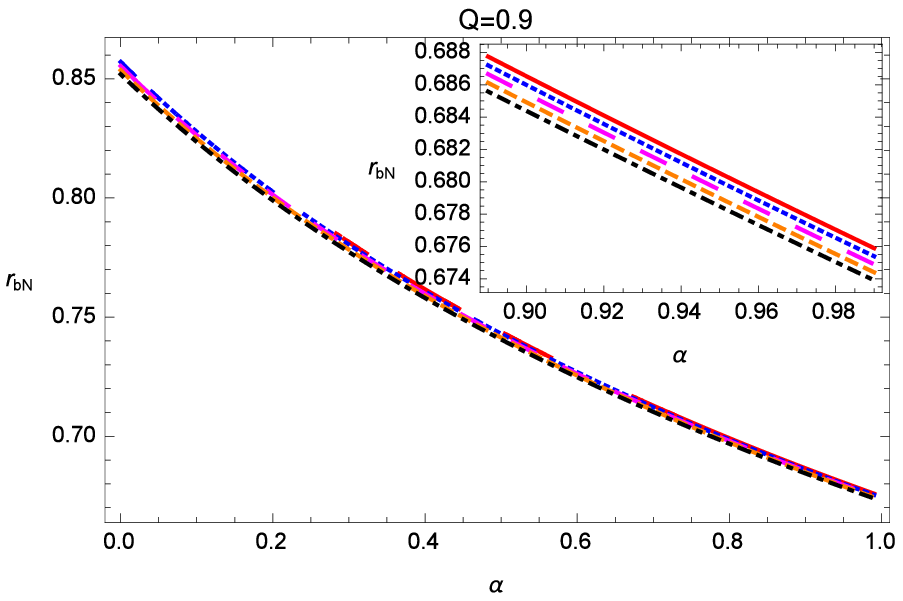, width=.32\linewidth,
height=2.02in}\epsfig{file=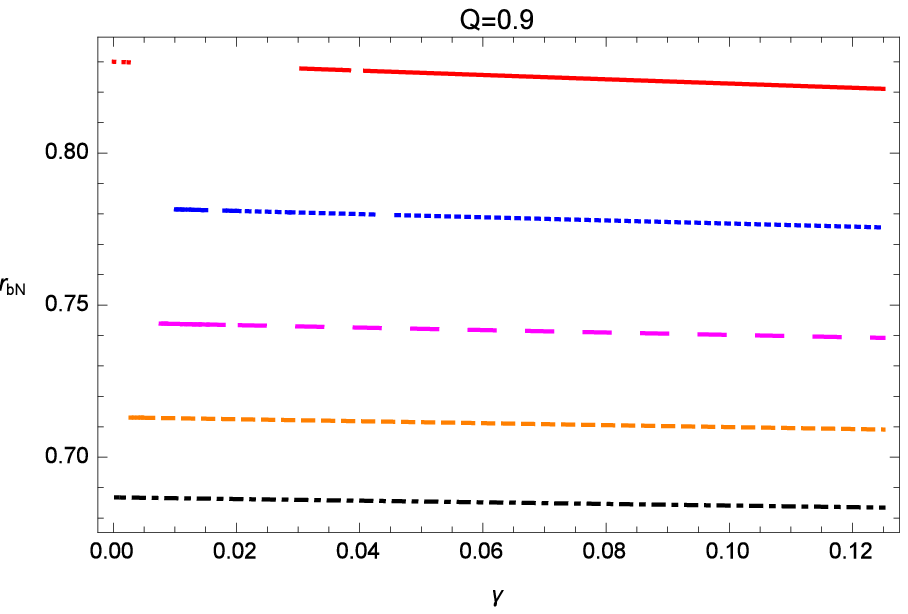, width=.32\linewidth,
height=2.02in}\epsfig{file=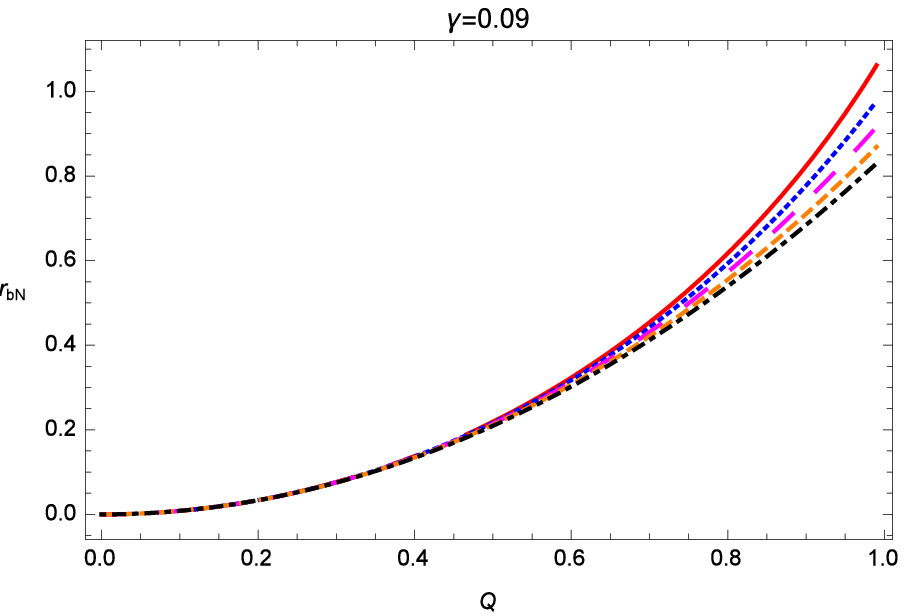, width=.32\linewidth,
height=2.02in}\caption{\label{fig7} Shows the behavior of $r_{bN}$. The left panel represents $\gamma=0.01(\textcolor{red}{\bigstar})$, $\gamma=0.03(\textcolor{blue}{\bigstar})$, $\gamma=0.05(\textcolor{magenta}{\bigstar})$, $\gamma=0.07(\textcolor{orange}{\bigstar})$, and $\gamma=0.09(\textcolor{black}{\bigstar})$, The middle part represents $\alpha=0.1(\textcolor{red}{\bigstar})$, $\alpha=0.3(\textcolor{blue}{\bigstar})$, $\alpha=0.5(\textcolor{magenta}{\bigstar})$, $\alpha=0.7(\textcolor{orange}{\bigstar})$, and $\alpha=0.9(\textcolor{black}{\bigstar})$, and right penal represents $Q=0.1(\textcolor{red}{\bigstar})$, $Q=0.3(\textcolor{blue}{\bigstar})$, $Q=0.5(\textcolor{magenta}{\bigstar})$, $Q=0.7(\textcolor{orange}{\bigstar})$, and $Q=0.9(\textcolor{black}{\bigstar})$.}
\end{figure}
The behavior of $r_{bN}$, can be seen from Fig. (\ref{7}), which shows that the radii of the photon circular orbits decrease with increasing $\alpha$ and $\gamma$, while they increase with increase in the charge $Q$ of the black hole.

\subsection{Timelike Case}

The effective potential for timelike case is expressed as:
\begin{equation}\label{25}
V_{eff}(r)= \left(-\alpha -\frac{2 M}{r}+\frac{Q ^2}{r^2}-\frac{\gamma }{r^{-1}}+1\right)\left(1+\frac{L^2}{r^2}\right)+\frac{q Q }{\mathfrak{m}r}.
\end{equation}
In the Figs. (\ref{8}-\ref{10}) it is shown that how the effective potential for a test particle varies with the parameters $\alpha$, $\gamma$ and charge $Q$ of the black hole.
\begin{figure}
\centering \epsfig{file=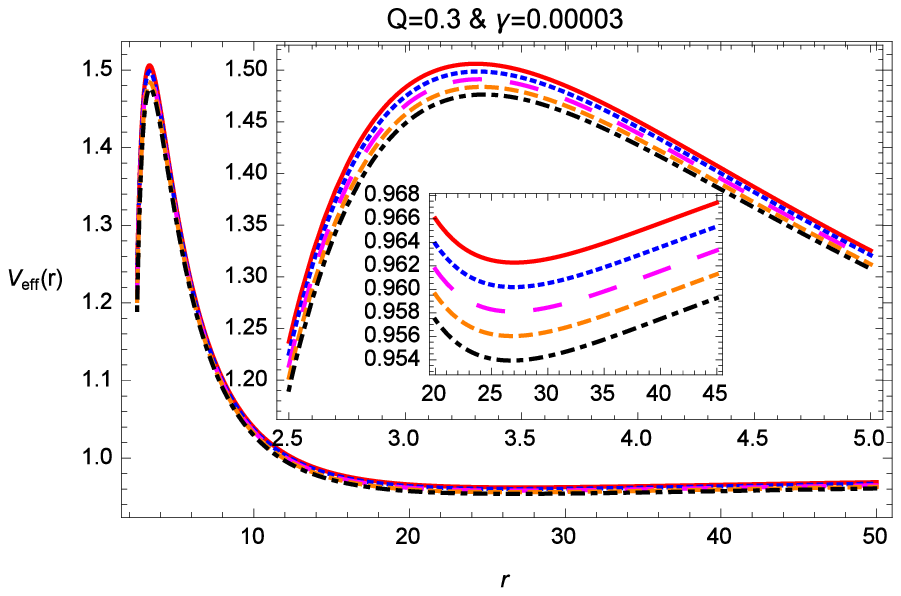, width=.32\linewidth,
height=2.02in}\epsfig{file=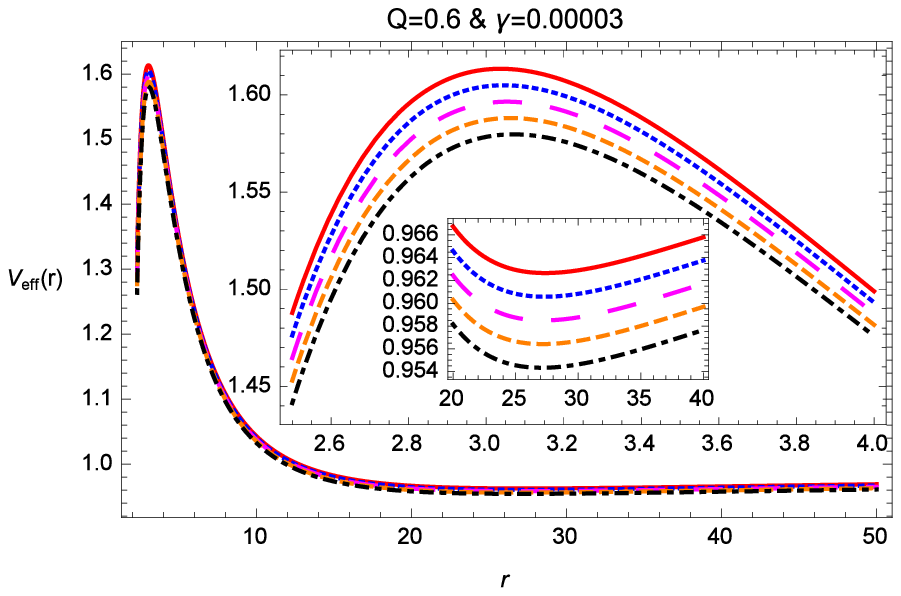, width=.32\linewidth,
height=2.02in}\epsfig{file=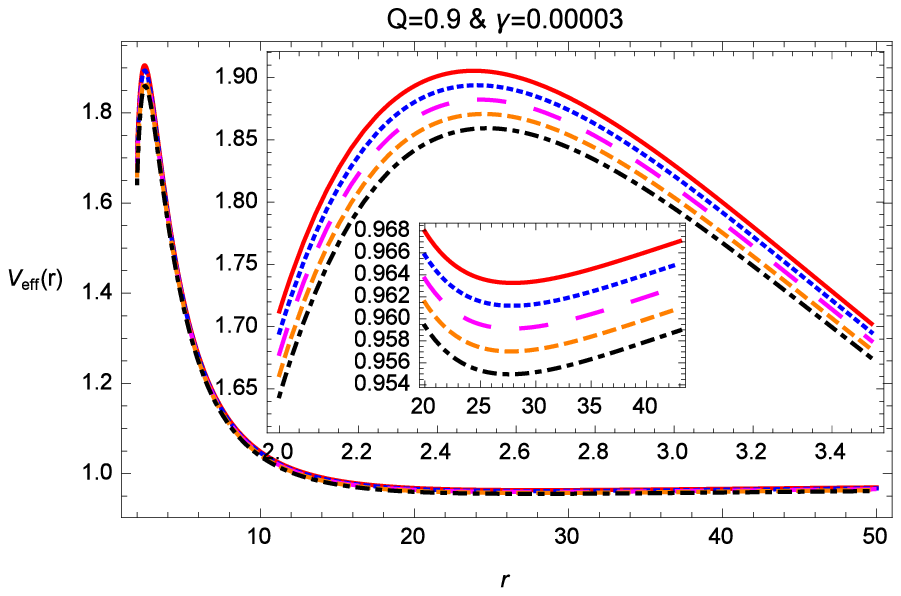, width=.32\linewidth,
height=2.02in}
\centering \epsfig{file=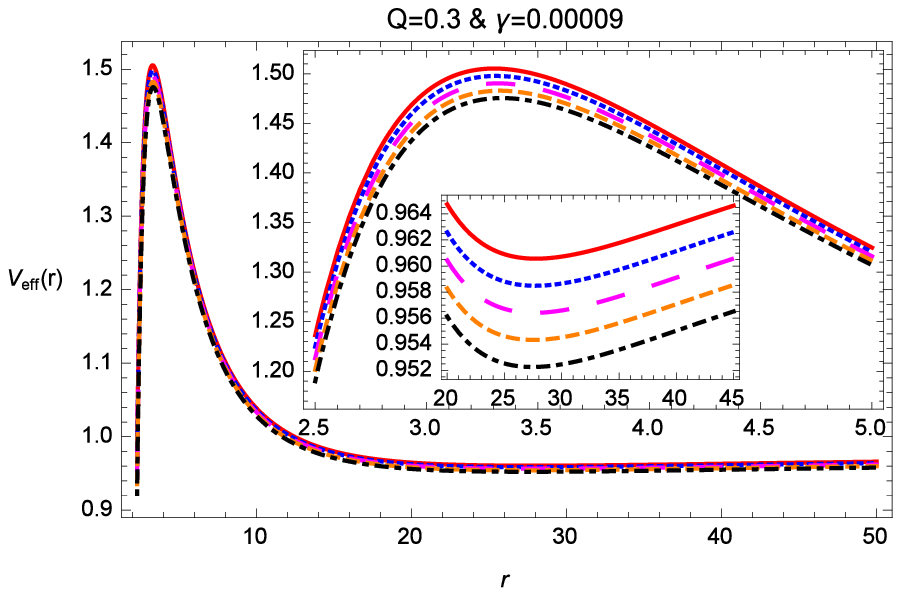, width=.32\linewidth,
height=2.02in}\epsfig{file=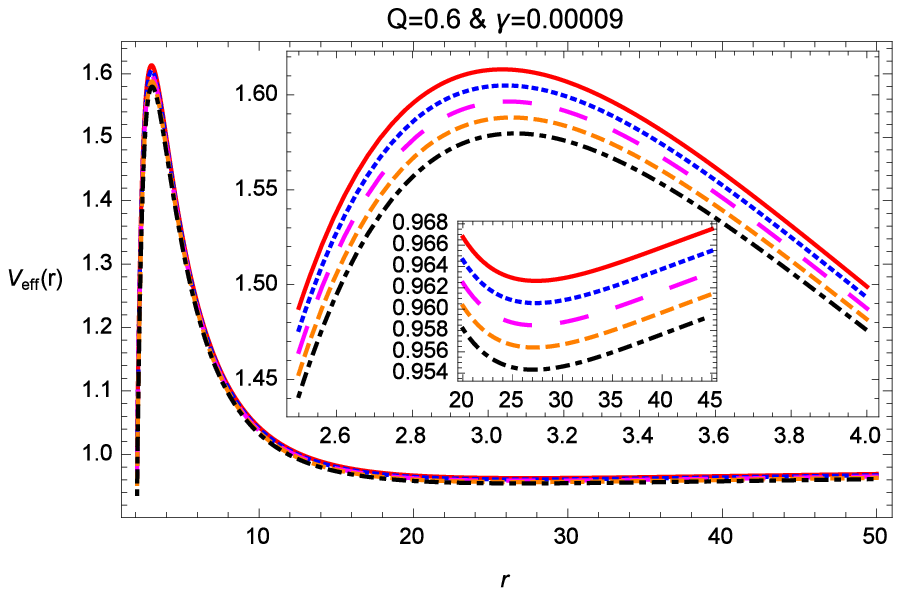, width=.32\linewidth,
height=2.02in}\epsfig{file=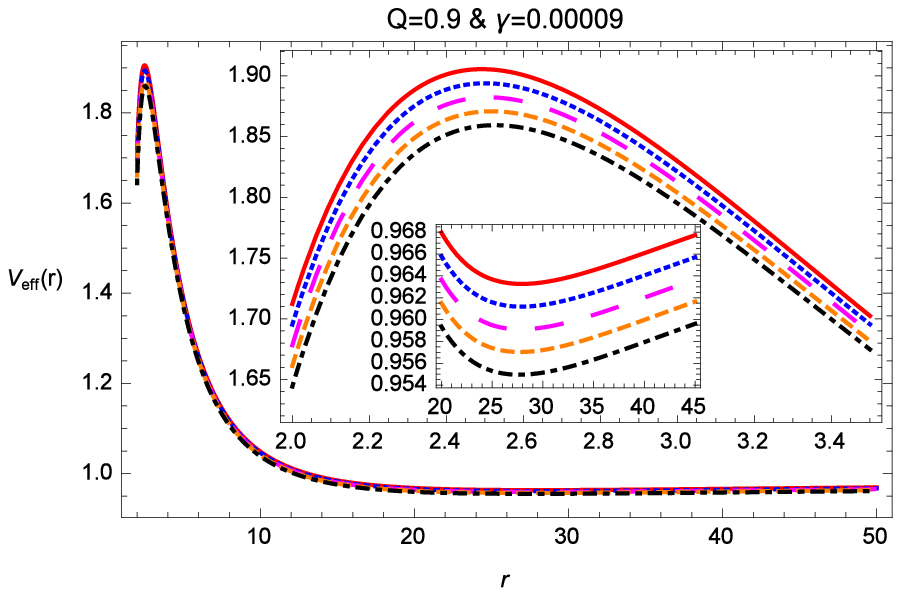, width=.32\linewidth,
height=2.02in} \caption{\label{fig8} Shows the behavior of $V_{eff}$ for timelike case. In the first and second row, we take different five values of cloud parameter, i.e., $\alpha=0.001(\textcolor{red}{\bigstar})$, $\alpha=0.003(\textcolor{blue}{\bigstar})$, $\alpha=0.005(\textcolor{magenta}{\bigstar})$, $\alpha=0.007(\textcolor{orange}{\bigstar})$, and $\alpha=0.009(\textcolor{black}{\bigstar})$.}
\end{figure}
\begin{figure}
\centering \epsfig{file=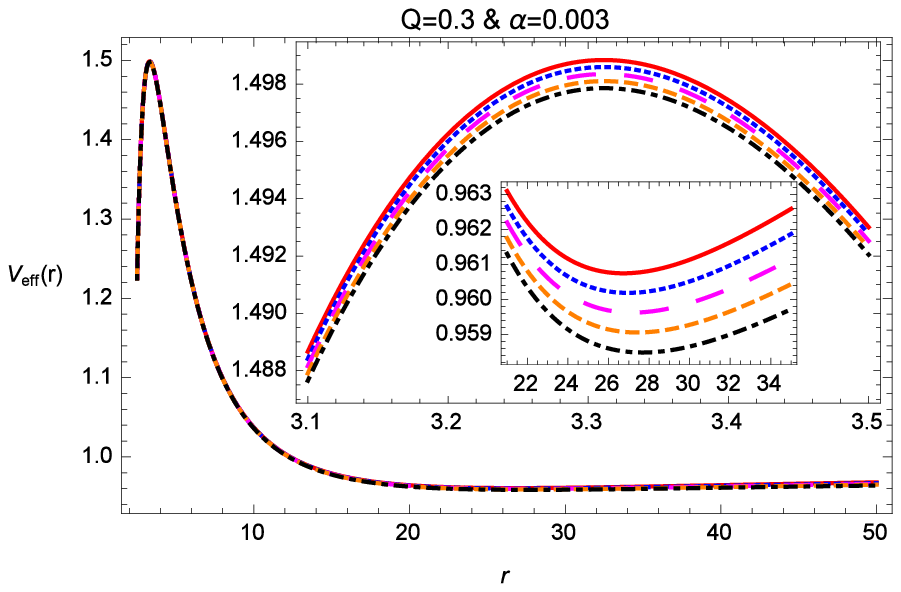, width=.32\linewidth,
height=2.02in}\epsfig{file=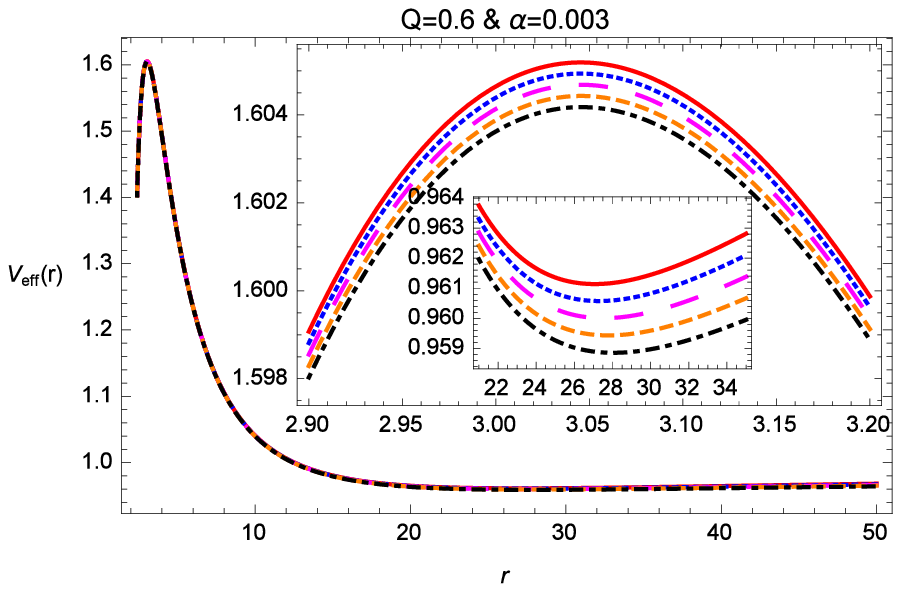, width=.32\linewidth,
height=2.02in}\epsfig{file=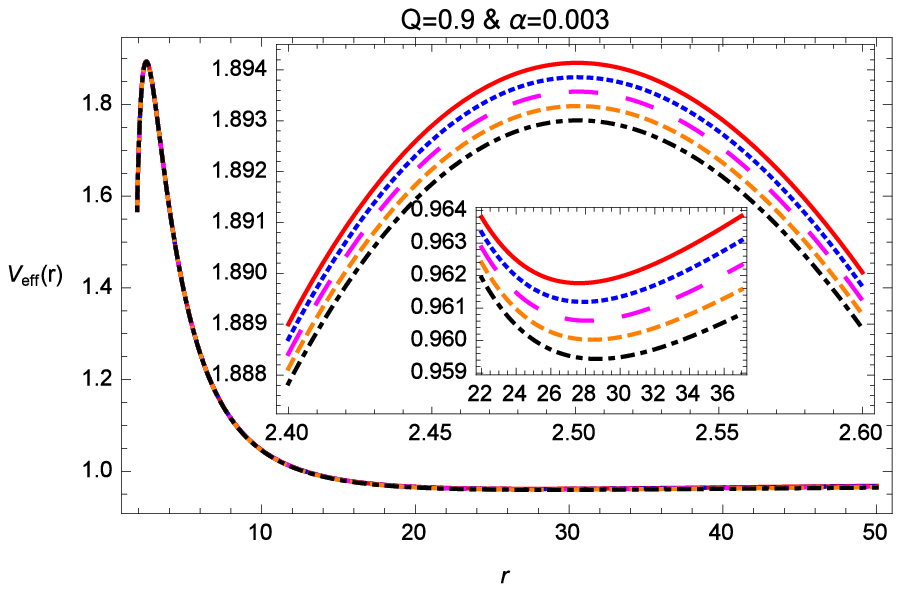, width=.32\linewidth,
height=2.02in}
\centering \epsfig{file=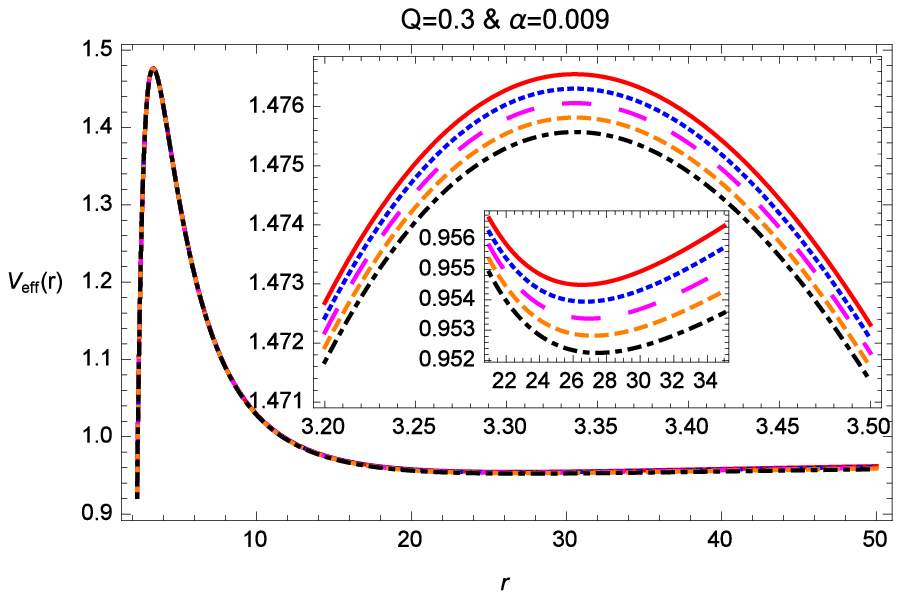, width=.32\linewidth,
height=2.02in}\epsfig{file=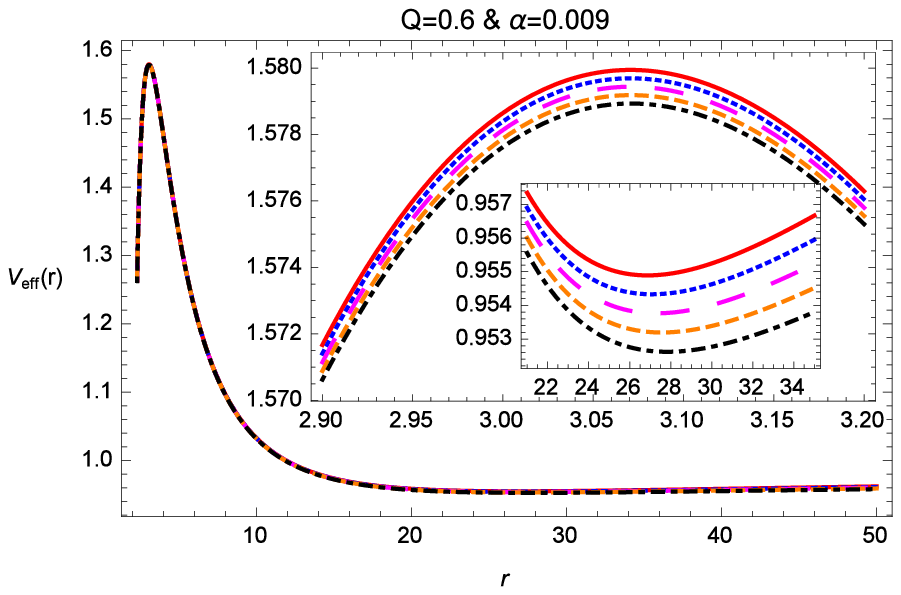, width=.32\linewidth,
height=2.02in}\epsfig{file=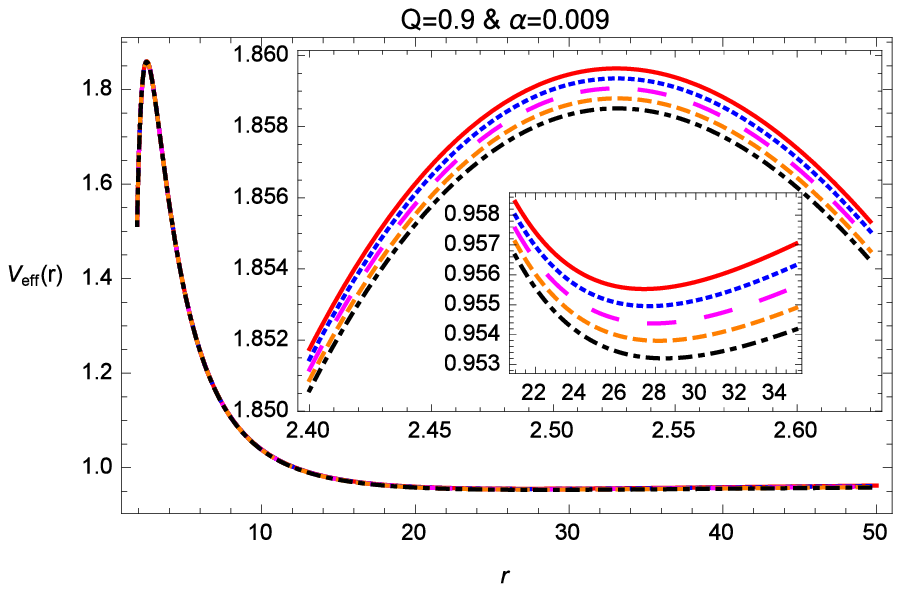, width=.32\linewidth,
height=2.02in} \caption{\label{fig9} Shows the behavior of $V_{eff}$ for timelike case. In the first and second row, we take different five values of cloud parameter, i.e., $\gamma=0.00001(\textcolor{red}{\bigstar})$, $\gamma=0.00003(\textcolor{blue}{\bigstar})$, $\gamma=0.00005(\textcolor{magenta}{\bigstar})$, $\gamma=0.00007(\textcolor{orange}{\bigstar})$, and $\gamma=0.00009(\textcolor{black}{\bigstar})$.}
\end{figure}
\begin{figure}
\centering \epsfig{file=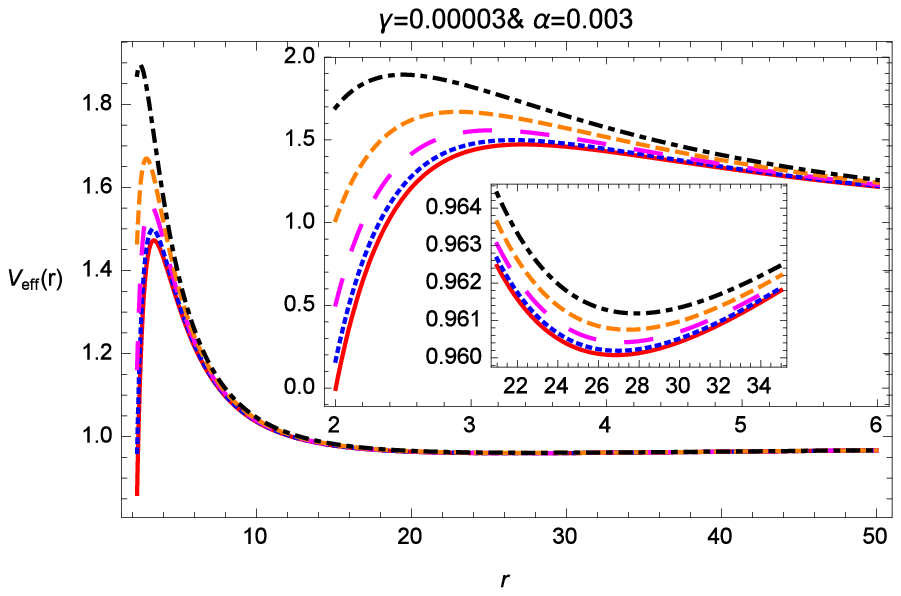, width=.32\linewidth,
height=2.02in}\epsfig{file=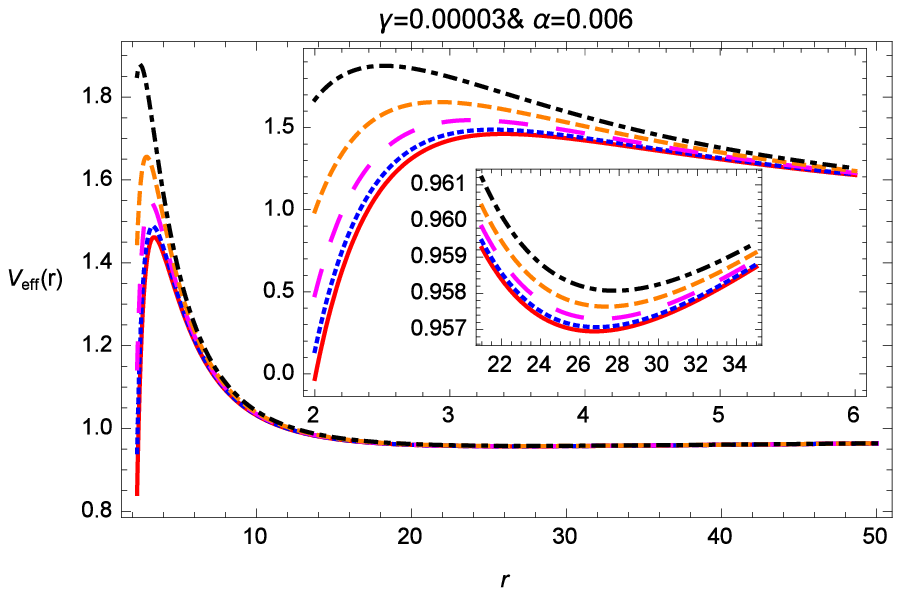, width=.32\linewidth,
height=2.02in}\epsfig{file=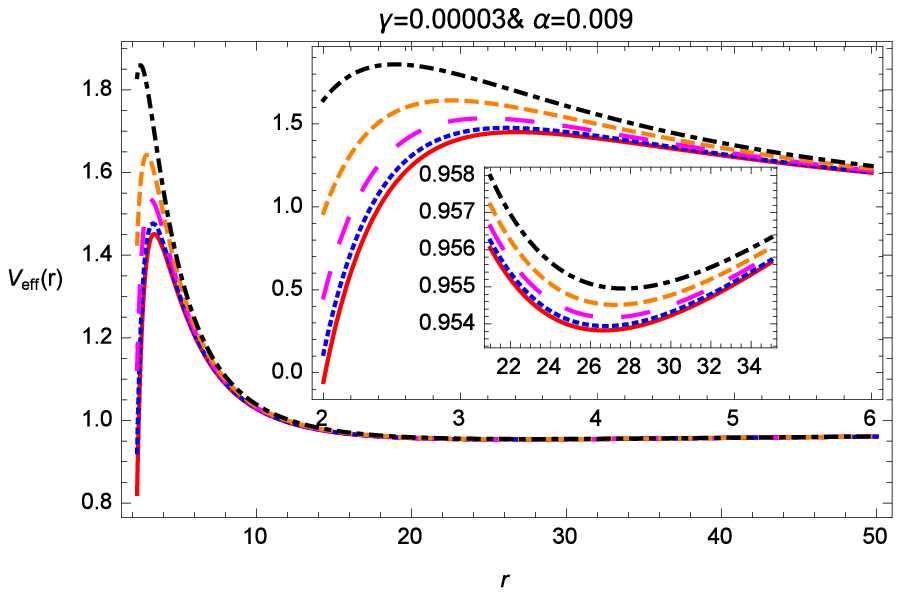, width=.32\linewidth,
height=2.02in}
\centering \epsfig{file=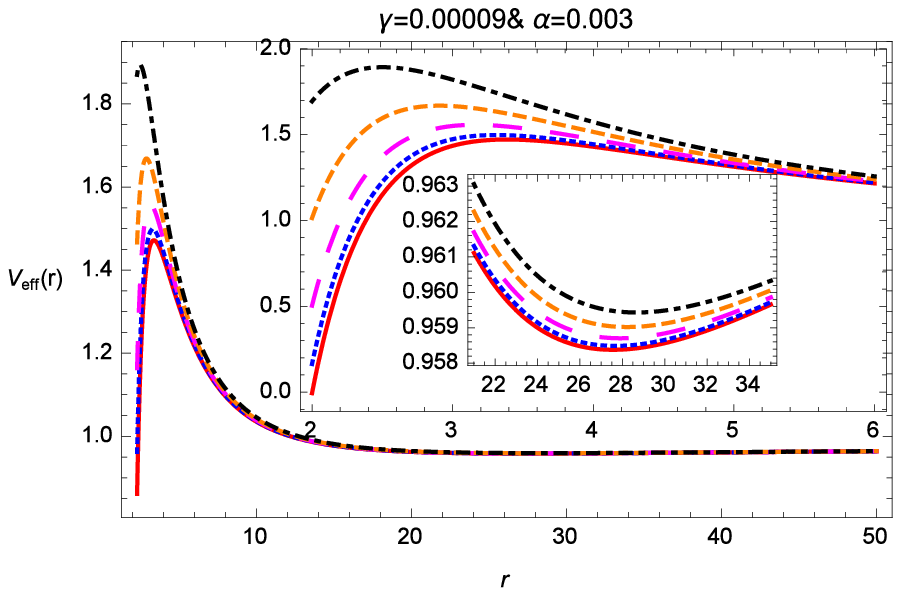, width=.32\linewidth,
height=2.02in}\epsfig{file=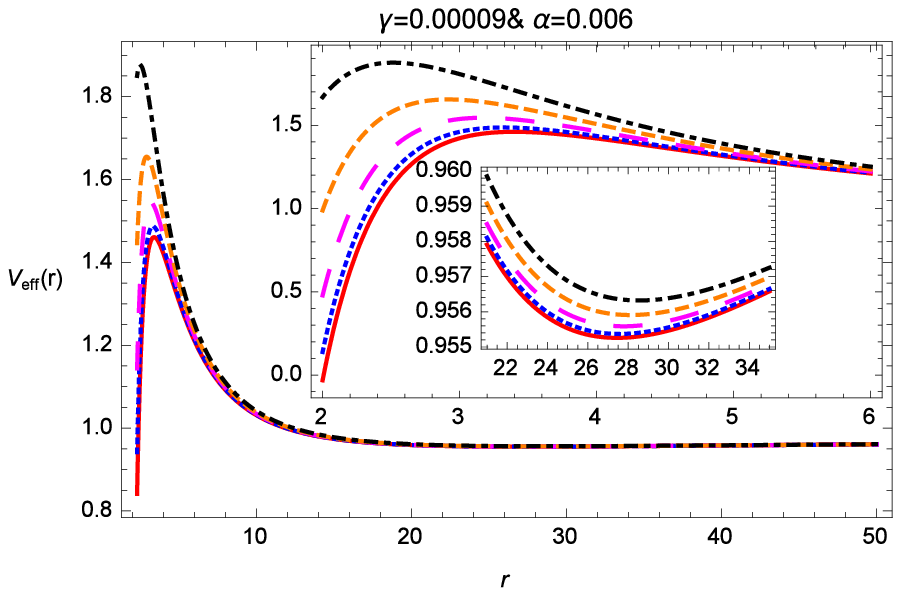, width=.32\linewidth,
height=2.02in}\epsfig{file=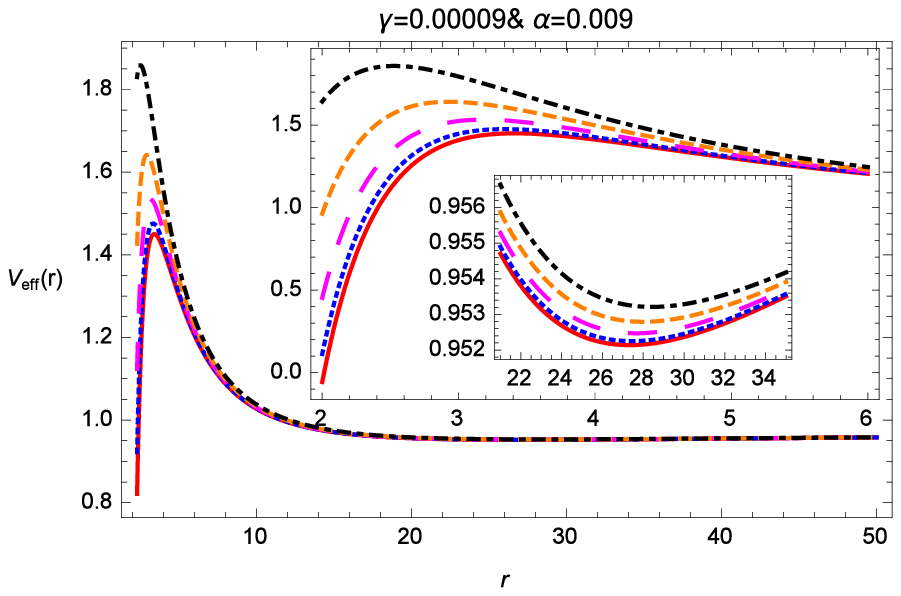, width=.32\linewidth,
height=2.02in} \caption{\label{fig10} Shows the behavior of $V_{eff}$ for timelike case. In the first and second row, we take different five values of cloud parameter, i.e., $Q=0.1(\textcolor{red}{\bigstar})$, $Q=0.3(\textcolor{blue}{\bigstar})$, $Q=0.5(\textcolor{magenta}{\bigstar})$, $Q=0.7(\textcolor{orange}{\bigstar})$, and $Q=0.9(\textcolor{black}{\bigstar})$.}
\end{figure}

The radii of the circular orbit for massive particles are calculated as
\begin{eqnarray}\label{26}
r_{bP}&&=-\frac{1}{\gamma  \left(L^2-Q \right)+2 M}\bigg((\alpha -1) L^2+\left(L^4 \left(\alpha ^2-2 \alpha -4 \gamma  M+1\right)+L^2 \left(-2 (\alpha -1) Q ^2-8 M^2+4 \gamma  M Q \right)\right.\nonumber\\&&+\left.Q ^4\right)-Q ^2 \times \bigg(L^6 \left(\alpha ^3-3 \alpha ^2-5 \gamma ^2 Q ^2+\alpha  (3-6 \gamma  M)+6 \gamma  M-1\right)+L^4 \left(Q ^2 \left(10 \gamma ^2 Q -3 (\alpha -1)^2\right)-12 \right.\nonumber\\&&\times\left.(\alpha -1) M^2-2 \gamma  M Q  (-3 \alpha +7 Q +3)\right)+L^2 Q ^2 \left(Q ^2 \left(3 \alpha -5 \gamma ^2-3\right)-8 M^2+14 \gamma  M Q \right)+(L^2 \left(\gamma  \left(L^2-Q \right)\right.\nonumber\\&&+\left.2 M\right)^2 \left(L^6 \left(5 Q ^2 \left(-2 \alpha ^3+6 \alpha ^2-6 \alpha +5 \gamma ^2 Q ^2+2\right)+64 \gamma  M^3-12 (\alpha -1)^2 M^2+60 (\alpha -1) \gamma  M Q ^2\right)+2 L^4\right.\nonumber\\&&\times\left. \left(5 Q ^4 \left(3 (\alpha -1)^2-5 \gamma ^2 Q \right)+64 M^4-32 \gamma  M^3 Q +72 (\alpha -1) M^2 Q ^2+10 \gamma  M Q ^3 (-3 \alpha +2 Q +3)\right)+L^2 Q ^4 \right.\nonumber\\&&\times\left.\left(5 Q ^2 \left(-6 \alpha +5 \gamma ^2+6\right)-32 M^2-40 \gamma  M Q \right)+10 Q ^8\right))^{1/2}-Q ^6\bigg)^{-1/3}+\big(L^6 \left(\alpha ^3-3 \alpha ^2-5 \gamma ^2 Q ^2+\alpha  \right.\nonumber\\&&\times\left.(3-6 \gamma  M)+6 \gamma  M-1\right)+L^4 \left(Q ^2 \left(10 \gamma ^2 Q -3 (\alpha -1)^2\right)-12 (\alpha -1) M^2-2 \gamma  M Q  (-3 \alpha +7 Q +3)\right)+L^2\nonumber\\&&\times Q ^2 \left(Q ^2 \left(3 \alpha -5 \gamma ^2-3\right)-8 M^2+14 \gamma  M Q \right)+(L^2 \left(\gamma  \left(L^2-Q \right)+2 M\right)^2 \left(L^6 \left(5 Q ^2 \left(-2 \alpha ^3+6 \alpha ^2-6 \alpha +5 \right.\right.\right.\nonumber\\&&\times\left.\left.\left. \gamma ^2 Q ^2+2\right)+64 \gamma  M^3-12 (\alpha -1)^2 M^2+60 (\alpha -1) \gamma  M Q ^2\right)+2 L^4 \left(5 Q ^4 \left(3 (\alpha -1)^2-5 \gamma ^2 Q \right)+64 M^4\right.\right.\nonumber\\&&-\left.\left.32 \gamma  M^3 Q +72 (\alpha -1) M^2 Q ^2+10 \gamma  M Q ^3 (-3 \alpha +2 Q +3)\right)+L^2 Q ^4 \left(5 Q ^2 \left(-6 \alpha +5 \gamma ^2+6\right)-32 M^2\right.\right.\nonumber\\&&-\left.\left.40 \gamma  M Q \right)+10 Q ^8\right))^{1/2}-Q ^6\bigg)\bigg)
\end{eqnarray}
\begin{figure}
\centering \epsfig{file=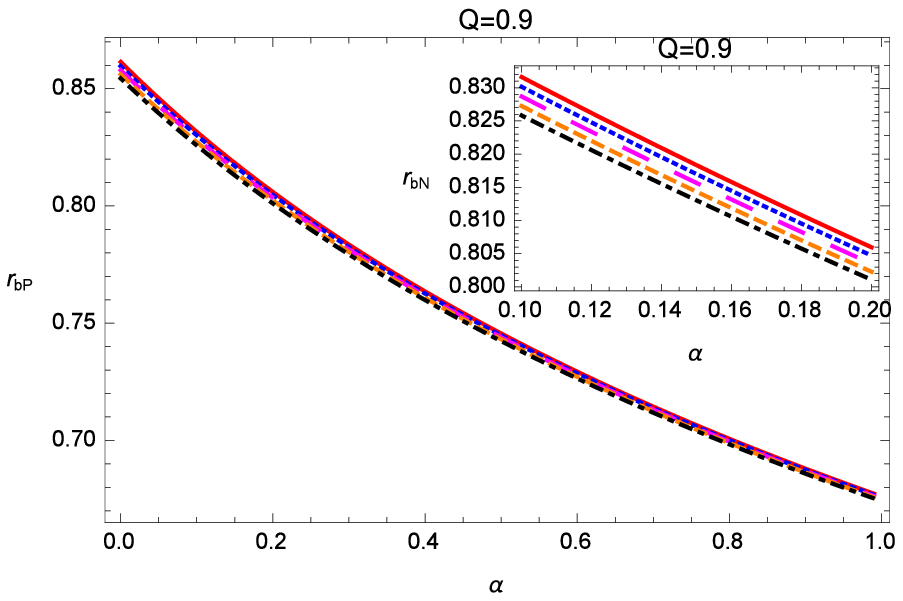, width=.32\linewidth,
height=2.02in}\epsfig{file=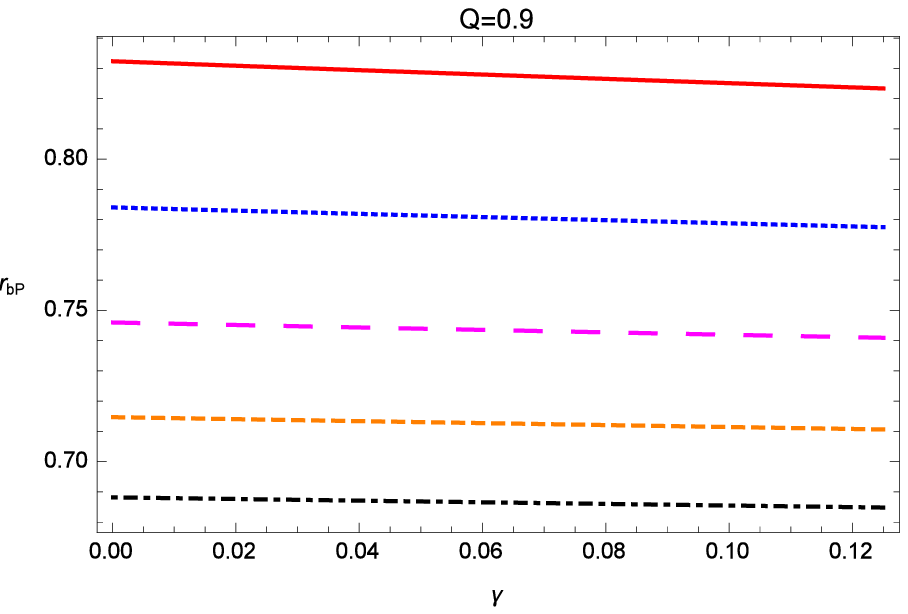, width=.32\linewidth,
height=2.02in}\epsfig{file=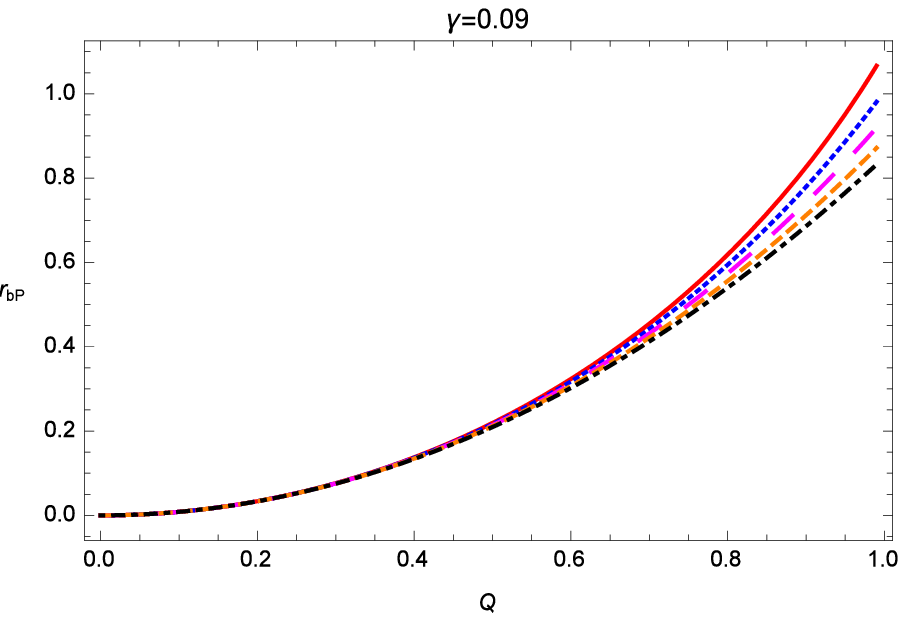, width=.32\linewidth,
height=2.02in}\caption{\label{fig11} Shows the behavior of $r_{bP}$ for timelike case. The left panel represents $\gamma=0.01(\textcolor{red}{\bigstar})$, $\gamma=0.03(\textcolor{blue}{\bigstar})$, $\gamma=0.05(\textcolor{magenta}{\bigstar})$, $\gamma=0.07(\textcolor{orange}{\bigstar})$, and $\gamma=0.09(\textcolor{black}{\bigstar})$, The middle part represents $\alpha=0.1(\textcolor{red}{\bigstar})$, $\alpha=0.3(\textcolor{blue}{\bigstar})$, $\alpha=0.5(\textcolor{magenta}{\bigstar})$, $\alpha=0.7(\textcolor{orange}{\bigstar})$, and $\alpha=0.9(\textcolor{black}{\bigstar})$, and right penal represents $Q=0.1(\textcolor{red}{\bigstar})$, $Q=0.3(\textcolor{blue}{\bigstar})$, $Q=0.5(\textcolor{magenta}{\bigstar})$, $Q=0.7(\textcolor{orange}{\bigstar})$, and $Q=0.9(\textcolor{black}{\bigstar})$.}
\end{figure}
From the Fig. (\ref{11}) one can infer that  the radii of the stable circular orbits for the charged test particles decrease with increasing  values of $\alpha$ and $\gamma$, and they increase with increase in the values of the charge $Q$ of the black hole.

\section{Oscillations of massive particles near the circular orbits}
Here we give the epicyclic frequencies of the QPOs of charged test particles moving in the close vicinity of stable circular orbits. For these frequencies the general formulas are given as follows (for detail one may see \cite{44}). The equatorial timelike circular orbits are characterized by three fundamental
frequencies namely (i) the Keplerian frequency or orbital frequency $\nu_{\phi}=\frac{\Omega _{\phi}}{2 \pi}$, (ii) the radial epicyclic frequency $\nu_r =\frac{\Omega _r}{2 \pi}$, and is the frequency of the radial oscillations in the surrounding of the mean orbit and (iii) the vertical epicyclic frequency $\nu_{\theta}=\frac{\Omega _{\theta}}{2 \pi}$, and is the frequency of the vertical oscillations around the mean orbit, were
\begin{eqnarray}
\Omega _{\phi }&&=\frac{d\phi }{dt}=\frac{-\frac{\partial g_{t\phi}}{\partial r}\pm\sqrt{\left(\frac{\partial g_{t\phi}}{\partial r}\right){}^2-\frac{\partial g_{tt}}{\partial r} \frac{\partial g_{\phi \phi }}{\partial r}}}{\frac{\partial g_{\phi \phi }}{\partial r}},\label{26}\\
\Omega _r^2&&=-\frac{1}{2 \dot{t}^{2} g_{rr}}\frac{\partial ^2V_{eff}}{\partial r^2},\label{27}\\
\Omega _{\theta }^2&&=-\frac{1}{2 \dot{t}^{2} g_{\theta \theta }}\frac{\partial ^2V_{eff}}{\partial \theta ^2}.\label{28}
\end{eqnarray}
For the metric presented by Eq. (\ref{1}) we get the following fundamental frequencies of the massive particles near the circular orbits
\begin{eqnarray}
\nu_{\phi}&&=\nu_{\theta}=\frac{\sqrt{r \left(-\gamma +\frac{2 M}{r^2}-\frac{2 Q^2}{r^3}\right)}}{2 \sqrt{2} \pi  r},\label{29}\\
\nu_{r}&&=\frac{\sqrt{-\frac{\left(-\alpha-\frac{2 M}{r}+\frac{Q^2}{r^2}-\gamma  r+1\right) \left(\frac{6 L^2 \left(-\alpha-\frac{2 M}{r}+\frac{Q^2}{r^2}-\gamma  r+1\right)}{r^4}-\frac{4 L^2 \left(-\gamma +\frac{2 M}{r^2}-\frac{2 Q^2}{r^3}\right)}{r^3}+\left(\frac{L^2}{r^2}+1\right) \left(\frac{6 Q^2}{r^4}-\frac{4 M}{r^3}\right)+\frac{2 q Q}{r^3}\right)}{2 \left(E-\frac{q Q}{r}\right)^2}}}{2 \pi}.\label{30}
\end{eqnarray}
The $\nu_{\theta}=\nu_{\phi}$ depend on the black hole's parameters $M$ and $Q$, the radial coordinate $r$ and quintessence parameter $\gamma$. It does not depend on the string cloud parameter $\alpha$. The $\nu_r$ along with $r$ and the parameters $M$, $Q$ and $\gamma$ also depends on the parameter $\alpha$. Besides, the expression for $\nu_r$ also involves the charge $q$, the energy $E$ and the angular momentum $L$ of the particle. In the Fig. (\ref{12}) we plot these frequencies for different values of $\alpha$, $\gamma$ and $Q$. The behavior $\nu_{\phi}=\nu_{\theta}$ remains same irrespective of the values of $\gamma$ and $Q$. The frequency $\nu_r$ becomes smaller as the values of the parameter $\alpha$ increase. For an increase in the values of $Q$ the frequency $\nu_r$ can be observed away from the central object for comparatively smaller values of the parameter $\alpha$. Further, it is seen that for bigger values of $\alpha$ i.e. close to one, the frequency $\nu_r$ does not change for different values of $Q$ and $\gamma$.
\begin{figure}
\centering \epsfig{file=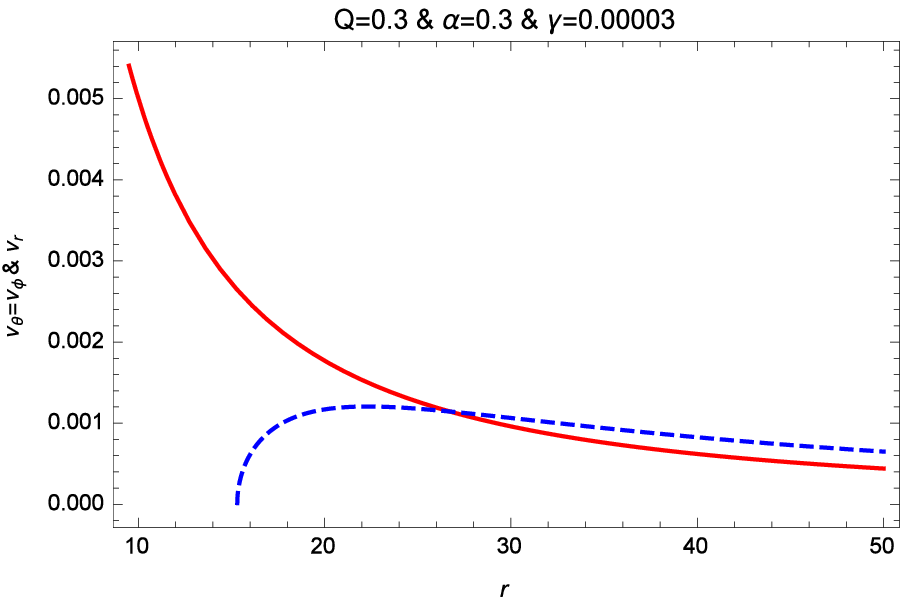, width=.32\linewidth,
height=2.02in}\epsfig{file=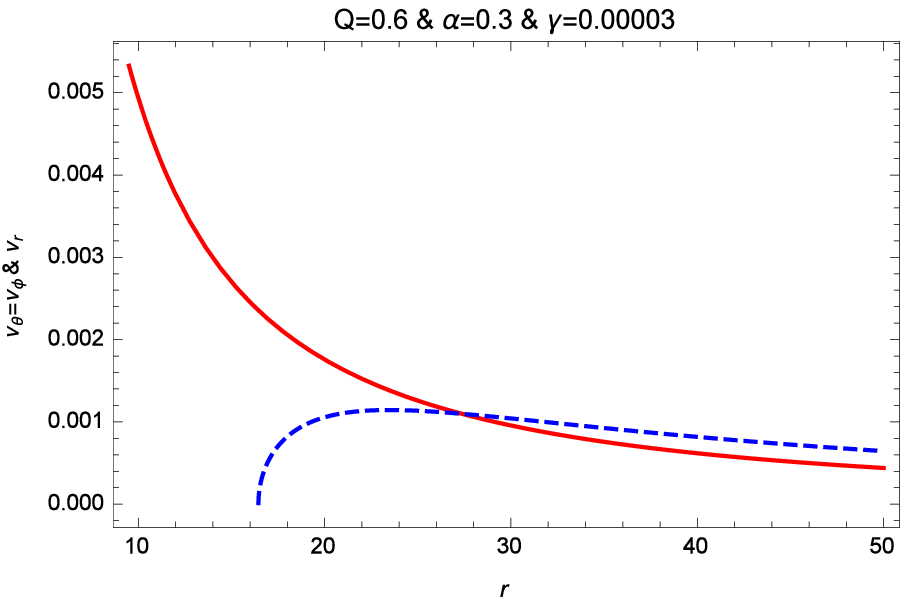, width=.32\linewidth,
height=2.02in}\epsfig{file=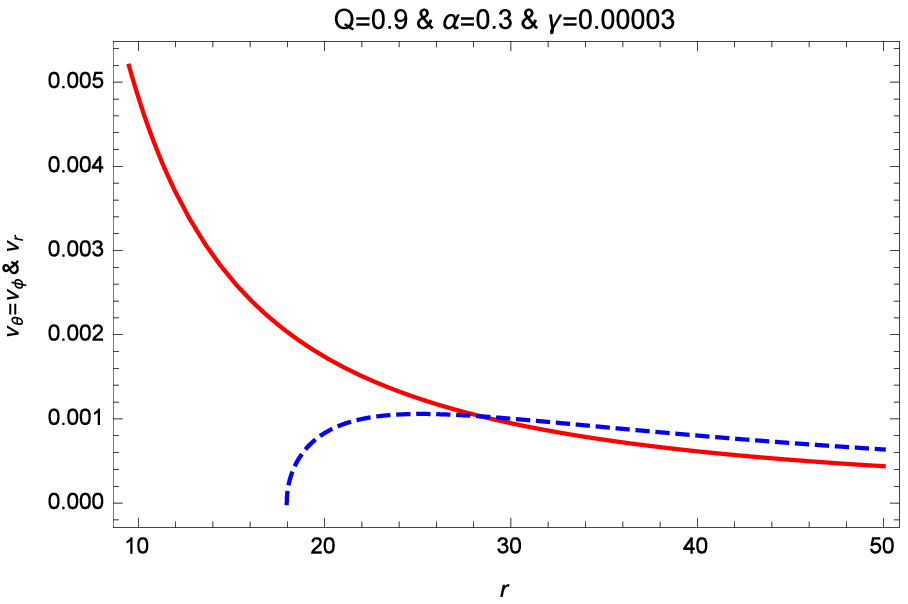, width=.32\linewidth,
height=2.02in}
\centering \epsfig{file=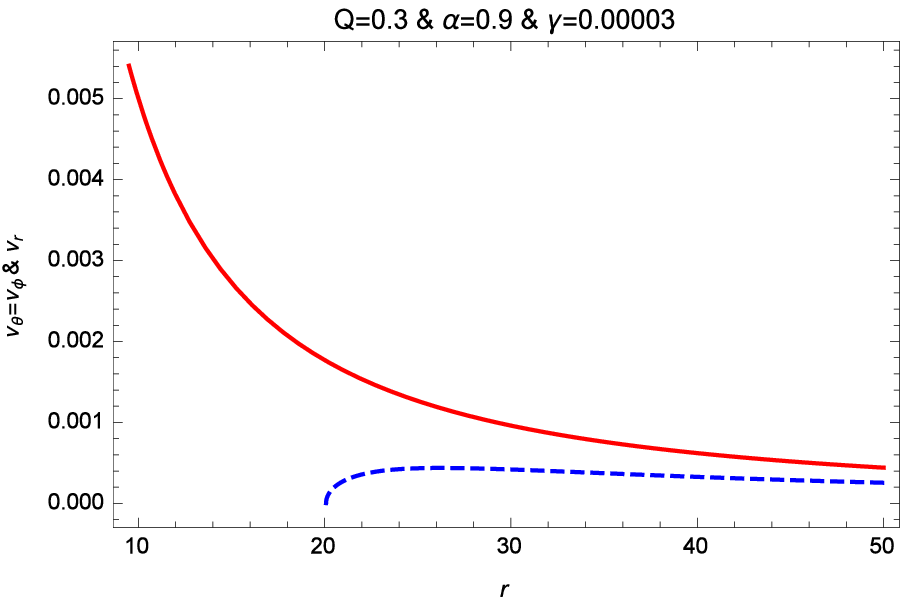, width=.32\linewidth,
height=2.02in}\epsfig{file=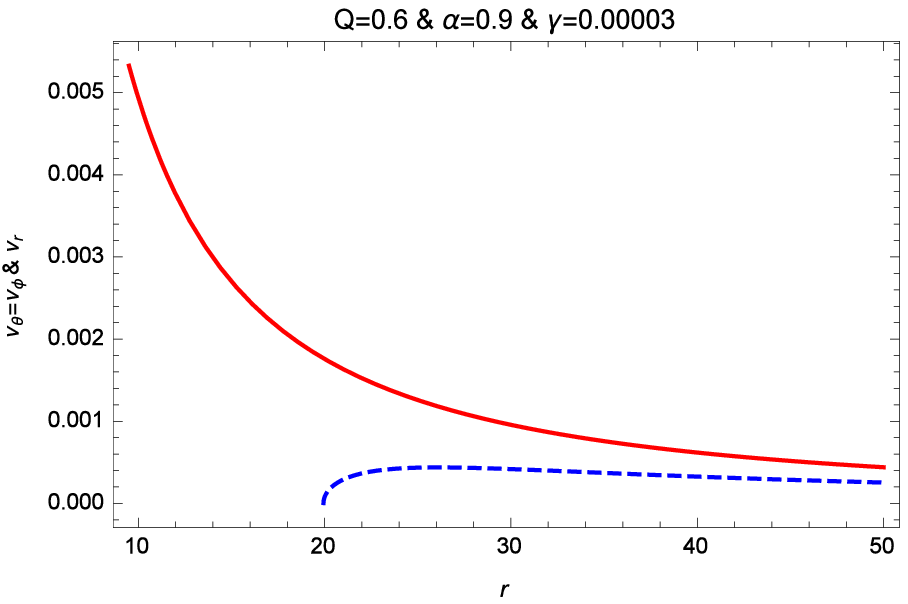, width=.32\linewidth,
height=2.02in}\epsfig{file=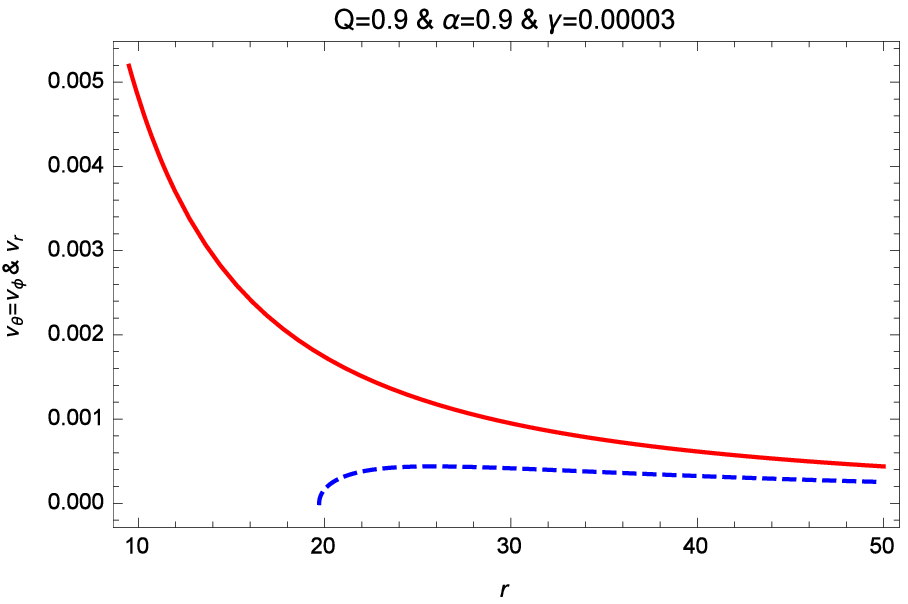, width=.32\linewidth,
height=2.02in}
\centering \epsfig{file=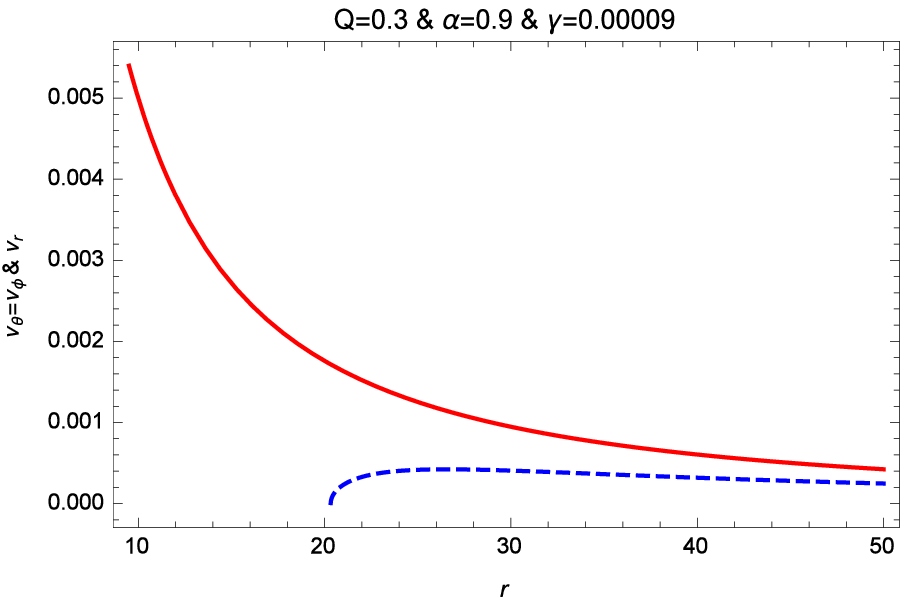, width=.32\linewidth,
height=2.02in}\epsfig{file=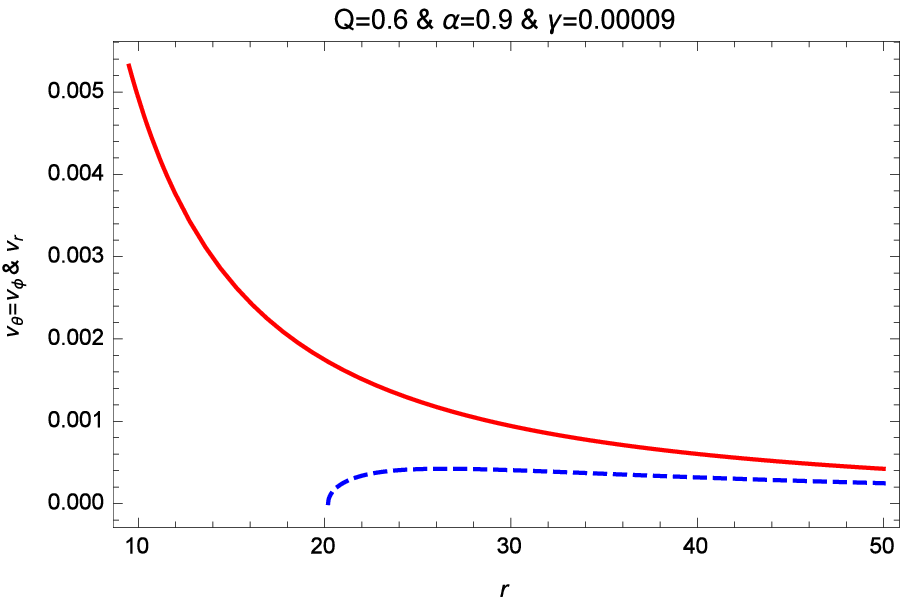, width=.32\linewidth,
height=2.02in}\epsfig{file=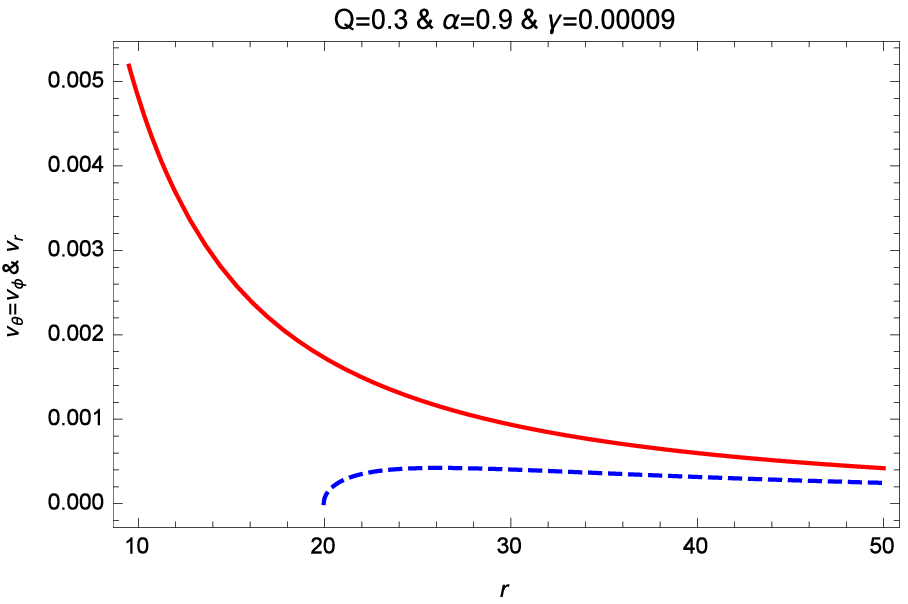, width=.32\linewidth,
height=2.02in}\caption{\label{fig12} Shows the behavior of $\nu_{\phi}=\nu_{\theta}$, and $\nu_{r}$.}
\end{figure}

\section{The red-blue shifts of the photons emitted by charged particles}
Here we present the red-blue shifts of the light coming from the orbiting charged test particles in the vicinity of the charged black hole in the background of quintessence and string clouds. In the spacetime under our consideration the angular momentum and photon energy are the conserved quantities, then the photon impact parameter can be obtained from the following formula \cite{33}
\begin{equation}\label{41}
b=\frac{L}{E}=\pm (-\frac{g_{\phi\phi}}{g_{tt}})^{1/2}.
\end{equation}
The frequency shift denoted by $z$ in the background of emission and detection of photons with circular geodesics $(\Pi^{r}=0)$ in equatorial motion $(\Pi^{\theta}=0)$ is calculated as
\begin{equation}\label{42}
1+z=\frac{\Pi^t_e-b_e \Pi^{\phi}_e}{\Pi^t_d-b_d\Pi^{\phi}_d}.
\end{equation}
The observational red shift is given in terms of the kinematic frequency shift, i.e., $z_{k}= z-z_{c}$, where $z_{c}$ is the shift at $b=0$, is given as
\begin{equation}\label{43}
1+z_c=\frac{\Pi^t_e}{\Pi^t_d}.
\end{equation}
The kinematic frequency shift can be realized as follows
\begin{equation}\label{44}
z_{kin}=\frac{\Pi^t_e\Pi^{\phi}_d b_d-\Pi^t_d\Pi^{\phi}_e b_e}{\Pi^t_d\left(\Pi^t_d-b_d\Pi^{\phi}_d\right)}.
\end{equation}
The detector is located far away from the black hole, therefore
\begin{equation}\label{45}
z=\Pi^{\phi}_eb_+\mid_{r_c}=\sqrt{-\frac{g_{\phi \phi}}{g_{tt}}}\left(\frac{L}{g_{\phi\phi}}-\frac{kA_{\phi}}{g_{\phi\phi}}\right)\mid_{r_c}.
\end{equation}
The redshift $z$, of the light emitted by charged particles in a stable circular orbit in the equatorial plane for the current analysis is calculated as
\begin{equation}\label{46}
z=\sqrt{\frac{r^2}{-\alpha-\frac{2 M}{r}+\frac{Q^2}{r^2}-\frac{\gamma }{r^{-1}}+1}}\times\frac{L}{r^2}\mid_{r_c}.
\end{equation}
\begin{figure}
\centering \epsfig{file=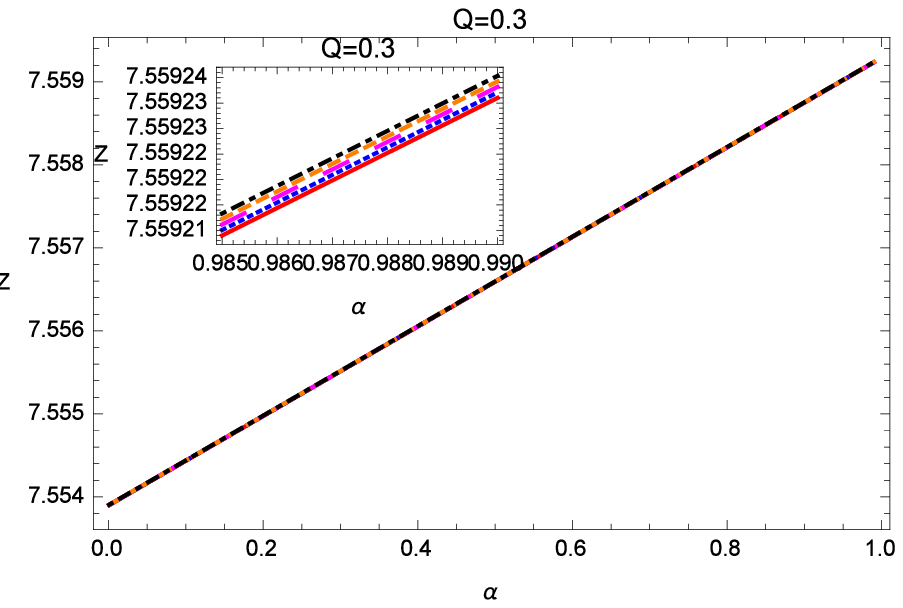, width=.32\linewidth,
height=2.02in}\epsfig{file=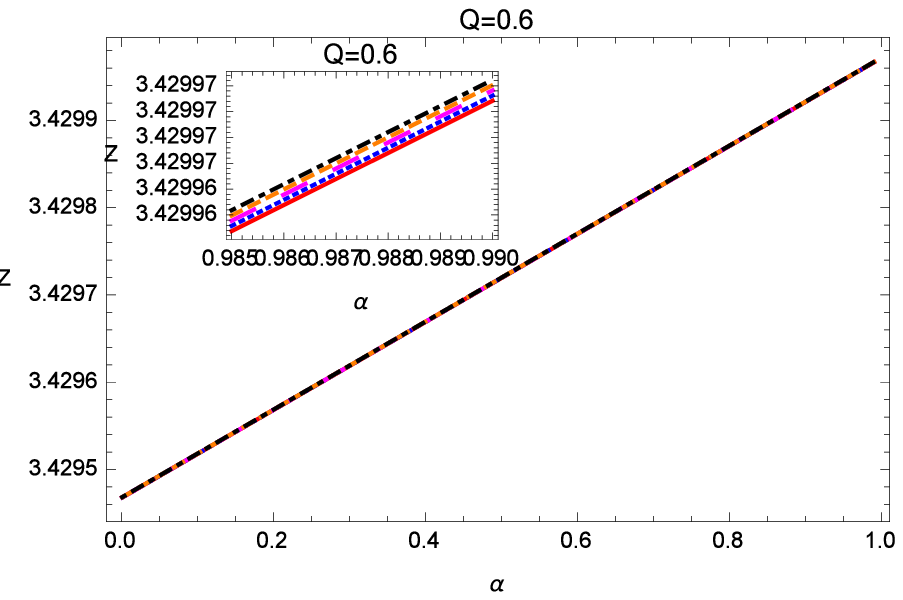, width=.32\linewidth,
height=2.02in}\epsfig{file=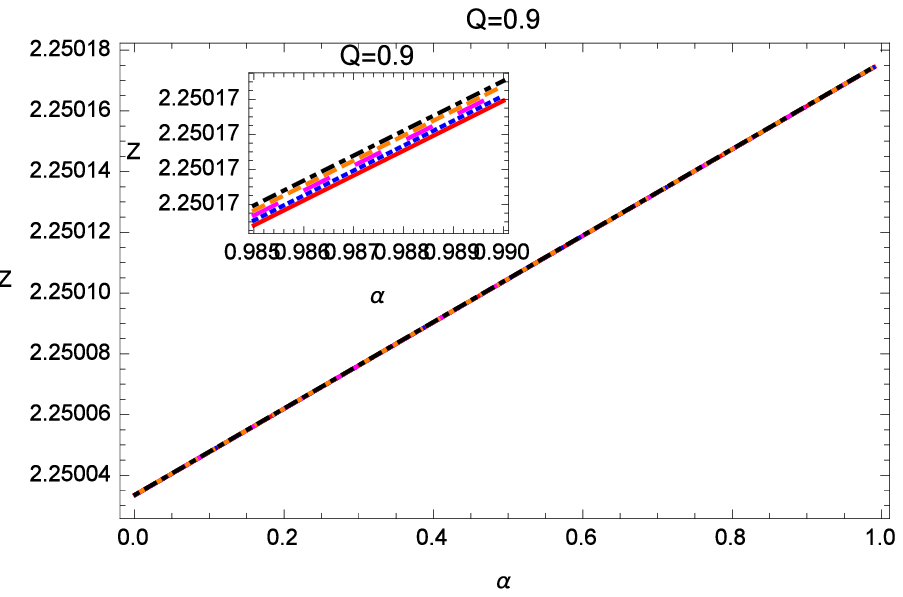, width=.32\linewidth,
height=2.02in}
\centering \epsfig{file=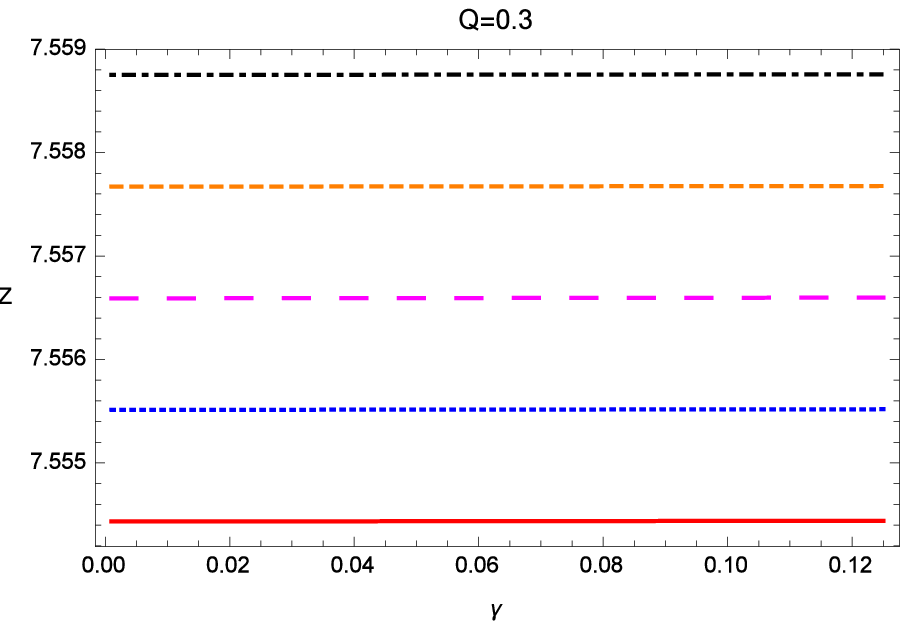, width=.32\linewidth,
height=2.02in}\epsfig{file=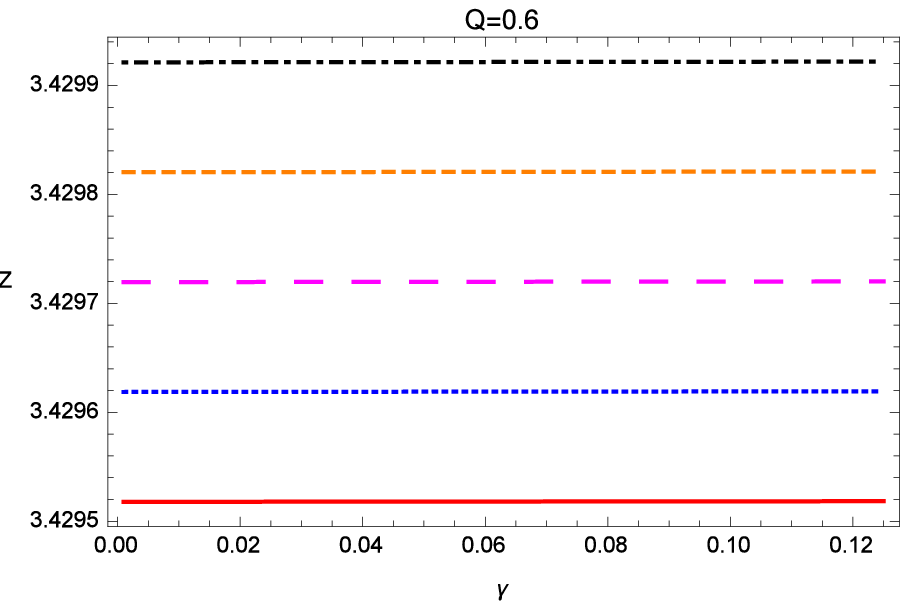, width=.32\linewidth,
height=2.02in}\epsfig{file=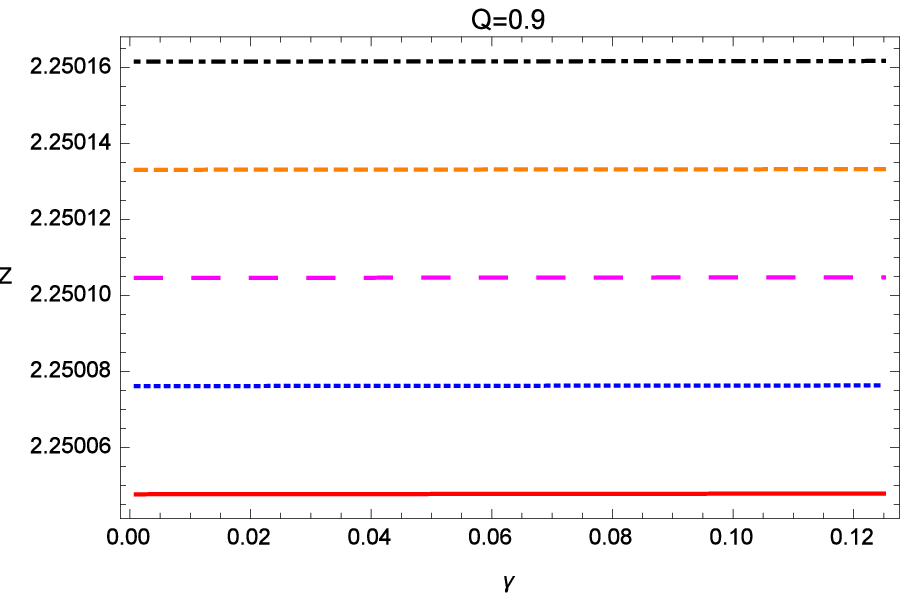, width=.32\linewidth,
height=2.02in}
\centering \epsfig{file=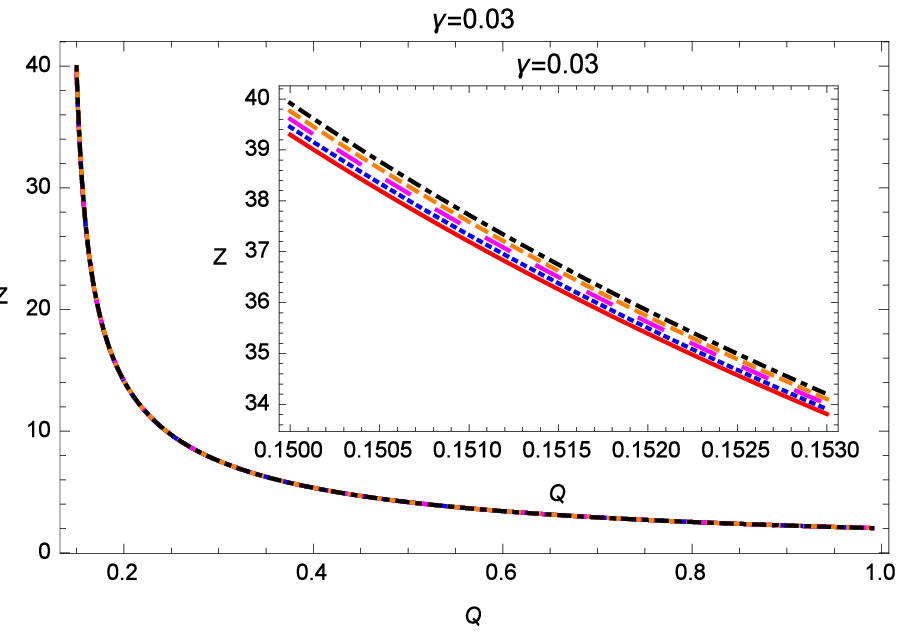, width=.32\linewidth,
height=2.02in}\epsfig{file=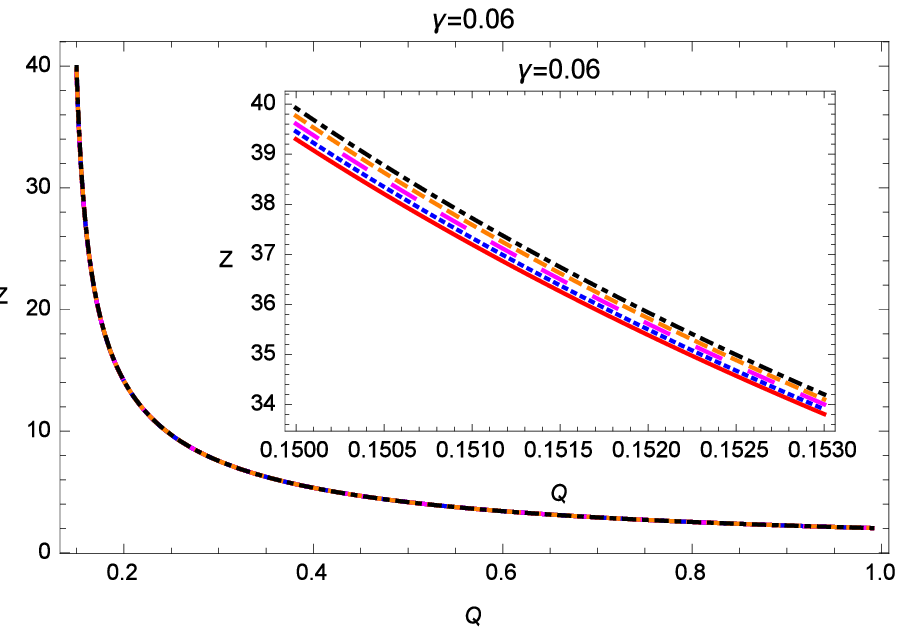, width=.32\linewidth,
height=2.02in}\epsfig{file=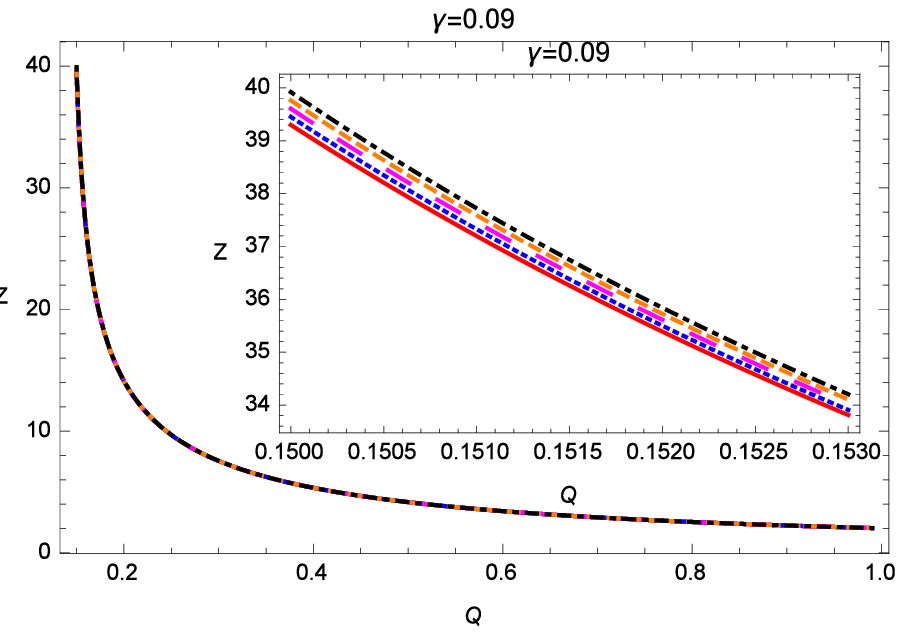, width=.32\linewidth,
height=2.02in}\caption{\label{fig13} Shows the behavior of red-shift parameter $z$. In the first row, we take different five values of quintessential parameter, i.e., $\gamma=0.0001(\textcolor{red}{\bigstar})$, $\gamma=0.0003(\textcolor{blue}{\bigstar})$, $\gamma=0.0005(\textcolor{magenta}{\bigstar})$, $\gamma=0.0007(\textcolor{orange}{\bigstar})$, and $\gamma=0.0009(\textcolor{black}{\bigstar})$. In the second row, we take different five values of cloud parameter, i.e., $\alpha=0.01(\textcolor{red}{\bigstar})$, $\alpha=0.03(\textcolor{blue}{\bigstar})$, $\alpha=0.05(\textcolor{magenta}{\bigstar})$, $\alpha=0.07(\textcolor{orange}{\bigstar})$, and $\alpha=0.09(\textcolor{black}{\bigstar})$. Again in the third row, we take different five values of cloud parameter, i.e., $\alpha=0.01(\textcolor{red}{\bigstar})$, $\alpha=0.03(\textcolor{blue}{\bigstar})$, $\alpha=0.05(\textcolor{magenta}{\bigstar})$, $\alpha=0.07(\textcolor{orange}{\bigstar})$, and $\alpha=0.09(\textcolor{black}{\bigstar})$.}
\end{figure}

From the plots presented in the Fig. (\ref{13}), it can be inferred that the parameter $z$ goes on increasing as the values of $\alpha$ increasing and $z$ goes on decreasing as the value of the charge $Q$ of the central source increases. While for the different values of the parameter $\gamma$ the behavior of $z$ remains same.

\section{Collision Energy in the Centre of Mass Frame}
In this section we discuss the center-of-mass energy of two radially falling particles from infinity in the close vicinity of the horizon of the charged black hole in the presence of clouds of strings and quintessence field. The centre-of-mass energy generally given by \cite{23}
\begin{equation}
E_{cm}=\sqrt{2\mathfrak{m}^2(1-g_{\mu \nu}\dot{x}^{\mu}\dot{x}^{\nu})}.
\end{equation}
The $\dot{t}$ is given in Eq. (\ref{16a}) and using the normalization condition we obtain the radial velocity of the particle is
\begin{equation}\label{34}
\dot{r}=-\frac{1}{\mathfrak{m}}\sqrt{\left(E-\frac{q Q}{r}\right)^2-\mathfrak{m}^2 f}.
\end{equation}
In the present study we get the following expression for the center-of-mass energy for tow charged colliding particles in radial motion, near the horizon of the black hole as
\begin{equation}\label{35}
E_{c\mathfrak{m}}=\mathfrak{m}\sqrt{2}\Big[1+\frac{1}{f \mathfrak{m}^2}\left(E_1-\frac{q_1 Q}{r}\right)\left(E_2-\frac{q_2 Q}{r}\right)\nonumber\\-\frac{1}{f \mathfrak{m}^2}\sqrt{\left(E_1-\frac{q_1 Q}{r}\right)^2-\mathfrak{m}^2 f}\sqrt{\left(E_2-\frac{q_2 Q}{r}\right)^2-\mathfrak{m}^2 f}\Big]^{\frac{1}{2}}.
\end{equation}
The centre-of-mass energy will go to infinity or will be arbitrarily large if $f\rightarrow 0$ and in such a situation the numerator in the expression for $E_{cm}$ becomes
\begin{equation}\label{36}
\left(E_1-\frac{q_1 Q}{r_+}\right)\left(E_2-\frac{q_2 Q}{r_+}\right)-\left(E_1-\frac{q_1 Q}{r_+}\right)\left(E_2-\frac{q_2 Q}{r_+}\right),
\end{equation}
where $r_{+}$ is the event horizon of the black hole. The quantity given in Eq. (\ref{37}) vanishes, therefore we are interested in the limiting value of the numerator in the expression for $E_{cm}$ and obtain the following expression for the centre-of-mass energy of two charged colliding particles in the vicinity of the horizon of the black hole
\begin{equation}\label{37}
E_{c\mathfrak{m}}=\mathfrak{m}\sqrt{2}\sqrt{1+\frac{1}{2}\left(\frac{q_2-\frac{E_2 r_+}{Q}}{q_1-\frac{E_1 r_+}{Q}}+\frac{q_1-\frac{E_1 r_+}{Q}}{q_2-\frac{E_2 r_+}{Q}}\right)},
\end{equation}
where $E_1$ and $E_2$ are the energies and $q_1$ and $q_2$ are the charges of the two colliding particles. It should be noted that this result obtained here for the charged black hole with quintessence and string clouds coincides with the already obtained results for other charged balk holes \cite{30,45}. From Eq. (\ref{37}) it can be seen that the centre-of-mass $E_{cm}$, which depends on the charge and mass of the particle and also on the charge of the central object, will go to infinity at the horizon of the black hole if one of the radially falling particles in collision has the critical value of the charge $q$
\begin{equation}\label{38}
q_c=\frac{Er_+}{Q},
\end{equation}
and the other one has some other value of the charge $q$. Since the centre-of-mass energy becomes arbitrarily large near the event horizon $r=r_{+}$, of the black hole, therefore, for the radially falling particle to reach the horizon, the square root in Eq. (\ref{37}) must have positive sign. As $f$ is less than zero for $r$ greater than  $r_+$, the square root in Eq. (\ref{37}) has positive sign arbitrarily close to the horizon of the black hole and hence it is possible for the radially falling  particles from infinity to reach the event horizon $r_+$ of the central black hole.

\section{Effective force}
The effective force which tells us whether the particle in the gravitational field is moving towards or away from the central source. The effective force on a test particles in the field of a gravitating source is given by \cite{40a}
\begin{equation}\label{39}
F=-\frac{1}{2}\frac{\partial V_{eff}}{\partial r}.
\end{equation}
For the spacetime taken in account in the present work this force can be calculated as
\begin{equation}\label{40}
F=-\frac{1}{2} \left(-\frac{2 L^2 \left(-\alpha-\frac{2 M}{r}+\frac{Q^2}{r^2}-\gamma  r+1\right)}{r^3}+\left(\frac{L^2}{r^2}+1\right) \left(-\gamma +\frac{2 M}{r^2}-\frac{2 Q^2}{r^3}\right)-\frac{\gamma Q}{r^2}\right).
\end{equation}
The effective force $F$ calculated here depends on the black hole parameters along with the string cloud and quintessence parameters. It also depends on the radial distance and on the angular momentum of the test particle moving in the spacetime field of the black hole. For the better understanding of the effective force we give the graphical representation of it in the Fig. (\ref{14}). From the plots one can see that for the changing values of $\alpha$ and for fixed values of $Q$ and $\gamma$ the force $F$ changes its behavior from attractive to repulsive as $\alpha$ increases, for a very short rang of the radial coordinate $r$ in the close vicinity of the central object. Then it becomes zero as one moves away from the central source. In the same plot it is evident that for comparatively smaller values of $\alpha$ this force remains attractive throughout. In the case of fixed $Q$ and $\alpha$, and varying $\gamma$ the force $F$ changes from repulsive to attractive with $r$ as the values of $\gamma$ decrease. For the fixed $\alpha$ and $\gamma$ the behavior of the force $F$ changes from attractive to neutral irrespective of the value of the charge $Q$ with the coordinate $r$.
\begin{figure}
\centering \epsfig{file=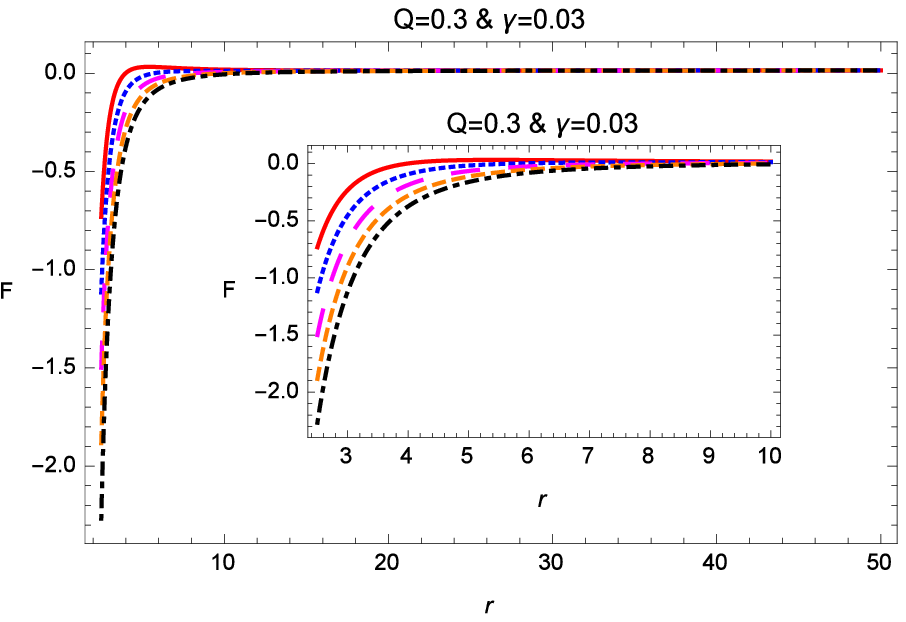, width=.32\linewidth,
height=2.02in}\epsfig{file=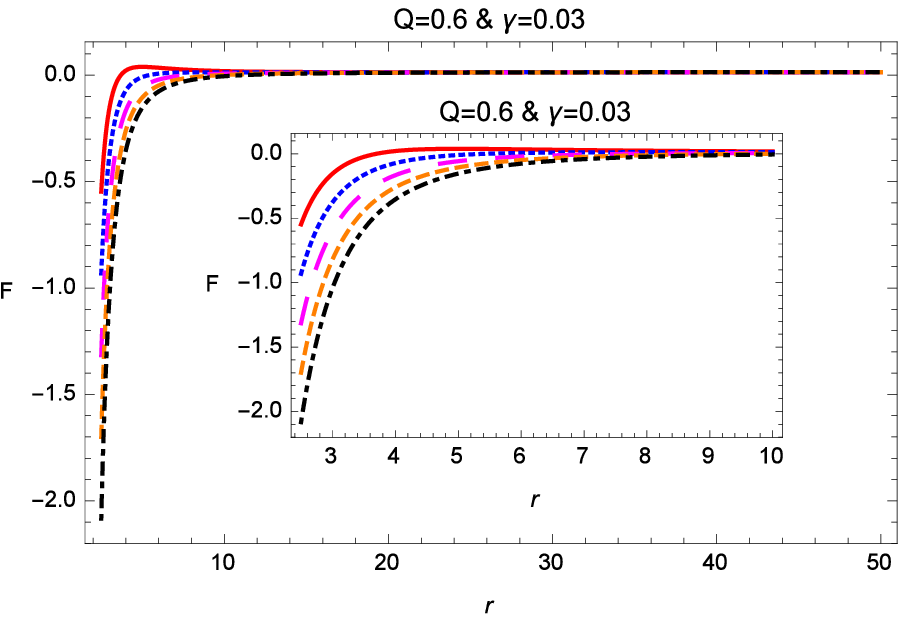, width=.32\linewidth,
height=2.02in}\epsfig{file=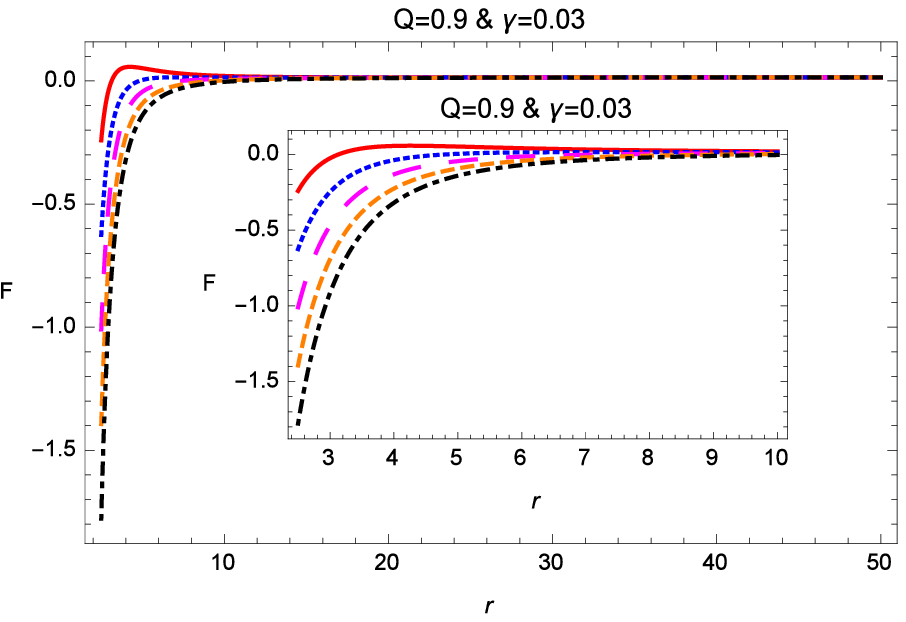, width=.32\linewidth,
height=2.02in}
\centering \epsfig{file=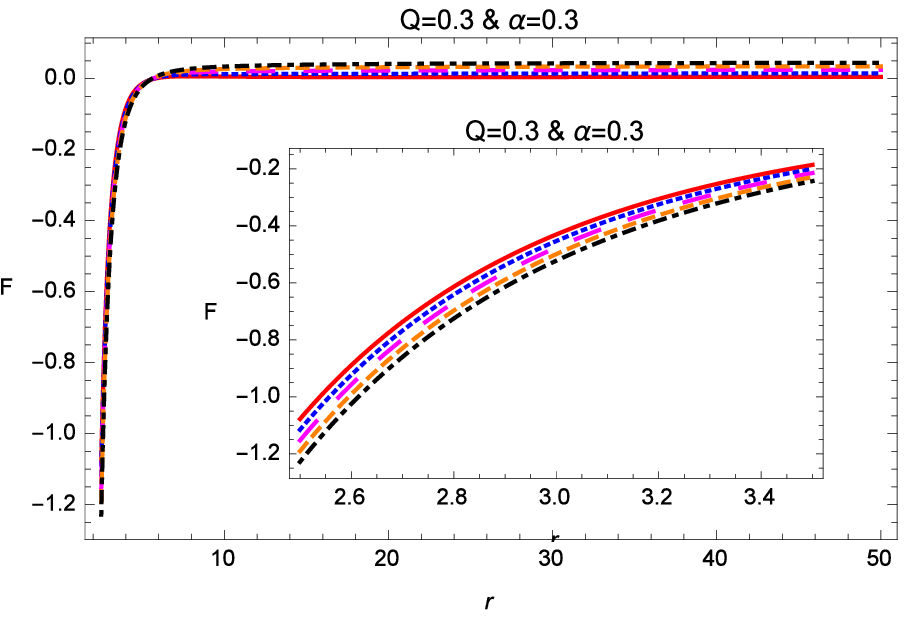, width=.32\linewidth,
height=2.02in}\epsfig{file=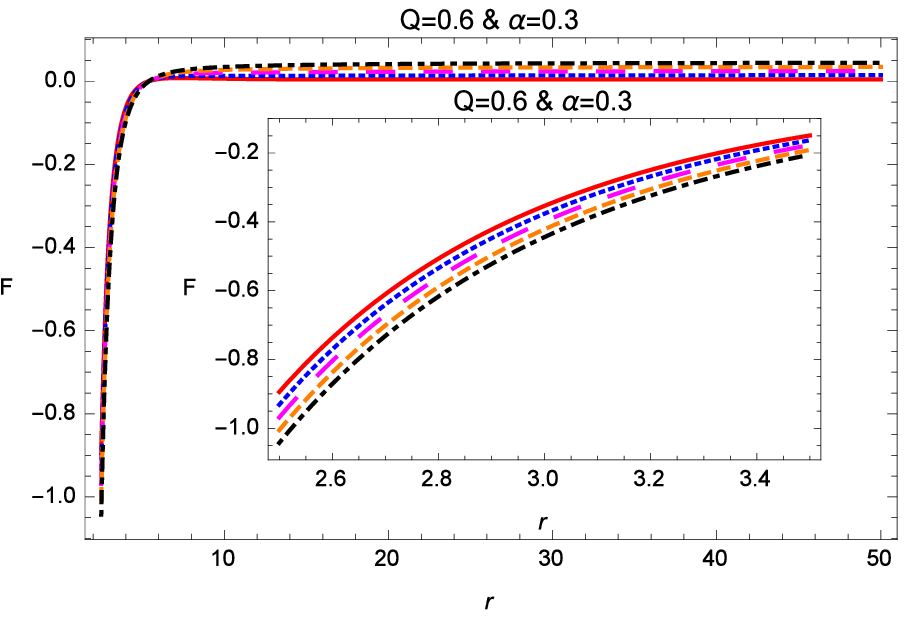, width=.32\linewidth,
height=2.02in}\epsfig{file=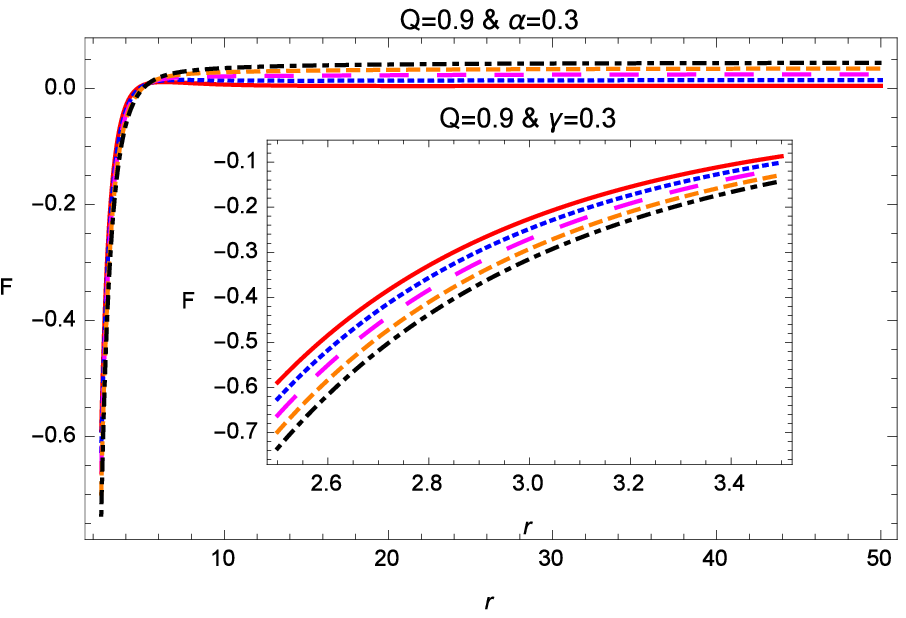, width=.32\linewidth,
height=2.02in}
\centering \epsfig{file=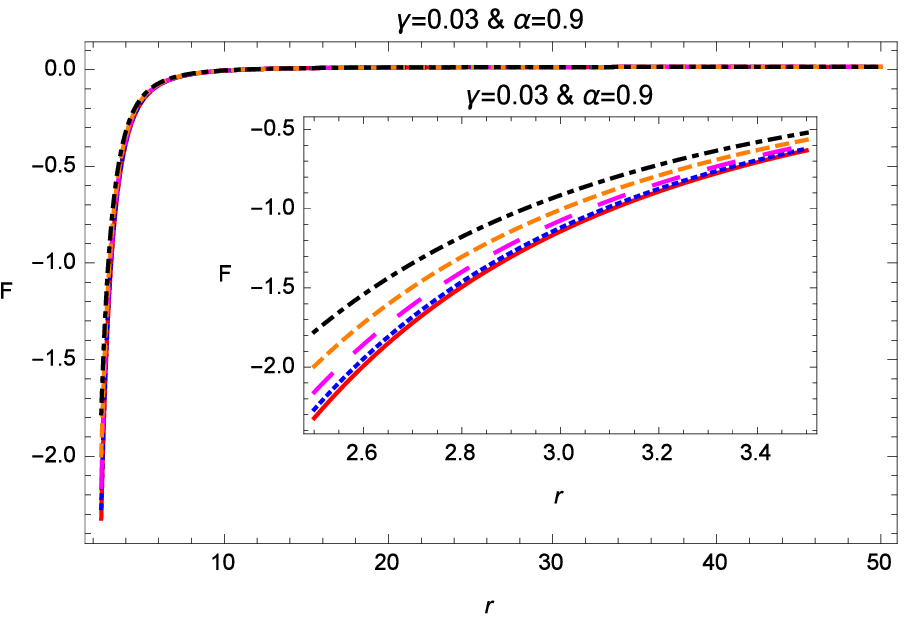, width=.32\linewidth,
height=2.02in}\epsfig{file=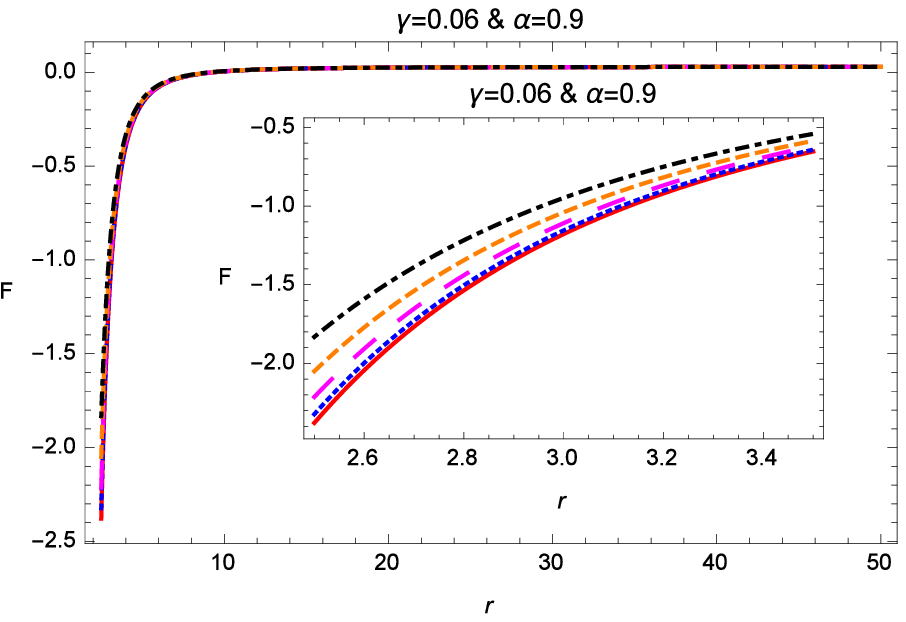, width=.32\linewidth,
height=2.02in}\epsfig{file=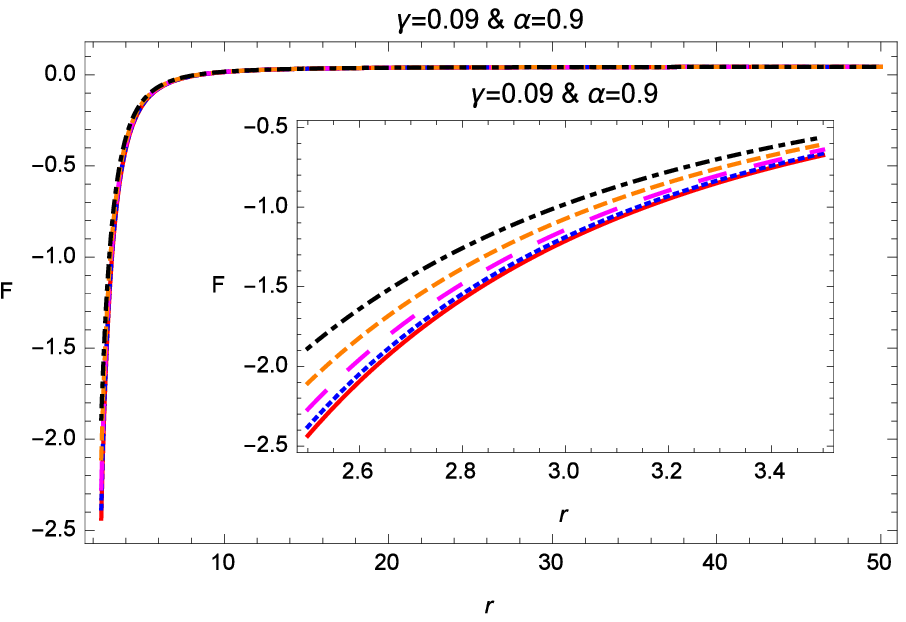, width=.32\linewidth,
height=2.02in}\caption{\label{fig14} Shows the behavior of effective force $F$.  In the first row, we take different five values of cloud parameter, i.e., $\alpha=0.01(\textcolor{red}{\bigstar})$, $\alpha=0.03(\textcolor{blue}{\bigstar})$, $\alpha=0.05(\textcolor{magenta}{\bigstar})$, $\alpha=0.07(\textcolor{orange}{\bigstar})$, and $\alpha=0.09(\textcolor{black}{\bigstar})$. In the second row, we take different five values of quintessential parameter, i.e., $\gamma=0.0001(\textcolor{red}{\bigstar})$, $\gamma=0.0003(\textcolor{blue}{\bigstar})$, $\gamma=0.0005(\textcolor{magenta}{\bigstar})$, $\gamma=0.0007(\textcolor{orange}{\bigstar})$, and $\gamma=0.0009(\textcolor{black}{\bigstar})$. In the third row, we take different five values of charge parameter, i.e., $Q=0.1(\textcolor{red}{\bigstar})$, $Q=0.3(\textcolor{blue}{\bigstar})$, $Q=0.5(\textcolor{magenta}{\bigstar})$, $Q=0.7(\textcolor{orange}{\bigstar})$, and $Q=0.9(\textcolor{black}{\bigstar})$.}
\end{figure}

\section{Summary}
In this article we have studied the circular motion of charged test particles in the vicinity of charged black hole with string clouds and quintessence field. We have found that the stable circular orbits exist for some very small values of the string cloud parameter $\alpha$ and the quintessence parameter $\gamma$, i.e. $0<\alpha<<1$ and $0<\gamma<<1$. Further we have obtained the radii of the circular orbits for  both the timelike and null particles which varies with the parameters $\alpha$ and $\gamma$ and charge of the black hole $Q$. We noticed that the charged test particle can more easily escape to infinity from the gravitational field of the central black hole as the values of the parameters $\alpha$ and $\gamma$ increase. While the effect of the charge $Q$ of the central black hole on the the effective potential is just opposite to that of the $\alpha$ and $\gamma$. It is concluded that in the comparison of pure Reissner-Nordstrom black hole it would be easy for test particle to leave the gravitational field of the central object if the string clouds and quintessence is present there i.e. the parameters $\alpha$ and $\gamma$ act as repulsive gravity.\\

Next we have studied fundamental frequencies of QPOs of test charged particles near the stable circular orbits in the spacetime geometry of the charged black hole with string clouds and quintessence field. We have found that with increase in the values of $Q$ the stable circular orbits get away from the central object, therefore lower epicyclic frequencies can be observed away from the black hole horizon. The stable circular orbits move towards the central source as $\alpha$ and $\gamma$ go on increasing and thus the epicyclic frequencies can be observed close to the horizon of the black hole and would be higher in comparison of the epicyclic frequencies observed for the pure Reissner-Nordstrom black hole i.e. in the absence of the quintessence field.\\

Another important astrophysical phenomenon which we have investigated here is about the red-blue shifts of the photons coming from the charged test particles moving in the stable orbits around the charged black hole with string clouds and quintessence scalar field. In this study we have found that the redshift parameter $z$ of the photons emitted by the charged test  particles moving in the stable circular orbits around the central source increases with increase
in the values of the parameter $\alpha$ and decreases with increase in the values of the charge $Q$ of the black hole. While the parameter $z$ remains same irrespective of the values of the parameter $\gamma$.\\

Then We have discussed the centre-of-mass energy in the vicinity of the event horizon of the black hole, of the radially falling charged particles from infinity. We have found that the centre-of-mass energy $E_{cm}$, would be as high as desired at the event horizon of the black hole if one of the colliding particles has the critical value of the charge $q_c$ given by Eq. (\ref{43}) and the other particle acquires any other value of the charge.\\

Finally, we have discussed the effective force acting on the test particles moving in the vicinity of the charged black hole in the presence of the quintessence field and string clouds. We have observed that the behavior of the effective force is sensible to the values of the parameters $\alpha$, $\gamma$ and $Q$. The behavior of the effective force for the parameters $\alpha$, $\gamma$ and $Q$ is well explained with help of graphs presented in Fig. (\ref{14}). It is observed that the magnitude of the effective force on the charged test particles decrease as the values of the parameters $\gamma$ and $\alpha$ increase and it increases with the increasing values of the charge $Q$.\\

\end{document}